\documentclass[12pt,a4paper]{article}
\pdfoutput=1
\usepackage{jheppub}
\usepackage{amsfonts}
\usepackage{bbm}
\usepackage{multirow}
\usepackage{enumitem}   
\usepackage{stmaryrd}
\usepackage{placeins}
\usepackage[none]{hyphenat}
\usepackage{makecell}
\setcellgapes{3pt}

\usepackage[dvipsnames]{xcolor}

\usepackage{tikz-cd}   
\usepackage{tikz}
\usepackage{pgfplots}
\usetikzlibrary{decorations.markings,decorations.pathmorphing,calc,intersections,through,backgrounds,shapes,snakes,shapes.geometric,knots}
\usepgfplotslibrary{fillbetween}

\def \Z{\mathbb{Z}}
\def \R{\mathbb{R}}
\def \C{\mathbb{C}}
\def \Q{\mathbb{Q}}

\def \sch{{\mathfrak{ch}_2}}
\newcommand{\crep}[1]{{\llbracket #1 \rrbracket}}

\def \CL{\mathcal{L}}

\def \CH{\mathcal{H}}

\def \pt{{\text{pt}}}
\def \lk{{\ell k}}

\def \Spin{{\mathrm{Spin}}}

\def \MCG{{\mathrm{MCG}}}

\def \M{\mathcal{M}}
\def \Tor{{\mathrm{Tor}\,}}
\def \Hom{{\mathrm{Hom}}}

\def \Arf{{\mathrm{Arf}}}

\def \r{{\mathrm{R}}}
\def \ns{{\mathrm{NS}}}

\def \c{{\mathrm{c}}}

\def \i{\iota}
\def \id{{\mathrm{id}}}
\def \tr{\tau_\varnothing}
\def \im{\mathrm{Im}}

\def \Spingf{\mathrm{Spin}\text{-}G^f}

\DeclareMathSymbol{\mh}{\mathord}{operators}{`\-}

\title{Anomalies of fermionic CFTs via cobordism and bootstrap}
\author[1, 2]{Andrea Grigoletto}

\affiliation[1]{SISSA, Via Bonomea 265, Trieste 34136, Italy }
\affiliation[2]{INFN, Sezione di Trieste, Via Valerio 2, 34127 Trieste, Italy}

\abstract{We study constraints on the space of $d=2$ fermionic CFTs as a function of non-perturbative anomalies exhibited under a fermionic discrete symmetry group $G^f$, focusing our attention also on cases where $G^f$ is non-abelian or presents a non-trivial twist of the $\mathbb{Z}^f_2$ subgroup. For the cases we selected, among our results we find that modular bootstrap consistency bounds predict the presence of relevant/marginal operators only for some groups and anomalies. From this point of view, the appearance in the analysis of several kinks around irrelevant operators with $\Delta>2$ means that for fermionic systems with increasingly larger symmetry groups modular bootstrap is able to give less constraining bounds than its bosonic counterpart. 
Within our analysis we show how the anomaly constraints on fermionic CFTs can be effectively recovered from the structure of the abelian subgroups of $G^f$. Finally, we extend the previous surgery description of bordism invariants that describe $3d$ abelian spin-TQFTs, in order to include the case of theories with $\Spin\text{-}\Z^f_{2^{l+1}}$ structures.}

\keywords{TQFT, SPT, Anomalies, Fermionic CFT, Bootstrap}

\begin{document}
\tikzset{->-/.style={decoration={
  markings,
  mark=at position .5 with {\arrow{>}}},postaction={decorate}}}
  
\maketitle

\vspace{1cm}

\section{Introduction}

't Hooft anomalies are one of the important concepts that allow us to explore QFTs from a non-perturbative point of view, giving us knowledge on physics behaviour that would not be accessible otherwise. The main feature that makes them so attractive is their invariance under RG flow. This means that flowing from a short-distance quantum system characterized by some symmetry $G_{\mathrm{UV}}$ and anomaly $\mathcal{A}_{\mathrm{UV}}$ to the IR description of its degrees of freedom, associated to a (often enhanced) symmetry $G_{\mathrm{IR}} \supseteq G_{\mathrm{UV}}$, the value of the anomaly is the same, $\mathcal{A}_{\mathrm{UV}}=\mathcal{A}_{\mathrm{IR}}$. Furthermore, their invariance under all possible continuous deformations means they are perfect tools to analyze in full generality many classes of theories and be able to infer information on the dynamical behaviour of physical systems simply based on a finite subset of their characterizing data.

For instance, it is well-known that an anomalous theory flowing to a gapped system in the IR must either present spontaneous symmetry breaking or be described by a TQFT that properly matches the value of $\mathcal{A}_{\mathrm{UV}}$.
In this work we are going to further analyze the consequences of a second possibility, namely when the theory flows to a gapless description in the IR, focusing on the particular case of $2$-dimensional fermionic systems with a global discrete symmetry $G^f$, complementing the analysis of the previous works \cite{Lin:2019kpn, Lin:2021udi, Benjamin:2020zbs,Grigoletto:2021zyv}. 

In this regard, it is interesting to infer the existence of relevant or marginal $G_{\mathrm{UV}}$-invariant operators for any candidate low-energy CFT. Indeed, the absence of such operators would mean that the CFT in analysis is a stable fixed point under RG flow and thus a satisfactory description of low-energy degrees of freedom of any system that approaches it under RG flow. In the particular example of $d=2$, we know a vast class of systems with fractional values of the central charge $c$ that are potentially interesting from this point of view, many of which even exhibit a dependence on a spin structure, namely the WZW models. However, understanding how relevant/marginal perturbations exist in general terms is a difficult task, see for example \cite{Ohmori:2018qza}. Therefore an alternative approach that quickly offers a way to gather this kind of information is appreciable. 
One way to try to answer this question is via modular bootstrap. The consistency constraints that modular transformations impose on a CFT have indeed already been proved to be effective in the bosonic case \cite{Collier:2016cls,Lin:2019kpn,Lin:2021udi,Benjamin:2016fhe, Friedan_2013,Hellerman:2010qd,Hellerman:2009bu,Qualls:2013eha,Keller:2012mr,Cho:2017fzo,Bae:2017kcl,Dyer:2017rul,Anous:2018hjh,Benjamin:2019stq,Pal:2020wwd}. For this reason, it is worth to analyze what else can we learn from it in the case of fermionic systems, focusing on the dependency of the bounds on the presence of \emph{non-perturbative} anomalies. This is precisely the goal of the present work, namely we are going to compute a set of upper bounds on the lightest symmetry-preserving scalar operators for various sets of fermionic theories with different discrete symmetry groups $G^f$ and anomalies.

Let us mention here that fermionic theories in principle can always be studied in terms of bosonic systems by means of bosonization and fermionization maps, see \cite{Gaiotto:2015zta,Novak:2015ela,Bhardwaj:2016clt,Thorngren:2018bhj,Karch:2019lnn,Fukusumi:2021zme,Thorngren:2021yso}. However, the resulting bosonic theory that emerges after $\Z^f_2$ gauging often exhibits non-trivial traits, like the presence of non-invertible symmetries \cite{Ji_2020}, and its relation with the anomaly of the original fermionic description might be subtle. Therefore studying fermionic theories without relying on their bosonization is more natural from this point of view, which indeed will be our approach.

While the analysis of perturbative anomalies is a standard textbook topic, our understanding of global anomalies has been greatly improved in recent years thanks to the development of a mathematical framework that allows to describe them in a more rigorous and general manner \cite{Freed:2016rqq,Yonekura:2018ufj,Witten:2019bou,Kapustin:2014dxa,Kapustin:2014tfa, Guo:2017xex,Guo:2018vij,Wan:2018bns,Wan:2019oax,Freed:2012hx,Freed:2014eja,Freed:2014iua,Monnier:2019ytc}. 
It is well known that the description of the non-perturbative case is connected to the non-invariance under large gauge transformations or diffeomorphisms of the partition function for the theory considered. Moreover, our understanding of such phases is given in terms of so called \emph{mapping class tori} in one dimension higher, to which the anomalous phases are properly associated by an invertible topological quantum field theory (iTQFT) \cite{Witten:1982fp,Witten:1983tw,Witten:1985xe,Witten:2015aba}.
However, the relation of global anomalies in $d$ dimensions with a $d+1$ dimensional TQFT can be made more explicit thanks to the concept of \emph{anomaly inflow}, namely the idea that an anomalous theory in $d$ dimension describes the boundary of an equally anomalous bulk $d+1$ dimensional theory, such that the two systems combine in a $d+1$ dimensional system with a restored non-anomalous symmetry \cite{Callan:1984sa}.
For our needs, this picture tell us that the groups of non-perturbative anomalies are captured by deformation classes of iTQFTs, which in turn are classified by cobordism groups, thus offering a map between such groups. This correspondence is what we are going to employ in our work.
On a side note, we mention that if we were to consider bosonic systems, the description of such iTQFTs could have been equivalently stated in terms of group cohomology \cite{Dijkgraaf:1989pz,Freed:1991bn}, which is usually more amenable. Instead, as it will be our case, the total cobordism structure appears explicitly and necessary when the fermionic nature of the system under consideration is manifest.


The rest of the paper is organized as follow. In Section \ref{sec:framework} we are going to recall the generic framework that allows to analyze non-perturbative anomalies of QFTs in any dimension. We will also review the picture in the particular case of $d=2$, paying a special attention to how it applies to the case of $2$-tori and recalling the results known for systems with abelian symmetries.  In Section \ref{sec:twist} we are instead going to focus on twisted symmetry groups, presenting the associated bordism invariants. In Section \ref{sec:mod-transf} we will apply the techniques aforementioned to determine the modular transformation of fermionic CFTs in terms of their anomaly, finding explicit results for some particular illustrative examples. In Section \ref{sec:mod-boot} we will briefly recall the numerical method used to extract data on the lightest symmetry-preserving scalar operators via modular bootstrap. After this, we are going to present the numerical analysis for several systems with different kinds of symmetries and anomalies, discussing the results that have been found. We anticipate that for the cases we selected only for the symmetry group $G^f=\Z^f_2\times\Z_2\times \Z_2$ we have been able to find an interval of central charges where the existence of symmetry-preserving relevant and marginal operators is implied. However, the bounds are still interesting, as the existence of several kinks for most of the symmetry groups in the non-anomalous case points to the presence of theories which saturate the bounds in terms of irrelevant operators and which might be worth to analyze in future investigations.

\subsubsection*{Notation} For ease of the reader, here we summarize few notation conventions that we are going to adopt in the rest of the work.
For a group homomorphism
\begin{equation*}
    f:G\rightarrow G'
\end{equation*}
we will denote by $f^*$, $\hat{f}_*$ and $\hat{f}^*$ the induced pullbacks and push-forward
\begin{align*}
    &f^*:RO(G')\rightarrow RO(G),\\
    &\hat{f}_*:\Omega^\Spin_d(BG)\rightarrow\Omega^\Spin_d(BG'),\\
    &\hat{f}^*:\Hom(\Omega^\Spin_d(BG'),U(1))\rightarrow\Hom(\Omega^\Spin_d(BG),U(1)).
\end{align*}
Moreover, we will denote as $\tr$ both the trivial maps between rings and between groups.

\section{Global anomalies and cobordism invariants}
\label{sec:framework}
 
In this section we are going to briefly review the description of non-perturbative anomalies of $d$-dimensional theories in terms of $d+1$ dimensional invertible TQFTs, focusing on the case of $d=2$ fermionic systems and discussing the tools at our disposal for their actual computation. For further details see for example \cite{Witten:2019bou, Yonekura:2018ufj, Freed:2016rqq,Grigoletto:2021zyv}.

\subsection{The general framework}

The presence of a global anomaly in a theory can be determined by the transformation properties of its partition function. More specifically, if one couples a theory on a $d$ manifold $X$ with a set of background fields $\{a_i\}$ such that under a large transformation\footnote{By large transformation we mean a large gauge transformation or a large diffeomorphism or a combination of both. Here $\{a_i\}$ includes both gauge fields and the metric $g$ on $X$.} $\psi$ its partition function gains an additional phase
\begin{equation}
    Z(X,\{a_i\}) \mapsto e^{i\alpha(\psi,\{a_i\})} Z(X,\{a_i\}),
    \label{eq:anom}
\end{equation}
then the theory is said to be anomalous if such phase cannot be cancelled by a local counterterm. This means that the partition function generally describes a non-trivial line bundle over the parameters space of the theory. In other terms, one can formalize this idea by looking at the partition function as a map that associates to a manifold with the proper tangential structure an element of a $1$-dimensional complex vector space and to a large transformation an automorphism of it. This is indeed what describes the idea of an invertible TQFT from the functorial point of view, which we are now going to make precise. 

From now on we are going to assume that the $d$ dimensional anomalous theory is fermionic and characterized by a global discrete symmetry group $G^f$. The fermionic parity $(-1)^F$ then will generate the naturally embedded $\Z^f_2\subset G^f$, while the quotient $G:=G^f/\Z^f_2$ will instead denote the bosonic symmetry group. Note that although one often assumes that $G^f$ is the trivial extension $G^f=\Z^f_2 \times G$, generally this in not case and we are not going to assume so if not explicitly stated. Instead, for simplicity we will assume that the theory does not present time-reversal symmetry, so that the symmetry group of spacetime is simply $SO(d)$. With these requirements our anomalous theories needs to be defined on $d$-manifolds $X$ equipped with an $H_d$ structure, where\footnote{Actually, one should more properly define $H_d:=\widetilde{\mathrm{GL}}_+(d,\R)\times_{\Z^f_2}G^f$, where $\widetilde{\mathrm{GL}}_+(d,\R)$ is the double cover of $\mathrm{GL}_+(d,\R)$ and deformation retracts to $\Spin(d)$. The requirement of $H_d$ to be non-compact is due to the requirement for the associated $d+1$ dimensional theory that will describe the anomaly to be topological. However, for ease of notation in the following we will use \eqref{Hd-str} even if we are actually assuming the structure to be its non-compact version, often saying that $X$ admits a $\Spin\text{-}G^f$ structure (or a $\Spin_G$ structure when $G^f=\Z^f_2 \times G$).}
\begin{equation}
    H_d := \Spin(d)\times_{\Z^f_2}G^f.
    \label{Hd-str}
\end{equation}
A manifold $X$ equipped with an $H_d$ structure actually denotes a triple of elements $(X,P,\theta)$, where $P$ is a principal $H_d$ bundle on $X$. Given the forgetful canonical homomorphism
\begin{equation}
    \rho_d : H_d \rightarrow SO(d),
\end{equation}
one defines the bundle map $\theta:P\rightarrow F_+X$ into the oriented orthonormal frame bundle $F_+X$ of $X$ that satisfies the property 
\begin{equation}
    \theta(p\cdot h) = \theta(p)\cdot \rho_d(h),\quad  \forall\, p \in P,\, h\in H_d,
\end{equation}
and that makes the following diagram commutative 
\begin{equation}
\begin{tikzcd}
    P \ar[r,"\theta"] \ar[rd,"\pi_P",swap] & F_+X \ar[d,"\pi_{F_+X}"]\\ & X
\end{tikzcd}.
\end{equation}
The set of $H_d$ manifolds can then be identified with the elements of a category $\mathrm{Bord}_{(d,d+1)}(H)$, while any morphism $Y:(X,P,\theta)\rightarrow (X',P',\theta')$ is a $d+1$ dimensional manifold $Y$ with a proper $H_{d+1}$ structure such that $\partial Y= X \sqcup \overline{X}'$. Here $\overline{X}'$ denotes the manifold $X'$ equipped with the opposite $H_d$ structure, for more details see \cite{Yonekura:2018ufj}. This category is actually the categorification of the abelian group $\Omega^H_d$, defined as the group of equivalence classes of $H_d$ $d$-manifolds. Here two manifolds $X$ and $X'$ will lie in the same equivalence class if the set of morphisms $\mathrm{hom}(X,X')$ in $\mathrm{Bord}_{(d,d+1)}(H)$ is not empty. The group structure on $\Omega^H_d$ is instead given by the disjoint union of manifolds. Note that this induces a natural monoid structure on $\mathrm{Bord}_{(d,d+1)}(H)$, which physically will be associated to the stacking of theories.

With these premises, a TQFT is a functor 
\begin{equation}
    \mathrm{TQFT}:\mathrm{Bord}_{(d,d+1)}(H)\rightarrow \mathrm{Vect}_\C
\end{equation} 
that associates to an element $(X,P,\theta)$ a complex vector space $\CH(X,P,\theta)$ and to a morphism $Y$ an isomorphism of vector spaces $\varphi_Y:\CH(X,P,\theta)\rightarrow \CH(X',P',\theta')$. 
In our work we will be interested to invertible TQFTs, which associate to each $d$-manifold a $1$-dimensional vector space. This will be the space over which the partition function of the $d$-dimensional anomalous theory lies. 
Consider now the set of equivalence classes $a_i$ of isomorphic $\Spin\textbf{-}G^f$ structures over $X$. From now on we will use $X_a$ and $\CH_a$ as shorthands respectively for the families of triples $(X,P_a,\theta_a)$ and vector spaces $\CH(X,P_a,\theta_a)$. Upon fixing a basis $e_a$ on $\CH_a$, the partition function of the anomalous theory on $X_a$ will define an element
\begin{equation}
    Z(X,a) e_a \in \CH_a.
\end{equation}
Since it will be useful for our purpose, we mention that from the explicit construction of $\CH_a$ the choice of the basis element $e_a$ for $X_a$ is equivalent to fixing a reference manifold $X_\alpha$ in the same bordism class of $X_a$, i.e. such that $[X_a]=[X_\alpha]\equiv \alpha \in\Omega^H_d$, and a reference bordism $Y_a:X_\alpha\rightarrow X_a$ \cite{Yonekura:2018ufj}.

The group of large transformations on $X$ is the extension $\widetilde{\MCG}_{G^f}(X)$ of the mapping class group $\MCG(X)$ by $G^f$,
\begin{equation}
    1 \longrightarrow G^f \longrightarrow \widetilde{\MCG}_{G^f}(X) \overset{\pi}{\longrightarrow} \MCG (X) \rightarrow 1.
\end{equation}
Then an element $[\tilde{\psi}]\in \widetilde{\MCG}_{G^f}(X)$ acts on the class $a$ via the map
\begin{equation}
    [\tilde{\psi}]: a \mapsto [\tilde{\psi}]\cdot a := \psi^{-1*} a,
    \label{mcg-action}
\end{equation}
where $\psi:X\rightarrow X$ is a diffemorphism representing $[\psi] = \pi([\tilde{\psi}])$, while the action of $[\tilde{\psi}]$ on $\CH_a$ is given by
\begin{equation}
    \widetilde{\Psi}: e_a \longmapsto C_{[\tilde{\psi}]}(a)e_{\psi^{-1*}a}.
\end{equation}
By unitarity one finds that $C_{[\tilde{\psi}]}(a)$ is actually a phase, as described in \eqref{eq:anom}. Moreover, a change of basis in $\CH_a$, which physically is equivalent to a redefinition of the partition function by the addition of local counterterms, shows us that this phase defines more correctly a cohomology class 
\begin{equation}
    [C]\in H^1(\widetilde{\MCG}_{G^f}(X),U(1)^{\Spingf(X)}),
    \label{def:anom-phases}
\end{equation}
where $U(1)^{\Spingf(X)}$ is regarded as a module of $\widetilde{\MCG}_{G^f}(X)$ with the action determined
by the action \eqref{mcg-action} on $\Spingf(X)$ structures.

To make our statement more precise, for a theory on $X$ free of perturbative anomalies, its global anomalies are associated to a deformation class of the torsion subgroup of all $(d+1)$-dimensional iTQFTs, where the group operation is given by stacking. Moreover, it is possible to show \cite{Yonekura:2018ufj} that with this structure there is an isomorphism 
\begin{equation}
    \mathrm{Tor}\{\text{ref. pos. }(d+1)\text{-dim}\;H_d\text{ -iTQFTs}\}/_{\text{def}} \cong \Hom(\Tor \Omega^{H}_{d+1},U(1)).
\end{equation}
In the case of $H_{d+1}=H'_{d+1} \times G'$ where $\Z^f_2 \subset H'_{d+1}$, the bordism groups can be simply understood as the set of manifolds with $H'_{d+1}$ structure and a map from $X$ to the classifying space $BG'$, 
\begin{equation}
    \Omega^{H}_{d+1} \cong \Omega^{H'}_{d+1}(BG').
\end{equation}
We will use the two notation interchangeably. Thus a $d$ dimensional anomalous theory will describe the boundary of an iTQFT, which in turn is identified with an element $\nu \in \Hom(\Omega^H_{d+1},U(1))$. In terms of the $d$ dimensional theory this means there is a map 
\begin{equation}
    \Hom(\Omega^H_{d+1},U(1)) \longrightarrow H^1(\widetilde{\MCG}_{G^f}(X),U(1)^{\Spingf(X)}) 
\end{equation}
that will associate an anomalous phase to each large transformation $[\tilde{\psi}]\in \widetilde{\MCG}_{G^f}(X)$ as a function of $\nu$. Without entering into further details, it is enough to know that this map can be found to be 
\begin{equation}
\begin{array}{rcl}
    \Phi: \Hom(\Omega^H_{d+1},U(1)) & \longrightarrow & H^1(\widetilde{\MCG}_{G^f}(X),U(1)^{\Spingf(X)}),   \\
    \nu & \longmapsto & \nu((\tilde{\psi}\circ Y_a)\cup \overline{Y_{\psi^{-1*}a}})
\end{array}
\end{equation}
where $Y_{a,\psi^{-1*}a}$ are the reference bordism previously mentioned for $X_a$ and $X_{\psi^{-1*}a}$.
Here the union denotes the manifolds joined together by merging their boundaries. In the following we will denote the resulting compact $3$-manifold associated to any transformation $\tilde{\psi}$ by
\begin{equation}
    \M(\tilde{\psi}) :=(\tilde{\psi}\circ Y_a)\cup \overline{Y_{\psi^{-1*}a}}.
    \label{def:map-tor}
\end{equation}
Therefore, the computation of the anomalous phases \eqref{def:anom-phases} boils down to evaluating bordism invariants for the closed manifolds \eqref{def:map-tor}.

\subsection{Peculiarities in $d=2$}

In the particular case of anomalous systems in $d=2$, the low dimensionality allows us to employ powerful techniques that otherwise would not be available to the same degree for the computation of \eqref{def:anom-phases}, as we are now going to discuss. From now on we will be specifically interested in theories defined on the $2$-torus, so we are going to assume $X=\mathbb{T}^2$ decorated with some structure $H_2$. 

An important tool for us is the sequence of maps 
\begin{equation}
    RO^f(G^f) \overset{\sch}{\longrightarrow} \Hom(\Omega^H_3,U(1)) \overset{\Phi}{\longrightarrow} H^1(\widetilde{\MCG}_{G^f}(\mathbb{T}^2),U(1)^{H(\mathbb{T}^2)}),
    \label{set-of-maps}
\end{equation}
where the first map $\sch$ is provided by free fermion theories. 
Here $RO^f(G^f)$ is defined as the subgroup of $RO(G^f)$, i.e. the free group generated by real irreps of $G^f$, that is generated by the image of the non trivial representation $\crep{1}$ of $\Z^f_2$ via the pullback of the inclusion $\xi:\Z^f_2 \hookrightarrow G^f$, i.e.
\begin{equation}
    RO^f(G^f):= (\xi^*)^{-1}(\Z\;\crep{1}).
\end{equation}
In the simple case $G^f=\Z^f_2\times G$ we have $RO^f(G^f)\cong RO(G)$.
Note that for a fixed $G^f$ the three spaces in \eqref{set-of-maps} are all determined as image of canonical functors. This means that the maps describe more properly a set of natural transformation between such functors, so that any homomorphism between groups create a commutative diagram from which we are able to gain important data about the anomaly, as we will see. To this regard, one needs to be careful to note that the middle term in these diagrams is correctly represented by $\Hom(\Omega^H_{3},U(1))$ only if one is allowed to forget about the perturbative part of the full anomaly group, that more precisely is non-canonically isomorphic to $\Hom(\Omega^H_{3},U(1))\oplus\Hom(\Omega^H_{4},\Z)$. Indeed, as we will explain more carefully in the next section, for a generic homomorphism of groups $f:G^f\rightarrow {G^f}'$ one may have a non-trivial interplay between the two kinds of anomalies, due to the fact that the splitting of the total anomaly group is non-canonical and might be different for $G^f$ and ${G^{f}}'$. A particular example of this subtlety playing a role is when $G^f=\Z^f_2\times_{\Z^f_2}G$ and ${G^f}'=\Z^f_2\times G$, while in the simplest case when both $G^f=\Z^f_2\times G$ and ${G^f}'=\Z^f_2\times G'$ one can correctly assume that the map is \eqref{set-of-maps} as already stated.

Given this premise, $\sch$ in \eqref{set-of-maps} plays the analogue role of the Chern character map
\begin{equation}
    c_2:RU(G) \longrightarrow H^{4}(BG,\Z)
\end{equation}
that describes bosonic anomalies. Therefore, the axiomatic properties of $\sch$ allow us to gain insight on the construction of the groups $\Hom(\Omega^H_{d+1},U(1))$ and we will make heavily use of this in Section \ref{sec:mod-transf}. 

Another useful property that comes at hand for the computation of anomalous phases and that will be central in our analysis is the super-cohomology description of the groups $\Hom(\Omega^H_3,U(1))$. For this aspect, it will be sufficient to restrict to the case of $G^f=\Z^f_2\times G$. From spectral sequence computations it is possible to see how the cobordism groups can be described by a set of cohomology layers with nontrivial differentials connecting them, namely
\begin{equation}
    \Hom(\Omega^\Spin_3,U(1))\overset{\mathrm{set}}{\cong} H^1(BG;\Z)\times SH^2(BG;\Z)\times H^3(BG;U(1)),
\end{equation}
where $SH^2(BG;\Z)$ is a certain subgroup of $H^2(BG;\Z_2)$ (kernel of $(\cdot \cup \cdot)$ operation composed with the canonical map $H^4(BG;\Z_2) \rightarrow H^4(BG;U(1))$).
Albeit for the computation of $\Hom(\Omega^\Spin_3,U(1))$ one could straightforwardly resort to them, by analyzing its structure from the super-cohomology point of view one gains also important knowledge on the subgroups data that helps to compute the anomalous phases \eqref{def:anom-phases}. For this reason, we are going to follow this approach in our analysis. See for example \cite{Delmastro:2021xox, cheng2018classification, Wang:2017moj, Wang:2018pdc} for other results based on this philosophy. 

Finally, once we select a particular large transformation $\tilde{\psi}$, it is necessary to compute the value of the cobordism groups evaluated for the proper closed $3$-manifold $\M(\tilde{\psi})$. To this regard, it is useful to translate the problem in terms of knot invariants. Indeed, being $\M(\tilde{\psi})$ always orientable, it can in particular be described by performing Dehn surgery on an appropriate framed link $\mathcal{L}$ in $S^3$, $\M(\tilde{\psi})=S^3(\mathcal{L})$. By framing of $\mathcal{L}$ we mean a choice of a non-vanishing normal vector field on each of its link components $\mathcal{L}_I\subset \mathcal{L}$, considered up to isotopy. Computationally, this corresponds to associating to each $\mathcal{L}_I$ a number $p_I \in \Z$ equal to its self-linking number. The Dehn surgery on $S^3$ is then performed by removing a tubular neighborhood of these components and gluing them back after performing a $T^{p_I}$ transformation on their meridians (i.e. their contractible cycles). Thus a description of $\M(\tilde{\psi})$ alone amounts to a link $\mathcal{L}$ on $S^3$ and its linking matrix
\begin{equation}
    B_{IJ}=\lk (\mathcal{L}_I,\mathcal{L}_J).
\end{equation}
To account also for its $\Spin\text{-}G^f$ structure, some other extra data is needed. For example, a spin structure on a $3$-manifold $Y$ describes an element in the set 
\begin{equation}
\Spin(Y) = \left\{ s \in \Z^L_2 \middle\vert \sum_J B_{IJ} s_J = B_{II} \mod 2\right\},
\label{def:spin}
\end{equation}
corresponding to a particular sublink $\mathcal{C}\subset \mathcal{L}$ \cite{kirby19913}. Moreover, a background gauge field with symmetry $G$ specifies a further element in
\begin{equation}
    H^1(Y;\Z_{n}) = \left\{a \in \Z^L_n \middle\vert \sum_{J} B_{IJ} a_J = 0 \mod n \right\}.
    \label{def:H}
\end{equation}

Starting from this, it is possible to compute the invariants that generate the cobordism groups for $G^f=\Z^f_2 \times G$, with $G$ abelian \cite{Grigoletto:2021zyv}. In the following we are going to focus also on other cases, namely when $G^f=\Z^f_2 \times G$ with $G$ non-abelian or $G^f=\Z^f_2\times_{\Z_2}\Z_{2^{l+1}}$, and we are going to employ two separate methods for their analysis. Here we briefly comment on the former case, while the latter will be analyzed more extensively in the next section.

Suppose then that $G^f=\Z^f_2\times G$, with $G$ non-abelian. In order to define a background gauge field on a torus one needs only a couple of commuting elements $g,h \in G$. This means that by acting with modular transformations $\psi\in \MCG (\mathbb{T}^2)$ on it, the new background gauge fields will always be described in terms of elements in the subgroup $G'=\left\langle g,h\right\rangle \overset{i}{\hookrightarrow} G$ generated by $g$ and $h$. In other words, considering the commutative diagram 
\begin{equation}
\begin{tikzcd}
   \Hom(\Omega^{\Spin}_3(BG'),U(1)) \ar[d] & \Hom(\Omega^{\Spin}_3(BG),U(1)) \ar[l,"\hat{i}^*",swap] \ar[d]\\
   H^1(\widetilde{\MCG}_{G^f}(\mathbb{T}^2),U(1)^{H(\mathbb{T}^2)}) &H^1(\widetilde{\MCG}_{G^f}(\mathbb{T}^2),U(1)^{H(\mathbb{T}^2)}) \ar[l]
\end{tikzcd}
\label{diag-comm}
\end{equation}
and restricting to transformations $\psi \in \MCG(\mathbb{T}^2)$, then by uncovering the map $\hat{i}^*$ and knowing $\Phi$ in \eqref{set-of-maps} for abelian groups is enough to extract all the information for the anomalous phases for the starting $G$ as well. This is exactly what we are going to do for some selected cases in Section \ref{sec:mod-transf}.

\section{Cobordism invariants for $\Spin$-$\Z^f_{2^{l+1}}$ structures}
\label{sec:twist}

In order to study the more complex case where $G^f$ has a non-trivial twist, the starting point is focusing on the case of $G^f$ abelian. Here we are going to work out the prototypical example, namely when $G^f=\Z^f_2 \times_{\Z_2}\Z_{2^{l+1}}\equiv \Z^f_{2^{l+1}}$. Its bordism group will indeed describe the $2$-torsion part of any bordism group with $G^f=\Z^f_2\times_{\Z_2}\Z_{2n}$, which is its most interesting component \cite{Guo:2018vij}. In fact, by decomposing $n=2^l\cdot k$ with $k$ odd, it holds $\Omega^{\Spin\text{-}\Z^f_{2n}}_3=\Omega^{\Spin\text{-}\Z^f_{2^{l+1}}}_3(B\Z_k)$.
We start by the known results \cite{Campbell:2017khc,Guo:2018vij}
\begin{equation}
    \Omega^{\Spin\text{-}\Z^f_{2^{l+1}}}_3\cong \Z_{2^{l-1}},\quad \Omega^{\Spin\text{-}\Z^f_{2^{l+1}}}_4\cong \Z.
\end{equation}
The question then is which kind of invariant generates $\Hom(\Omega^{\Spin\text{-}\Z^f_{2^{l+1}}}_3,U(1))\cong \Z_{2^{l-1}}$. To address it we are going to employ the anomaly interplay between $G^f=\Z^f_{2^{l+1}}$ and $G^f=U(1)$, corresponding to $\Spin^c$ structures. We will need to consider the total anomaly group for both cases, which is a generalized cohomology group $H^4_{I\Z}(MTH)$ that satisfies the short exact sequence \cite{Freed:2016rqq,Freed:2014eja}
\begin{equation}
    1 \longrightarrow \Hom(\Tor \Omega^H_{3},U(1))\longrightarrow H^4_{I\mathbb{Z}}(MTH) \longrightarrow \Hom(\Omega^H_4,\mathbb{Z})\longrightarrow 1.
    \label{tota-anom-group}
\end{equation}
Moreover, this sequence is split, albeit non canonically, so that one can identify it with the direct sum of the non-perturbative anomaly part and the perturbative one as we already anticipated.

From the functorial properties, in a similar manner to what we discussed for the sequence of maps \eqref{set-of-maps},  an homomorphism of groups $f:G^f\rightarrow {G^f}'$ induces the commutative diagram
\begin{equation}
\begin{tikzcd}
    1 \ar[r] & \Hom(\Tor \Omega^H_{3},U(1)) \ar[r] & H^4_{I\Z}(MTH) \ar[r]& \Hom(\Omega^H_4,\mathbb{Z}) \ar[r] & 1\\
    1 \ar[r] & \Hom(\Tor \Omega^{H'}_{3},U(1)) \ar[r] \ar[u] &H^4_{I\Z}(MTH') \ar[r] \ar[u,"\tilde{f}^*",swap] & \Hom(\Omega^{H'}_4,\mathbb{Z}) \ar[r] \ar[u] & 1
\end{tikzcd}.
\label{anom-interplay}
\end{equation}
This kind of diagrams describe the anomaly interplay between different symmetry groups, see \cite{Davighi:2020bvi, Lee:2020ojw} for other interesting applications of it. The utility of these diagrams lies mainly in the particular case of when one of the groups on the left or right side of the sequences is trivial. Then the knowledge of the map $\tilde{f}^*$ in \eqref{anom-interplay} allows to deduce important information on the other groups present in the diagram, as we will see in a moment. For our case of interest $\Hom (\Omega^{\Spin^c}_3,U(1))$ is trivial, while $\Hom(\Omega^{\Spin^c}_4,\Z)=\Z\oplus \Z$. Therefore we have 
\begin{equation}
\begin{tikzcd}
    1 \ar[r] & \Z_{2^{l-1}} \ar[r] & H^4_{I\Z}(MT\Spin\text{-}\Z^f_{2^{l+1}}) \ar[r]& \Z \ar[r] & 1\\
    1 \ar[r] & 1 \ar[r] \ar[u] &H^4_{I\Z}(MT\Spin^c) \ar[r] \ar[u,"\tilde{\i}^*",swap] & \Z\oplus\Z \ar[r] \ar[u] & 1
\end{tikzcd}.
\label{SpinZ2n:interplay}
\end{equation}
To see how $\Spin^c$ structure invariants in $d=4$ are translated into $d=3$ $\Spin$-$\Z^f_{2^{l+1}}$ ones, it is necessary to uncover the torsion part of the map $\tilde{\i}^*:\Z\oplus \Z \rightarrow \Z_{2^{l-1}}\oplus \Z$. 

Consider now the series of manifolds quotients of the unit sphere bundle of the Whitney sum of the tensor square of the complex Hopf line bundle $H$ over $\mathbb{CP}^1$,
\begin{equation}
    Y(2^{l};\crep{a}) := S(\mathbb{H}\otimes \mathbb{H})/\crep{a}.
\end{equation}
Here $l$ and $a$ are both integers, $\crep{a}$ is the representation $\crep{a}:\lambda \mapsto \lambda^{a}$ of $\Z_{2^l}$ in $U(1)$ and its action on the associated unit sphere bundle is fixed-point free \cite{Chang:2018iay}. 
It is possible to show \cite{BBG,hha/1175791076} that they admit a natural $\Spin$-$\Z^f_{2^{l+1}}$ structure, i.e. $Y(2^{l};\rho_a)\in \Omega^{\Spin\text{-}\Z^f_{2^{l+1}}}_3$. Moreover, in this case the $\eta$ invariant gives an homomorphism 
\begin{equation}
    \eta(\rho):\Omega^{\Spin\text{-}\Z^f_{2^{l+1}}}_3 \rightarrow \R/\Z,
\end{equation}
for any $\rho\in RU_0(\Z_{2^{l}})$, where $RU_0(\Z_{2^{l}})$ is the augmentation ideal of $RU(\Z_{2^{l}})$ \cite{hha/1175791076}. We recall that 
\begin{equation}
    RU(\Z_{2^{l}}) = \oplus_{q=0}^{2^{l-1}} \Z \cdot \crep{q},
\end{equation}
with $\crep{q}:\lambda \mapsto \lambda^q$ the $1$-dimensional complex irreps of $\Z_{2^{l}}$, while 
\begin{equation}
    RU_0(\Z_{2^{l}}) = \left\{ \rho=\sum_{q=0}^{2^l-1}b_k \crep{q} \in RU(\Z_{2^l}) \middle\vert \sum_{q=0}^{2^l-1} b_q = 0\right\}.
\label{RU0}
\end{equation}
An alternative and useful description of \eqref{RU0} is given by
\begin{equation}
    RU_0(\Z_{2^l}) \cong ([0]-\crep{1})\cdot \sum_{k=0}^{2^l-2}c_k\crep{k},
\end{equation}
where, after fixing $b_{2^l-1}$ using the constraint of \eqref{RU0} one has the relation
\begin{equation}
b_q=\sum_{k=0}^{2^l-2}A_{qk}c_k,\quad A = \begin{pmatrix}1&\\-1&1\\&&\ddots\\&&-1&1
\end{pmatrix}, \quad A^{-1} = \begin{pmatrix}1&\\1&1\\&&\ddots\\1&1&1&1
\end{pmatrix}.
\end{equation}
For this particular class of manifolds the $\eta$ invariant can be computed explicitly via the formula \cite{hha/1175791076}
\begin{equation}
    \eta(X(2^{l};\rho_a))(\rho)=\frac{1}{2^{l}}\sum_{\lambda \in \Z_{2^{l}}, \lambda \neq 1} \mathrm{Tr}(\rho(\lambda))\frac{(1+\lambda^a) }{(1-\lambda^a)^2}\lambda^{(a+1)/2} \mod \Z.
\end{equation}
Therefore $X(2^{l};\crep{1})$ is a generator of $\Omega^{\Spin\text{-}\Z^f_{2^{l+1}}}_3$, since\footnote{Use the identity
\begin{equation*}
    \sum_{\lambda \in \Z_{2^{l}},\lambda \neq 1} \frac{\lambda^n}{1-\lambda} =-\sum_{\lambda \in \Z_{2^{l}},\lambda \neq 1} \sum_{k=0}^{n-1}\lambda^k+\frac{1}{2}\sum_{\lambda \in \Z_{2^{l}},\lambda \neq 1} \left(\frac{1}{1-\lambda}+\frac{1}{1-\bar{\lambda}}\right)=  n+2^{l-1}-\frac{1}{2} \mod 2^l.
\end{equation*}}
\begin{align}
\eta(X(2^l;\crep{1}))\left(([0]-\crep{1})\cdot\crep{s}\right) = \frac{1}{2^l}\sum_{\lambda\in \Z_{2^l},\lambda \neq 1} \frac{\lambda^{s+1}+\lambda^{s+2}}{1-\lambda}= \frac{s+1}{2^{l-1}} \mod \Z.
\end{align}
For $\rho = ([0]-\crep{1})\cdot \sum_{q} c_q \crep{q}$ we arrive to
\begin{equation}
    \eta(X(2^l;\crep{1}))(\rho) = \sum_{q=0}^{2^l-2} \frac{c_q(q+1)}{2^{l-2}} \mod \Z = -\sum_{q=0}^{2^l-2}b_q\frac{q(q+1)}{2} \mod \Z.
    \label{eta:spinZ2m}
\end{equation}
Note that $\crep{q}$ here denotes the irreps of charge $q$ of $\Z_{2^l}$. The relation between the charges of this $\Z_{2^l}$ and $\Z_{2^{l+1}}$ is \cite{Hsieh:2018ifc}
\begin{equation}
    \tilde{q} = 2q+1 \mod 2^{l+1},
    \label{charge-rel}
\end{equation}
where $\tilde{q}$ in instead the charge of $\Z_{2^{l+1}}$, consistently with the fact that fermions are characterized by odd charges under $\Z_{2^{l+1}}$.

Formula \eqref{eta:spinZ2m} tells us the anomaly coefficient for the torsion part of $H^4_{I\mathbb{Z}}(MT\Spin\text{-}\Z^f_{2^{l+1}})$. Now we focus on the coefficients for the anomaly group of $\Spin^c$. Being its anomalies perturbative, we can determine them by computing the anomaly polynomial for a $d=4$ manifold $W$. If $W$ is equipped with a line bundle $\mathfrak{v}$ corresponding to a $U(1)$ gauge field and tangent space $TW$ then its value is 
\begin{equation}
    \mathcal{P}|_{4\mathrm{-vol}} = \frac{1}{2}c^2_1(\mathfrak{v})-\frac{1}{24}p_1(TW)\overset{\Spin^c}{\leadsto} \frac{1}{8}c^2_1(\mathfrak{s})-\frac{1}{8}\sigma (TW),
    \label{anom-pol}
\end{equation}
where $\sigma(TW)$ is the signature of $W$, related to the Pontryagin class by $\sigma(TW)=p_1(TW)/3$.
The second expression holds if $W$ is a $\Spin^c$ manifolds with a determinant line bundle $\mathfrak{s}$. This can be found by formally considering  $\mathfrak{v}$ the virtual bundle that satisfy $\mathfrak{v}^{\otimes 2}=\mathfrak{s}$, together with the properties of the Chern classes under tensor product. 
The APS index theorem guarantees that \eqref{anom-pol} is always an integer, so that\footnote{From a more geometrical point of view, one can use as generators of $\Hom(\Omega_4^{\Spin^c},\Z)\cong \Z \oplus \Z$ the pair of maps $\varphi_1 =\sigma$, $\varphi_2=(\sigma - \Sigma\cdot \Sigma)/8)$. Here $\c^2_1(\mathfrak{s})=\Sigma \cdot \Sigma$, with $\Sigma$ a characteristic surface determined by the bundle $\mathfrak{s}$. It is possible to show \cite{Scorpan2005TheWW} that in this base a good set of generators $(W,\Sigma)$ of $\Omega^{\Spin^c}_4$ is given by $(W,\Sigma)_1=(\mathbb{CP}^2,\mathbb{CP}^1)$ and $(W,\Sigma)_2=(\mathbb{CP}^2\#\overline{\mathbb{CP}}^2,\#_3 \mathbb{CP}^1\#\overline{\mathbb{CP}}^1)$, as $\varphi_i((W,\Sigma)_j)=\delta_{i,j}$.}
\begin{equation}
    \Hom(\Omega^{\Spin^c}_4,\Z) \cong \Z c^2_1(\mathfrak{s}) \oplus \Z \frac{c^2_1(\mathfrak{s})-\sigma(TW)}{8}.
    \label{base:anom-pol}
\end{equation} 
Suppose our $4$-manifold $W$ is fixed. Then a set of fermions $\{\psi_i\}_{i\in\mathcal{I}}$ with charges $\tilde{q}_i$ represent an element  
\begin{equation}
\omega = \left(\sum_i \frac{\tilde{q}^2_i -1}{8},1\right)=\left(\sum_i \frac{q_i(q_i+1)}{2},1\right).
\end{equation}
Note that up to a sign the first entry is exactly the value \eqref{eta:spinZ2m} of the $\eta$ invariant for the generators of $\Omega^{\Spin\text{-}\Z^f_{2^{l+1}}}_3$, thus showing us what is the torsion part of the map $\tilde{\i}^*$ in \eqref{SpinZ2n:interplay}. Therefore we know that the bordism invariant that describes $\Hom(\Omega^{\Spin\text{-}\Z^f_{2^{l+1}}}_3,U(1))\cong \Z_{2^{l-1}}$ arises from the bordism invariant $c^2_1(\mathfrak{s})$ of $\Spin^c$ $4$-manifolds.

To determine how the Chern class induces an invariant in $3$ dimensions, we now consider the case of a $4$-manifold $W$ with non-trivial boundary $\partial W=Y$. We suppose $W$ to be simply connected and $Y$ a rational homology sphere, so in particular $H_2(W;\Z)$ will be torsion-free and isomorphic to $\Z^L$ for some $L$. The matrix $B$ will be invertible over $\Q$ and $H_1(Y;\Z)\cong\Tor H_1(Y;\Z)$. Then $Y$ will be represented via Dehn surgery on $S^3$ by a link $\CL$ with $L$ components, $Y=S^3(\CL)$. Under the appropriate basis the linking form $B$ of $\CL$ describes the intersection form $\cap:H_2(W;\Z)\times H_2(W;\Z)\rightarrow \Z$ and induces the cohomological pairing\footnote{Since $H_2(W;\Z)$ is torsionless one can actually think of it in terms of de Rahm cohomology, where the pairing is given by the wedge product.}
\begin{equation}
\begin{array}{rcl}
B^{-1}: H^2(W;\Z)\otimes H^2(W;\Z)& \rightarrow &\Q/\Z,\\
\alpha \otimes \beta & \mapsto & \alpha^T B^{-1} \beta
\end{array}
\label{pairing-D}
\end{equation}
so that $c^2_1(\mathfrak{s})=c^T B^{-1}c$ for\footnote{In the case of a manifold $W$ with boundary one might ask whether $c_1$ defines an element in $H^2(W,Y;\Z)$ rather than $H^2(W;\Z)$. From the physical point of view it is clear that the latter is the correct one, as an element in the relative cohomology would imply a trivial theory on the boundary $X$.} $c_1=c\in \Z^L\cong H^2(W;\Z)$.
Actually \eqref{pairing-D} induces also a pairing for the homology class $H_1(Y;\Z)$ by recalling the long exact sequence
\begin{equation}
    \ldots \rightarrow H_2(W;\Z) \xrightarrow{B} H_2(W,X;\Z) \xrightarrow{\partial} H_1(Y;\Z) \rightarrow H_1(W;\Z) \cong 1,
\end{equation}
and the Poincar\'e-Lefschetz duality $H^2(W;\Z)\cong H_2(W,Y;\Z)$. As a result
\begin{equation}
    c^2_1(\mathfrak{s})=c^TB^{-1}c \mod \Z\mapsto \partial(c)^TB^{-1}\partial(c) \mod \Z = \tilde{c}^T B \tilde{c} \mod \Z.
    \label{projection-c1square}
\end{equation}
Here $\partial(c)\in H_1(Y;\Z)\cong \mathrm{coker} B$, while $\tilde{c}\in H^1(X;\Q/\Z)\cong \Hom (H_1(Y;\Z),\Q/\Z)$ is its image under the isomorphism given by the linking form
\begin{equation}
    \begin{array}{rcl}
    B^{-1}:H_1(Y;\Z) & \rightarrow & \Hom(H_1(Y;\Z),\Q/\Z)\\
    x & \mapsto & x^TB^{-1}   
    \end{array}.
\end{equation}
Now one needs to understand which kind of element $\tilde{c}$ is determined by a $\Spin$-$\Z^f_{2^{l+1}}$ structure on $Y$. First we know there is a commutative diagram of tangential structures given by
\begin{equation}
\begin{tikzcd}
\Spin(3)\times \Z_{2^{l+1}} \ar[d,"\psi_1"] \ar[r,"\varphi_2"] & \Spin(3)\times_{\Z_2}\Z_{2^{l+1}} \ar[d,"\psi_2"]\\
SO(3) \times \Z_{2^{l+1}} \ar[r,"\varphi_1"] & SO(3)\times \Z_{2^l}
\end{tikzcd}.
\label{tangential-structures-diagram}
\end{equation}
We already know from \eqref{def:spin}-\eqref{def:H} that a spin structure and a $\Z_{2^{l+1}}$ gauge field on $Y$ define a couple $(s,a)\in \Spin(Y)\times H^1(Y;\Z_{2^{l+1}})$.
Then 
\begin{equation}
\begin{array}{rcl}
    \psi_1:\Spin(Y) \times H^1(X;\Z_{2^{l+1}})&\rightarrow &H^1(Y;\Z_{2^{l+1}}),\\
    (s,a) & \mapsto & a\\
    \varphi_1:H^1(Y;\Z_{2^{l+1}})& \rightarrow & H^1(Y;\Z_{2^l}).\\
     a & \mapsto & \tilde{a} =a \mod 2^l
\end{array}
\end{equation}
The set of $\Spin\text{-}\Z^f_{2^{l+1}}$ structures is instead described by
\begin{equation}
    \Spin\text{-}{\Z^f_{2^{l+1}}}(Y)=\left\{b \in \Z^L_{2^{l+1}} \middle\vert \sum_J B_{IJ}b_J= 2^l B_{II} \mod 2^{l+1}\right\},
    \label{def:spinZ2m}
\end{equation}
while the maps $\psi_2,\varphi_2$ are
\begin{align}
\begin{array}{rcl}
         \varphi_2:\Spin(Y)\times H^1(Y;\Z_{2^{l+1}}) & \rightarrow & \Spin\text{-}{\Z^f_{2^{l+1}}}(Y),\\
         (s,a) & \mapsto & a+2^l s\\
         \psi_2:\Spin\text{-}{\Z^f_{2^{l+1}}}(Y) & \rightarrow & H^1(X;\Z_{2^l}).\\
         b & \mapsto & \tilde{a}=b \mod 2^l
\end{array}
\end{align}
Note that we have the property $\varphi_2(s+\delta,a-2^l \delta) = \varphi_2(s,a)$ for any $\delta \in H^1(Y;\Z_2)$. Via $\psi_2$ we know that there exists a $\Z_{2^l}$ bundle for any $\Spin$-$\Z^f_{2^{l+1}}$ structure, which under the inclusion map describes an element
\begin{equation}
    \begin{array}{rcl}
        H^1(Y;\Z_{2^l}) &\rightarrow & H^1(Y;\Q/\Z).\\
        \tilde{a} & \mapsto & \frac{1}{2^l}\tilde{a}
    \end{array}
\end{equation}
Thus, by restricting the tangential structure of $Y$ to be described by $b \in \Spin\text{-}{\Z^f_{2^{l+1}}}(Y)$, we get the surgery description of the $\eta$ invariant
\begin{align}
 \tilde{c}^T B \tilde{c} = \frac{\tilde{a}^T B\tilde{a}}{2^{2l}} \mod \Z \equiv \frac{1}{2^{l-1}}\tilde{\lk}(\tilde{a},\tilde{a}),
\label{bBb}
\end{align}
where $\tilde{a}=b \mod 2^l$ and
\begin{equation}
     \tilde{\lk} (\tilde{a},\tilde{a}) :=\frac{\tilde{a}^T B\tilde{a}}{2^{l+1}} \mod 2^{l-1}.
\end{equation}
Given a couple $(s,a)$ such that $\psi_2\circ \varphi_2(s,a) = \tilde{a}$ and substituting\footnote{Here the identity is to be understood under the lifting $\Z_n \rightarrow \Z$.} 
\begin{equation}
    \tilde{a} = 2^l\delta - a, \quad \delta \in H^1(Y;\Z_2),
\end{equation}
it is easy to prove that $\tilde{\lk} (\psi_2\circ \varphi_2(s,a)) = \tilde{\lk} (\psi_2\circ \varphi_2(s+\delta,a-2^l \delta))$ and that $\tilde{\lk}$ is always an integer, so that such invariant is well defined in terms of the $\Spin\text{-}\Z^f_{2^{l+1}}$ structures \eqref{def:spinZ2m} and is precisely the generator of $\Hom(\Omega^{\Spin\text{-}\Z^f_{2^{l+1}}}_3,U(1))$.
Note that the commutativity of  \eqref{tangential-structures-diagram} tells us that for the purpose of computing $\tilde{\lk}$ one can formally think of $b$ as a $\Z_{2^{l+1}}$ gauge field and $\tilde{a}$ as its $\mod 2^l$ reduction.

\section{Fermionic CFTs and modular transformations}
\label{sec:mod-transf}

In this section we are going to discuss the anomalous modular transformations of fermionic CFTs on a torus, representing a generic theory in the IR. In particular, we are going to focus on the unitary case with $c=\bar{c}>1$.

By assuming a discrete symmetry group $G^f$, the transformation rules of the fields are implemented in the theory by the presence of topological defect lines (TDLs). These defects $\hat{g}$ are such that sweeping them past some local operator, be it $\phi(x)$, then they are transformed into $\phi'(x)= g \cdot \phi(x)$, where $g\in G^f$. Their presence is particularly useful for the definition of the twisted Hilbert spaces. 
Indeed, considering partition functions on a torus with a TDL $\hat{g}$ parallel to the space direction\footnote{We are working in the Euclidean so the notion of time and space direction is arbitrary. We will work by pretending that the direction parallel to the defect line of \eqref{def:TwistPartFunct} is the space direction.} is equivalent to twisting the boundary condition on the time direction by the action of $g$. In pictorial terms
\begin{equation}
    Z^g(\tau,\bar{\tau}) := \mathrm{Tr}_{\CH_\mathbb{I}}[\hat{g}\,q^{L_0-c/24}\bar{q}^{\bar{L}_0-\bar{c}/24}]
    \quad \sim \quad 
    \begin{tikzpicture}[scale=0.4,baseline=(current  bounding  box.center),>=stealth]
    \draw (0,0) rectangle (4,4);
    \draw[ForestGreen,thick,->-] (0,2) -- (4,2);
    \node[left] at (0,2) {$g$};
    \node[below] at (2,0) {$\mathbb{I}$};
    \end{tikzpicture}\quad.
    \label{def:TwistPartFunct}
\end{equation}
Instead, the partition function defined over the twisted Hilbert spaces $\CH_g$ corresponds to the insertion of a line defect along the time direction:
\begin{equation}
    Z_g(\tau,\bar{\tau}) := \mathrm{Tr}_{\CH_g}[q^{L_0-c/24}\bar{q}^{\bar{L}_0-\bar{c}/24}] 
    \quad \sim \quad 
    \begin{tikzpicture}[scale=0.4,baseline=(current  bounding  box.center),>=stealth]
    \draw (0,0) rectangle (4,4);
    \draw[ForestGreen,thick,->-] (2,0) -- (2,4);    
    \node[left] at (0,2) {$\mathbb{I}$};
    \node[below] at (2,0) {$g$};
    \end{tikzpicture}\quad.
    \label{def:DefctPartFunct}
\end{equation}
Here one clarification is in order. Since we are focusing on the case of spin-theories, for us the untwisted boundary condition defined along the space and time directions and associated to $\mathbb{I}\in G^f$ is assumed to be $\ns$, namely anti-periodic. Thus for us in any Hilbert space $\CH_{g\neq \mathbb{I}}$ the fermions exhibit $\ns$ boundary condition \emph{plus} a twist by $g\in G^f$. In particular, an $\r$ boundary condition correspond to $\ns$ plus an additional twist by $(-1)^F$.

TDLs describe the $G^f$ action over the untwisted Hilbert space $\CH_\mathbb{I}$, which admits the natural grading 
\begin{equation}
    \CH _{\mathbb{I}} = \bigoplus_{\Gamma\,\mathrm{irreps}} \CH^{(\Gamma)}_{\mathbb{I}}.
\end{equation}
Therefore the untwisted partition function can be expressed as a sum
\begin{align}
    Z(\tau,\bar{\tau}) &= \mathrm{Tr}_{\CH_{\mathbb{I}}}[q^{L_0-c/24}\bar{q}^{\bar{L}_0-\bar{c}/24}]=\mathrm{Tr}_{\CH_{\mathbb{I}}}\left[\sum_{\Gamma} \Pi_\Gamma q^{L_0-c/24}\bar{q}^{\bar{L}_0-\bar{c}/24}\right]\equiv \sum_\Gamma Z^\Gamma(\tau,\bar{\tau}),
\end{align}
where $\Pi_\Gamma$ are the projectors over the sectors $\CH^{(\Gamma)}_{\mathbb{I}}$, defined by
\begin{equation}
    \Pi_\Gamma = \frac{d_\Gamma}{|G^f|}\sum_{g\in G^f}\chi^*_\Gamma(g) \hat{g}.
\end{equation}
Here $d_{\Gamma}$ is the dimension of the irrep $\Gamma$.
Using \eqref{def:TwistPartFunct} the sectors $Z^\Gamma$ that compose $Z(\tau,\bar{\tau})$ are
\begin{equation}
Z^\Gamma(\tau,\bar{\tau}) = \frac{d_\Gamma}{|G^f|}\sum_{C_{G^f}}\chi^*_\Gamma(C_{G^f}) Z^{C_{G^f}}(\tau,\bar{\tau}),\qquad Z^{C_{G^f}}(\tau,\bar{\tau}):=\sum_{g \in C_{G^f}}Z^g(\tau,\bar{\tau}),
\end{equation}
where $\{C_{G^f}\}$ are the conjugacy classes of $G^f$.

\subsection{$S$ transformation}
Our final goal is to study constraints on the sector of the trivial irrep (that we generally denote as $\Gamma_0$) on $\CH_\mathbb{I}$, where the symmetry-preserving scalar operators lie. Thus, in order to make use of the modular bootstrap technique one needs to consider all the defect partition functions $Z^g$ and $Z_g$, since under $S$ transformation
\begin{equation}
    S[Z^g](\tau,\bar{\tau}) \equiv Z^g(-1/\tau,-1/\bar{\tau})=  e^{2\pi i \vartheta_S(g)} Z_g(\tau,\bar{\tau}).
\end{equation}
A priori, the angle $\vartheta_S$ can be non-zero for non-trivial anomalies. However, in the cases we are going to consider we can fix it to $\vartheta_S(g)=0$ for any $g\in G^f$. To motivate it, consider initially the basic case with $G^f = \Z^f_2 \times G$ and then the general one with $G^f = \Z^f_2 \times_{\Z_2} G$. 

In the first case, one can immediately note that the partition functions we are interested in always exhibit either a time or space direction with boundary condition\footnote{Here when $G^f=\Z^f_2 \times G$ we denote by $\ns h$ ($\r h$) the boundary condition where fermions are antiperiodic (periodic) and further twisted by $h\in G$.} $\ns \mathbb{I}$. This means that we can always contract such direction to a point in order to create a solid torus $\mathcal{W}_{\mu,\beta}$ such that $\partial \mathcal{W}_{\mu,\beta}=\mathbb{T}^2_{\mu,\beta}$. Here the subscripts $\mu,\beta$ represent the second direction on $\mathbb{T}^2_{\mu,g}$ ($\mu=0$ the time one and $\mu=1$ the space one) and its boundary condition $\beta$. This means that $\mathbb{T}^2_{\mu,\beta}$ always lies in the trivial bordism class $[\varnothing]\in \Omega^\Spin_2(BG)$. Moreover, it is clear that when $\beta \neq \ns\mathbb{I}$, then under $S$ the bordism $\mathcal{S}:\mathbb{T}^2_{\mu,\beta}\mapsto\mathbb{T}^2_{\mu+1\mod 2,\beta}$ is such that
\begin{equation}
\mathcal{S}\circ \mathcal{W}_{\mu,\beta}=\mathcal{W}_{\mu+1\mod 2,\beta}.    
\end{equation}
Thus by choosing these manifolds as the proper base of the Hilbert spaces describing the partition functions, we can always set $\vartheta_S(g\neq \mathbb{I})=0$. The only slightly different case is when $g=\mathbb{I}$, since a priori $[(\mathcal{S}\circ\mathcal{W}_{0,\ns \mathbb{I}})\cup \overline{\mathcal{W}_{1,\ns \mathbb{I}}}]\neq [\varnothing]$ even if $\mathbb{T}^2_{0,\ns \mathbb{I}}=\mathbb{T}^2_{1,\ns \mathbb{I}}$. However, since more accurately $(\mathcal{S}\circ\mathcal{W}_{0,\ns \mathbb{I}})\cup \overline{\mathcal{W}_{1,\ns \mathbb{I}}}\in \Omega_3^\Spin(\pt)\subset \Omega^\Spin_3(BG)$ and $\Omega_3^\Spin(\pt)=\varnothing$, we can set again $\vartheta_S(\mathbb{I})=0$.

In the twisted case $G^f = \Z^f_2 \times_{\Z_2} G$ we can use the knowledge we have from the argument above. Indeed, we have always a map 
\begin{equation}
    \sigma:\Spin(2) \times G \rightarrow \Spin(2)\times_{\Z_2} G 
\end{equation}
which induces the bordism map
\begin{equation}
    \hat{\sigma}_*:\Omega^\Spin_2(BG) \rightarrow \Omega^{\Spin\text{-}G^f}_2.
    \label{map-bords}
\end{equation}
This means that the torus with a $\ns \mathbb{I}$ boundary condition in one of the two directions and viewed as an element of $\Omega^{\Spin\text{-}G^f}_2$ admits in its fiber under $\hat{\sigma}_*$ the trivial bordism class $[\varnothing]\in \Omega^\Spin_2(BG)$. Being \eqref{map-bords} a homomorphism, it follows that the angle $\vartheta_S(g)$ can be set again to zero.

This means that for the cases we are going to consider it is always possible to choose an appropriate bases of the Hilbert spaces of the partition functions such that all the information of the global anomalies can be recasted into the $T$ transformation properties and, ultimately, the spin selection rule for $\CH_{g\neq \mathbb{I}}$. 

\subsection{$T$ transformation and spin selection rules}
We now proceed to compute the spin selection rule for the twisted Hilbert space as a function of the anomaly of $G^f$.  We consider the case $G^f=\Z^f_2\times G$ since the second case will follow analogously. 
Under a $T$ transformation the torus associated to $\CH_g$ will be mapped as
\begin{equation}
    \mathcal{T}:
    \begin{tikzpicture}[scale=0.4,baseline=(current  bounding  box.center),>=stealth]
    \draw (0,0) rectangle (4,4);
    \draw[ForestGreen,thick,->-] (2,0) -- (2,4);
    \node[left] at (0,2) {$\mathbb{I}$};
    \node[below] at (2,0) {$g$};
    \end{tikzpicture}\quad \longmapsto \quad
    \begin{tikzpicture}[scale=0.4,baseline=(current  bounding  box.center),>=stealth]
    \draw (0,0) rectangle (4,4);
    \draw[ForestGreen,thick,->-] (2,0) .. controls (2,1) and (1,2) .. (0,2);
    \draw[ForestGreen,thick,->-] (4,2) .. controls (3,2) and (2,3) .. (2,4);
    \node[left] at (0,2) {$(-1)^F g$};
    \node[below] at (2,0) {$g$};
    \end{tikzpicture}
    \label{T-twist}.
\end{equation}
This means that generally, considering the smallest integer $n_g$ such that $((-1)^F g)^{n_g}=\mathbb{I}$, then $T^{n_g}$ is an automorphism on the Hilbert space related to the partition function $Z_g$, described by a phase $e^{2 \pi i \vartheta_T(g)}$. Moreover, we know that for a Virasoro module associated to a primary of weight $(h,\bar{h})$ the $T$ transformation multiply it by a phase given by the spin $s=h-\bar{h}$,
\begin{equation}
    q^{L_0 -c/24}\bar{q}^{\bar{L}_0-\bar{c}/24}|h,\bar{h}\rangle \overset{T}{\longmapsto} e^{2\pi i s}q^{L_0 -c/24}\bar{q}^{\bar{L}_0-\bar{c}/24}|h,\bar{h}\rangle.
\end{equation}
Therefore, one gets that the spin selection rule for $\mathcal{H}_g$ is given by
\begin{equation}
    T^{n_g}[Z_g](\tau,\bar{\tau}) = e^{2 \pi i \vartheta_T(g)}Z_g(\tau,\bar{\tau}) \quad \leftrightarrow \quad \{s\}_{\CH_g} \subseteq \frac{\vartheta_T(g)}{n_g} + \frac{1}{n_g}\Z.
    \label{phase:Ttransf}
\end{equation}
For $\CH_\mathbb{I}$ we are more specifically interested to the spin rule for each of the various sectors $\CH^{(\Gamma)}_\mathbb{I}$. In this case we have that under $T$ any TDL $\hat{g}$ parallel to the time direction is mapped into $(-1)^F\hat{g}$ and, since $\Z_2^f$ sits in the center of $G^f$, that
\begin{align}
    T[Z^\Gamma](\tau,\bar{\tau})=\frac{d_\Gamma}{|G^f|}\sum_{g\in G^f} \chi^*_\Gamma(g(-1)^F) \frac{\chi^*_\Gamma((-1)^F)}{\chi^*_\Gamma(1)}Z^{(-1)^Fg}(\tau,\bar{\tau})=\pm Z^\Gamma(\tau,\bar{\tau}).
\end{align}
Thus the spin selection rule is not affected by the anomaly of $G^f$, but is simply given by 
\begin{equation}
\{s\}_{\CH^{(\Gamma)}_\mathbb{I}} \subseteq \frac{\chi_\Gamma(1)-\chi_\Gamma((-1)^F)}{4 \chi_\Gamma(1)} + \Z.
\end{equation}
The $3$-manifolds 
\begin{equation}
    \M(T^{n_g})=\mathcal{T}^{n_g}\circ \mathcal{W}_{1,g} \cup \overline{\mathcal{W}_{1,g}}
\end{equation} 
associated to the phases \eqref{phase:Ttransf} admit a simple description in terms of an unknot with framing number $n_g$, so that its linking matrix is simply equal to $B=(n_g)$. Moreover, the vectors $s,\,a,\,b$ that define the various possible structures \eqref{def:spin}, \eqref{def:H} and \eqref{def:spinZ2m} of $\mathcal{M}(T^{n_g})$ are fixed by the element $g$. Finally, we note that the group $G'$ that describes the background gauge fields generated by the modular transformation in this case is just $\left\langle g \right\rangle =\Z_{n_g}$. Therefore the value of $\vartheta_T(g)$ is completely fixed via the commutative diagram \eqref{diag-comm}.

We now proceed to compute explicitly the spin selection rules for a few selected cases, namely when
\begin{equation}
    G^f=\Z^f_2\times\Z_4,\,\Z^f_2\times\Z_2\times\Z_2,\,\Z^f_2\times S_3,\,\Z^f_2\times S_4,\,\Z^f_2\times D_8,\, \Z^f_8.
    \label{Gf-chosen}
\end{equation}
In the following we are going to use the notation $\mathfrak{ch}^G_2$ to denote the maps 
\begin{equation}
    \mathfrak{ch}^G_2 : RO(G) \longrightarrow \Hom(\Omega^\Spin_3(BG),U(1))
\end{equation}
when $G^f=\Z^f_2 \times G$.

\subsubsection{Case 1: $G^f=\Z^f_2\times \Z_4$}

\begin{table}[tb]
    \centering
    \begin{tabular}{|c||c|c|}
        \hline
        $(s,a)$ & $\vartheta_T(g)$  & $\displaystyle \{s\}_{\CH_{(s,a)}}$\\
        \hline \hline
         $(0,0)$&  0 & $\Z/2$\\
         \hline
         $(0,1)$& $\nu_\gamma/8$ & $\nu_\gamma/32 + \Z/4$\\
         \hline
         $(0,2)$& $\nu_\omega /2 +\nu_\gamma/4 $ & $\nu_\omega /4 +\nu_\gamma/8 + \Z/2 $\\
         \hline
         $(0,3)$& $\nu_\gamma/8$ & $\nu_\gamma/32 + \Z/4$\\
         \hline
         $(1,0)$& $0$ & $\Z$\\
         \hline
         $(1,1)$& $\nu_\omega/2+5\nu_\gamma/8 $ & $\nu_\omega/8+5\nu_\gamma/32 +\Z/4 $\\
         \hline
         $(1,2)$& $\nu_\omega/2+3\nu_\gamma/4 $ & $\nu_\omega/4+3\nu_\gamma/8 +\Z/2$\\
         \hline
         $(1,3)$& $\nu_\omega/2+5\nu_\gamma/8$ & $\nu_\omega/8+5\nu_\gamma/32 +\Z/4 $\\
         \hline
    \end{tabular}
    \caption{Spin selection rule for $G^f=\Z^f_2\times\Z_4$.}
    \label{phases:Z4}
\end{table}

In this case we know explicitly the invariants that describe $\Hom(\Omega_3^\Spin(B\Z_4),U(1))$ from \cite{Grigoletto:2021zyv}, namely 
\begin{align}
    & \gamma_s(a) = \frac{a^T B a}{4} + s^TBa \mod 8,\\
    & \omega_s(a) = \frac{1}{4} \beta_s(a \mod 2) - \frac{1}{2}\gamma_s(a) \mod 2,
\end{align}
where 
\begin{align}
    \beta_s(a) = \frac{(s+a)^T B (s+a) -s^T B s}{2} + 4(\Arf(\mathcal{C}_{s+a})+\Arf(\mathcal{C}_s)) \mod 8.
\end{align}
Note that the $\Arf$ invariant for the links that represent $\M(T^n)$ are always zero for any $n$ \cite{jones1997polynomial, murakami1986recursive}, so we can drop its term. Moreover, here $(s+a)$ is intended to take values in $\Z_2$. 
The generic phase $\vartheta_T(g)$ will be equal to 
\begin{equation}
    \vartheta_T(g)=\frac{\nu_\omega}{2} \omega_s(a) + \frac{\nu_\gamma}{8} \gamma_s(a),
\end{equation}
where $(\nu_\omega,\nu_\gamma) \in \Z_2 \times \Z_8$. Here and later on the couple $(s,a)$ describes the element $g\in \Z^f_2 \times G$ that determine the twist along the space direction. In particular, the cases $s=0$ and $s=1$ will correspond respectively to the cases of $\ns$ and $\r$ boundary condition. 
The results are given in Table \ref{phases:Z4}.

\subsubsection{Case 2: $G^f=\Z^f_2\times \Z_2\times \Z_2$}

Let differentiate the two $\Z^{(i)}_2$ subgroups by the index $i=1,2$. Then the cobordism group is described by the triple $(\nu_1,\nu_2,\nu_\delta)\in \Z^{(1)}_8\times\Z^{(2)}_8\times \Z_4$. The generators are given by the invariants $\beta_s(a_i)$ and $\delta(a_1,a_2)$, where
\begin{align}
\delta(a_1,a_2)&= \frac{\beta_s(a_1+a_2)-\beta_s(a)-\beta_s(b)}{2} \mod 4
\end{align}
and $a_i\in\Z^{(i)}_2$. 
The spin selection rules can be found in Table \ref{phases:Z2Z2}.
\begin{table}[t]
    \centering
    \begin{tabular}{|c||c|c|}
    \hline
        $(s,a_1,a_2)$& $\vartheta_T(g)$ & $\{s\}_{\CH_{(s,a_1,a_2)}}$\\
        \hline\hline
        $(0,0,0)$   & $0$  & $\Z/2$  \\
        \hline
        $(0,0,1)$   & $\nu_2/8$  & $\nu_2/16+\Z/2$  \\
        \hline
        $(0,1,0)$   & $\nu_1/8$  & $\nu_1/16+\Z/2$  \\
        \hline
        $(0,1,1)$   & $\nu_1/8+\nu_2/8-\nu_\delta/4$  & $\nu_1/16+\nu_2/16-\nu_\delta/8+\Z/2$  \\
        \hline
        $(1,0,0)$   & $0$  & $\Z$  \\
        \hline
        $(1,0,1)$   & $-\nu_2/8$  & $-\nu_2/16+\Z/2$ \\
        \hline
        $(1,1,0)$   & $-\nu_1/8$  & $-\nu_1/16+\Z/2$  \\
        \hline
        $(1,1,1)$   & $-\nu_1/8-\nu_2/8+\nu_\delta/4$  & $-\nu_1/16-\nu_2/16+\nu_\delta/8+\Z/2$  \\        
    \hline    
    \end{tabular}
    \caption{Spin selection rule for $G^f=\Z^2_f\times\Z_2\times \Z_2$.}
    \label{phases:Z2Z2}
\end{table}

\subsubsection{Case 3: $G^f=\Z^f_2\times S_3$}

We recall that the commutative diagram of $S_3$
\begin{equation}
\begin{tikzcd}
    \Z_2 \ar[r,"i_a",swap] \ar[rr, bend left =30,"\text{id}",dashed] & S_3 \ar[r,"\pi",swap] & \Z_2 \\
    \Z_3 \ar[ru,"i_b",swap]  & &
\end{tikzcd}
\label{S3-pullbacks-diagram}
\end{equation}
completely fixes $\Hom (\Omega_3^\Spin(BS_3),U(1))\cong \Z_8 \times \Z_3$ and that the map $\hat{\pi}^*$ is injective. By means of the commutative diagram
\begin{equation}
\begin{tikzcd}
 RO(\Z_3) \ar[d,"\mathfrak{ch}^{\Z_3}_2"] & RO(S_3)  
\ar[d,"\mathfrak{ch}^{S_3}_2"] \ar[l,"i^*_b"]
& RO(\Z_2) \ar[l,"\pi^*"]  \ar[d,"\mathfrak{ch}^{\Z_2}_2"] \\
\Z_3 & \Z_8\times \Z_3 \ar[l,"\hat{i}^*_b"] & \Z_8 \ar[l,"\hat{\pi}^*"] 
\end{tikzcd}.
\end{equation}
one can check that indeed the short sequence given by $\hat{\pi}^*$ and $\hat{i}^*_b$ is exact and split. Therefore we can simply define the elements $(p,q)\in \Z_8 \times \Z_3 \cong \Hom (\Omega^\Spin_2(BS_3),U(1))$ by the maps
\begin{equation}
    \hat{\pi}^*(p)=(p,0), \qquad \hat{i}^*_a(p,q)=p, \qquad \hat{i}^*_b(p,q) = q.
\end{equation}
The two non-trivial conjugacy classes of $S_3$ are generated by its elements of order $2$ and $3$, corresponding to the two subgroups $\Z_2$ and $\Z_3$. Having figured out the maps $\hat{i}^*_{a,b}$, the spin selection rules follow easily, see Table \ref{phases:S3}. Note that the spin selection rule is defined uniquely for elements in the same conjugacy class in $G^f$, as one expects from consistency.
\begin{table}[t]
    \centering
    \begin{tabular}{|c||c|c|}
        \hline
        $\{(s,g)\}$& $\vartheta_T(g)$ & $\{s\}_{\CH_{(s,g)}}$\\
        \hline\hline
        $\{(0,())\}$ & $0$ &$\Z/2$ \\
        \hline
        $\{(0,(ab))\}$ & $p/8$ &$p/16 +\Z/2$ \\
        \hline
        $\{(0,(abc))\}$ & $-q/3$ &$-q/18 + \Z/6$\\
        \hline
        $\{(1,())\}$ & $0$ &$\Z$ \\
        \hline
        $\{(1,(ab))\}$ & $-p/8$ &$-p/16+\Z/2$\\
        \hline
        $\{(1,(abc))\}$ &  $q/3$ &$q/9 + \Z/3$\\
        \hline
    \end{tabular}
    \caption{Spin selection rule for $G^f=\Z^f_2\times S_3$. The couples $\{(s,g)\}$ represent the various conjugacy classes of $G^f$. The elements of $S_3$ here are represented by the standard cycle notation used to describe permutation of three elements.}
    \label{phases:S3}
\end{table}

\FloatBarrier
\subsubsection{Case 4: $G^f=\Z^f_2\times S_4$}

In this case the cobordism group is not yet known, so we start by looking at the relevant homomorphisms from and into $S_4$:
\begin{equation}
\begin{tikzcd}[>=stealth]
\Z_2 \ar[d,"i_4"]\ar[rr,dashed,bend left=30,"\phi"] \ar[r,"i_{1,2,d}",swap] & \Z_2\times\Z_2 \ar[d,"j_1"] & \Z_2\\
\Z_4 \ar[r,"j_2"] & S_4 \ar[rd,"\pi_1"] \ar[ru,"\pi_4"] & \\
A_4 \ar[ru,"j_4"] & S_3 \ar[u,"j_3"] \ar[r,dashed,"\id"] & S_3 \ar[uu,"\tilde{\pi}"]\\
\end{tikzcd}.
\label{S4-pullbacks-diagram}
\end{equation}
The maps are defined by
\begin{align}
&i_1(1) = (1,0), && i_2(1)=(0,1), && i_d(1)=(1,1), \\
& i_4(1)=2, &&j_1(n,m) = (ab)^n (cd)^m, && j_2(n) = (acbd)^n,
\end{align}
while $j_{3,4}$ are the natural inclusions. The projections are defined by the property $\ker\pi_i = \im j_i$. The diagram is commutative if we drop $i_{1,2}$, so that we are left with $i_d$. In this case $\phi = \tr$. Instead, if we consider the maps $i_{1,2}$ and drop $i_d,i_4$ then the commutativity of the diagram is restored with $\phi=\id$.

From the super-cohomology description of the cobordism groups we are able to constrain the order of $\Hom(\Omega_3^\Spin(BS_4),U(1))$.
In particular we have that 
\begin{equation}
  H^3(BS_4;U(1))=\Z_2 \times \Z_4 \times \Z_3,\quad H^2(BS_4;\Z_2)=\Z_2 \times \Z_2,\quad H^1(BS_4;\Z_2)=\Z_2,
\end{equation}
so $2^4\cdot 3 \le |\Hom(\Omega_3^\Spin(BS_4),U(1))|\le 2^6\cdot 3$.
The order of the group can be actually computed by means of \eqref{S4-pullbacks-diagram}. The first thing to notice is that by functoriality $\pi_1 \circ j_3=\id_{S_3}$ implies $\hat{\pi}^*_1:\Z_8\times \Z_3\rightarrow\Hom(\Omega_3^\Spin(BS_4),U(1))$ is injective and thus $\Z_2\times \Z_8\times \Z_3 \subseteq \Hom(\Omega_3^\Spin(BS_4),U(1))$.

Next we focus on the $\Z_2\times\Z_2$ and $\Z_4$ subgroups of $S_4$. By looking at the character tables (see Appendix \ref{app:A}) one finds that $j^*_1:RO(S_4)\rightarrow RO(\Z_2 \times \Z_2)$ is defined by 
\begin{equation}
\begin{array}{rlrl}
    j^*_1(\rho_a)&=[1,1],&j^*_1(\rho_b)&=[0,0]+[1,1],\\
    j^*_1(\rho_c)&=[0,1]+[1,0]+[0,0],&j^*_1(\rho_d)&=[0,1]+[1,0]+[1,1]. 
\end{array}
\end{equation}
It follows also that the map $j^*_2:RO(S_4) \rightarrow RO(\Z_4)$ is surjective, as it maps
\begin{equation}
\begin{array}{rlrl}
j^*_2(\rho_a)&=[2], & j^*_2(\rho_b)&=[0]+[2],\\
j^*_2(\rho_c)&=[2]+\crep{1}, & j^*_2(\rho_d)&=[0]+\crep{1}.
\end{array}
\end{equation}
The maps $\mathfrak{ch}^{\Z_2\times\Z_2,\Z_4}_2$ are defined by 
\begin{equation}
\begin{array}{rclcrcl}
    \mathfrak{ch}^{\Z_2\times\Z_2}_2:RO(\Z_2\times \Z_2) & \rightarrow & \Z_8\times \Z_8\times \Z_4&,&  \mathfrak{ch}^{\Z_4}_2:RO(\Z_4) & \rightarrow & \Z_2\times \Z_8.\\
    \left[1,0\right] &\mapsto & (1,0,0) && \crep{1}&\mapsto & (0,1)\\
    \left[0,1\right] &\mapsto & (0,1,0) &&     [2]&\mapsto & (1,2)\\
    \left[1,1\right] &\mapsto & (1,1,1)
\end{array}
\end{equation}
Recall that in both cases we have at our disposal the commutative diagram 
\begin{equation}
\begin{tikzcd}
RO(G_i) \ar[d,"\mathfrak{ch}^{G_i}_2"] & RO(S_4) \ar[d,"\mathfrak{ch}^{S_4}_2"] \ar[l,"j^*_i"]\\
\Hom(\Omega_3^\Spin(BG_i),U(1)) &  \Hom(\Omega_3^\Spin(BS_4),U(1)) \ar[l,"\hat{j}^*_i"]
\end{tikzcd},
\end{equation}
where $G_1=\Z_2\times\Z_2$ and $G_2=\Z_4$. By denoting $f_i:=\hat{j}^*_i \circ \mathfrak{ch}^{S_4}_2= \mathfrak{ch}^{G_i}_2\circ j^*_i$, we have  
\begin{align}
f_1 &=(n_a+n_b+n_c+2n_d \mod 8, n_a+n_b+n_c+2n_d \mod 8, n_a + n_b + n_d\mod 4),\\
f_2 &= (n_a+n_b+n_c \mod 2, 2(n_a+n_b+n_c)+n_c+n_d\mod 8).
\end{align}
Here $n_\rho$ represent the multiplicity of the generators $\crep{\rho}$ of $RO(S_4)$.
It follows that $\hat{j}^*_2$ is surjective and that $\im(\hat{j}^*_1)\supseteq\im(f_1)=\Z_8\times\Z_4$. 

Next we focus on the kernels of $\hat{j}^*_{1,2}$. We know that they satisfy
\begin{equation}
    |\ker \hat{j}^*_i|\ge |\mathfrak{ch}^{S_4}_2(\ker f_i)| \ge |f_k(\ker f_i)|,\qquad i\neq k.
\end{equation}
and from the super-cohomology description we already know that $|\ker \hat{j}^*_{k}| \le 2^k$. We have
\begin{equation}
    \ker f_1:\begin{cases}
    n_a+n_b+n_d=4N_1,\\
    n_c+n_d=8M_1+4N_1,
    \end{cases}\qquad
    \ker f_2:\begin{cases}
    n_a+n_b+n_c=2N_2,\\
    n_c+n_d=8M_2+4N_2.
    \end{cases}
\end{equation}
Therefore 
\begin{align*}
    f_1(\ker f_2) &=(2(N_2+n_d) \mod 8,2(N_2+n_d) \mod 4)\cong \Z_4,\\
    f_2(\ker f_1) &= (0 \mod 2, 4(N_1+n_c) \mod 8) \cong \Z_2.
\end{align*}
This means there are two exact sequences 
\begin{align*}
    1 \rightarrow \Z_2 \xrightarrow{} \mathcal{E}(\Hom(\Omega_3^\Spin(BS_4),U(1))) \xrightarrow{} \Z_8 \times \Z_4\rightarrow 1,\\
    1 \rightarrow \Z_4 \xrightarrow{} \mathcal{E}(\Hom(\Omega_3^\Spin(BS_4),U(1))) \xrightarrow{} \Z_8 \times \Z_2\rightarrow 1,
\end{align*}
where $\mathcal{E}(\cdot)$ denotes the $2$-torsion subgroup. 
This implies that the map $\mathfrak{ch}^{S_4}_2$ is surjective and that the cobordism group is one between $\Z_8\times\Z_8\times\Z_3$, $\Z_{16}\times\Z_4\times\Z_3$ and $\Z_8\times \Z_4 \times \Z_2\times\Z_3$. For the computation of the spin selection rule it is enough to be able to describe it from a set point of view, under which we can identify it by
\begin{equation}
    \Hom(\Omega_3^\Spin (BS_4),U(1)) \overset{\mathrm{set}}{\cong} \Z_8 \times \Z_4\times \Z_2\times \Z_3.
\end{equation}
\begin{table}[t]
    \centering
    \begin{tabular}{|c||c|c|c|}
    \hline
        $\{(s,g)\}$ & $\vartheta_T(g)$ & $\{s\}_{\CH_{(s,g)}}$  \\
    \hline \hline
        $\{(0,())\}$ & $0$ & $\Z/2$ \\ 
    \hline
        $\{(0,(ab))\}$ &$p/8$ & $p/16+\Z/2$\\ 
    \hline
        $\{(0,(ab)(cd))\}$ &$(p-q)/4$ &$(p-q)/8+\Z/2$ \\ 
    \hline
        $\{(0,(abcd))\}$ &$(3q-p)/8+r/2$&$(3q-p)/32+r/8+\Z/4$ \\
    \hline
        $\{(0,(abc))\}$ &$-t/3$ & $-t/18+\Z/6$\\        
    \hline
        $\{(1,())\}$ & $0$ & $\Z$\\ 
    \hline
        $\{(1,(ab))\}$ &$-p/8$ & $-p/16+\Z/2$\\ 
    \hline
        $\{(1,(ab)(cd))\}$ &$(q-p)/4$ & $(q-p)/8+\Z/2$ \\ 
    \hline
        $\{(1,(abcd))\}$ &$-(p+q)/8+r/2$ & $-(p+q)/32+r/8+\Z/4$\\
    \hline
        $\{(1,(abc))\}$ &$t/3$ & $t/9+\Z/3$\\        
    \hline    
    \end{tabular}
    \caption{Spin selection rule for $G^f=\Z^f_2\times S_4$.}
    \label{phases:S4}
\end{table}
To describe its elements we will use the $4$-tuple $(p,q,r,t)\in\Z_8 \times \Z_4\times \Z_2\times \Z_3$. The first two terms can be identified with the generators of $\im f_1$, while $r$ and $t$ describe respectively the even and odd torsion part of its kernel. In particular, we have 
\begin{align}
    f^{-1}_1(p,q)&=\begin{cases}
    n_a+n_b+n_d = 4 N_1 + q,\\
    n_c+n_d=8M_1 -4N_1 +p-q,
    \end{cases}\\
    f_2(f^{-1}_1(p,q)) &= (p \mod 2, 3q - p +4(n_c+N_1) \mod 8).
\end{align}
Therefore from the set point of view\footnote{Here the maps $\hat{j}^*_i$ are to be intended only as maps between sets, since we are describing $\Hom(\Omega^\Spin_3(BS_4),U(1))$  from that point of view.} we see that $\hat{j}^*_2(p,q,r,t)=(p \mod 2, 3q-p+4r \mod 8)$. 
Following the same line of thought and looking at the commutative diagram \eqref{S4-pullbacks-diagram} we get the following set of maps between cobordism groups\footnote{Via the commutative diagram associated to $i_d:\Z_2^{(d)}\hookrightarrow \Z_2\times \Z_2$ and $i_4:\Z^{(d)}_2\hookrightarrow\Z_4$ one finds that 
\begin{equation}
\begin{array}{rcl}
    \hat{i}^*_d:\Z_8\times\Z_8\times\Z_4 & \rightarrow & \Z_8\\
    (a,b,c)&\mapsto& a+b-2c \mod 8
\end{array},\qquad
\begin{array}{rcl}
    \hat{i}^*_4:\Z_2\times\Z_8 & \rightarrow & \Z_8\\
    (a,b)&\mapsto& 2b-4a \mod 8
\end{array}.
\end{equation}}
\begin{equation}
\begin{tikzcd}[>=stealth]
& {\Z_8\times\Z_4\times \Z_2\times \Z_3}^{(p,q,r,t)} \ar[ld,"\hat{j}^*_3",swap] \ar[d,"\hat{j}^*_1"] \ar[rd,"\hat{j}^*_2"] & \\
{\Z_8\times \Z_3}^{(p,t)} \ar[d,"\hat{i}^*_b"] \ar[dr,"\hat{i}^*_b",swap] & {\Z_8\times\Z_8\times\Z_4}^{(p,p,q)} \ar[d,"\hat{i}^*_{1,2}"] \ar[rd,"\hat{i}^*_d",swap] & {\Z_2\times \Z_8}^{(p,3q-p+4r)} \ar[d,"\hat{i}^*_4"]\\
{\Z_3}^{(t)} & {\Z_8}^{(p)} & {\Z_8}^{(2p-2q)}
\end{tikzcd}.
\label{S4-maps}
\end{equation}
Here the elements in the parenthesis specify elements in the corresponding groups. 
These results allow us to find the spin selection rule for all the Hilbert spaces twisted by any elements of any conjugacy class, see Table \ref{phases:S4}.

\subsubsection{Case 5: $G^f=\Z^f_2\times D_8$}

Like before, we start again by looking at the super-cohomology description in order to gain knowledge on the order of $\Hom(\Omega^\Spin_3(BD_8),U(1))$. We have 
\begin{equation*}
    H^3(BD_8;U(1)) = \Z_2 \times \Z_2 \times \Z_4, \quad H^2(BD_8;\Z_2) = \Z_2 \times \Z_2 \times \Z_2, \quad H^1(BD_8;\Z_2)=\Z_2 \times \Z_2.
\end{equation*}
This means that $2^6\le|\Hom(\Omega_3^\Spin(BD_8),(1))|\le 2^9$. We start by considering a set of different homomorphisms all described by the same sequence
\begin{equation}
    \Z_2 \xrightarrow{i_x} \Z_2 \times \Z_2 \xrightarrow{j_y} D_8 \xrightarrow{\pi_z} \Z_2.
    \label{seqD8}
\end{equation}
Here $i_x$ and $j_y$ are defined by 
\begin{equation}
i_x(1)=x,\quad j_y(1,0)=(y,1),\quad j_y(0,1)=(2,0),
\end{equation}
where we are using the notation $(n,m)\in D_8\cong \Z_4 \rtimes \Z_2$. Instead, $\pi_z:D_8\rightarrow\Z_2$ defines the homomorphism with $\ker\,\pi_z =\langle (z,1), (2,0)\rangle$. The inequivalent set of choices of sequences \eqref{seqD8} are given by the possible values $y,z=0,1$.

By definition $\pi_y\circ j_y\circ i_{(1,0)}= \tr$, while $\pi_z\circ j_{y}\circ i_{(1,0)}= \id$ for $z\neq y$. Using the latter condition we get that the maps $\hat{i}^*_{(1,0)}\circ \hat{j}^*_x$ are surjective, while $\hat{\pi}^*_z$ are injective. Moreover $\Z_8^{(x)}\equiv \im (\hat{\pi}^*_x) \subseteq \ker (\hat{i}^*_{(1,0)}\circ \hat{j}^*_x)$ and $\Z_8^{(x)} \subseteq \im (\hat{i}^*_{(1,0)}\circ \hat{j}^*_y)$, so the two $\Z_8^{(0)}$ and $\Z_8^{(1)}$ components are disjoint.

More precisely consider now the commutative diagram related to \eqref{seqD8}. We have
\begin{equation}
\begin{tikzcd}
RO(\Z_2\times \Z_2) \ar[d,"\mathfrak{ch}^{\Z_2\times\Z_2}_2"] & RO(D_8) \ar[d,"\mathfrak{ch}^{D_8}_2"] \ar[l,"j^*_x"] & RO(\Z_2) \ar[l,"\pi^*_x"] \ar[d,"\mathfrak{ch}^{\Z_2}_2"] \ar[ll,bend right=20,"\tr",dashed]\\
\Z_8 \times \Z_8 \times \Z_4 &  \Hom(\Omega_3^\Spin(BS_4),U(1)) \ar[l,"\hat{j}^*_x",swap]&\Z_8 \ar[l,"\hat{\pi}^*_x",swap] \ar[ll,bend left=20,"\tr",dashed,swap]
\end{tikzcd}.
\end{equation}
Since $\mathfrak{ch}^{\Z_2}_2$ is surjective, then $\hat{j}^*_x\circ \hat{\pi}^*_x=\tr$ and $\im (\hat{\pi}^*_x) \subseteq \ker \hat{j}^*_x$. Moreover $\im (\hat{j}^*_x)\supseteq \im (\hat{j}^*_x \circ \mathfrak{ch}^{D_8}_2)=\im(\mathfrak{ch}^{\Z_2\times\Z_2}_2\circ j^*_x)$. 

To figure out how the maps $j^*_x$ are defined we need the character table of $D_8$, see Table \ref{char:D8}. We have 
\begin{align}
    &\begin{array}{crcl}
        j^*_0: & RO(D_8) & \rightarrow & RO(\Z^{(0)}_2 \times \Z_2),\\
         & \rho_a & \mapsto & [0,0]\\
         & \rho_b & \mapsto & [1,0]\\
         & \rho_c & \mapsto & [1,0]\\
         & \rho_d & \mapsto & [0,1]+[1,1] 
    \end{array},&&
    \begin{array}{crcl}
        j^*_1: & RO(D_8) & \rightarrow & RO(\Z^{(1)}_2\times\Z_2).\\
         & \rho_a & \mapsto & [1,0]\\
         & \rho_b & \mapsto & [1,0]\\
         & \rho_c & \mapsto & [0,0]\\
         & \rho_d & \mapsto & [0,1]+[1,1]
    \end{array}
\end{align}
This means that defining $f_i := \mathfrak{ch}^{\Z_2\times\Z_2}_2\circ j^*_i$, then
\begin{align}
    f_0 &= (n_b+n_c+n_d \mod 8, 2 n_d \mod 8, n_d \mod 4),\quad \im (f_0) \cong \Z_8 \times \Z_4,\\
    f_1 &= (n_a+n_b+n_d \mod 8, 2 n_d \mod 8, n_d \mod 4),\quad \im (f_1) \cong \Z_8 \times \Z_4.
\end{align}
In particular
\begin{equation}
f_1(\ker f_0) = (n_a-n_c \mod 8, 0 \mod 8,0 \mod 4) \cong \Z_8.
\end{equation}
We then consider the inclusion map $j_2: \Z_4 \hookrightarrow D_8$ finding
\begin{equation}
        \begin{array}{crcl}
        j^*_2: & RO(D_8) & \rightarrow & RO(\Z_4)\\
         & \rho_a & \mapsto & [2]\\
         & \rho_b & \mapsto & [0]\\
         & \rho_c & \mapsto & [2]\\
         & \rho_d & \mapsto & \crep{1}
    \end{array}.    
\end{equation}
Then the image of $f_2:=\mathfrak{ch}^{\Z_4}_2\circ j^*_2$ is such that 
\begin{align}
    f_2 &= (n_a+n_c \mod 2, 2(n_a+n_c)+n_d \mod 8)\\
    f_2(\ker f_0 \cap \ker f_1) & = (0 \mod 2, 4(N+n_b) \mod 8)\cong \Z_2,
    \label{D8:remainingZ2subgroup}
\end{align}
where $n_d=4N$.
Therefore 
\begin{equation}
|\ker \hat{i}^*_0| \ge |\mathfrak{ch}^{D_8}_2(\ker f_0)|= |f_1(\ker f_0)|\cdot |\ker  f_1|_{\ker f_0}|\ge |f_1(\ker f_0)|\cdot |f_2(\ker f_0 \cap \ker f_1)| =2^4.
\end{equation}
But we already know that it cannot be higher that $2^4$. This means that the inequalities are saturated and from the set-point of view we have
\begin{equation}
    \Hom (\Omega^\Spin_3(BD_8),U(1)) \overset{\mathrm{set}}{\cong} \Z_8^{(0)}\times \Z_8^{(1)} \times \Z_4\times\Z_2.
\end{equation}
We denote the elements as $(p,q,r,t)$. In this case the diagram of the cobordism subgroups is\footnote{Here $\hat{i}^*_2$ is the pullback associated to $i_2:\Z_2 \hookrightarrow \Z_4$.}
\begin{equation}
\begin{tikzcd}[>=stealth]
& {\Z^{(0)}_8\times\Z^{(1)}_8\times \Z_4 \times \Z_2}^{(p,q,r,t)} \ar[ld,"\hat{j}^*_0",swap] \ar[d,"\hat{j}^*_2"] \ar[rd,"\hat{j}^*_1"] & \\
{\Z_8\times\Z_8\times\Z_4}^{(p+r,2r,r)} \ar[rd,"\hat{i}^*_{(0,1)}",swap] \ar[d,"i^*_{(1,0)}"] & {\Z_2\times\Z_8}^{(p+q,2(p+q)+r+4t)} \ar[d,"\hat{i}^*_2"] & {\Z_8\times\Z_8\times\Z_4}^{(q+r,2r,r)}  \ar[ld,"\hat{i}^*_{(0,1)}"] \ar[d,"i^*_{(1,0)}"]\\
{\Z_8^{(0)}}^{(p+r)}&{\Z_8}^{(2r)}&{\Z_8^{(1)}}^{(q+r)}
\end{tikzcd}.
\label{D8-maps}
\end{equation}
We took as definition of $(p,q,r,t)=(1,0,0,0)$ and $(p,q,r,t)=(0,0,1,0)$ the generators of $\im (j^*_0) \cong \Z_8\times \Z_4 $, i.e. $j^*(p,q,r,t)=(p+r\mod 8,2r\mod 8,r\mod 4)$. The subgroup $\Z_8 \subset \im (j^*_1)$ is instead generated by $q$, while $t$ defines the remaining subgroup \eqref{D8:remainingZ2subgroup}. The resulting spin selection rule is presented in Table \ref{phases:D8}. 

\begin{table}[tb]
    \centering
    \begin{tabular}{|c||c|c|c|}
    \hline
        $\{(s,g)\}$ & $\vartheta_T(g)$ & $\{s\}_{\CH_{(s,g)}}$  \\
    \hline \hline
        $\{(0,(0,0))\}$ & $0$ & $\Z/2$ \\ 
    \hline
        $\{(0,(1,0))\}$ & $-(p+q)/4+r/8+t/2$ & $-(p+q)/16+r/32+t/8+\Z/4$ \\ 
    \hline
        $\{(0,(2,0))\}$ & $r/4$ & $r/8+\Z/2$ \\ 
    \hline
        $\{(0,(0,1))\}$ & $(p+r)/8$ & $(p+r)/16+\Z/2$ \\ 
    \hline
        $\{(0,(1,1))\}$ & $(q+r)/8$ & $(q+r)/16+\Z/2$ \\ 
    \hline
        $\{(1,(0,0))\}$ & $0$ & $\Z$ \\ 
    \hline
        $\{(1,(1,0))\}$ & $(p+q)/4+5r/8+t/2$ & $(p+q)/16+5r/32+t/8+\Z/4$ \\ 
    \hline
        $\{(1,(2,0))\}$ & $-r/4$ & $-r/8+\Z/2$ \\ 
    \hline
        $\{(1,(0,1))\}$ & $-(p+r)/8$ & $-(p+r)/16+\Z/2$ \\ 
    \hline
        $\{(1,(1,1))\}$ & $-(q+r)/8$ & $-(q+r)/16+\Z/2$ \\ 
    \hline
    \end{tabular}
    \caption{Spin selection rule for $G^f=\Z^f_2\times D_8$. Here the conjugacy classes are represented for convenience by a single representative.}
    \label{phases:D8}
\end{table}

As a final comment, we notice that $D_8\subset S_4$. From our analysis we see that the induced map 
\begin{equation}
    \Hom(\Omega^\Spin_3(BS_4),U(1))\rightarrow \Hom(\Omega^\Spin_3(BD_8),U(1))
\end{equation} 
cannot be surjective since 
\begin{equation}
|\Hom(\Omega^\Spin_3(BS_4),U(1))|<|\Hom(\Omega^\Spin_3(BD_8),U(1))|.
\end{equation} 
Thus, from 
\begin{equation}
\begin{tikzcd}
RO(D_8) \ar[d] & RO(S_4) \ar[d] \ar[l] \\
\Hom(\Omega^\Spin_3(BD_8),U(1)) &  \Hom(\Omega_3^\Spin(BS_4),U(1)) \ar[l]
\end{tikzcd}
\end{equation}
it follows that also $\mathfrak{ch}^{D_8}_2$ cannot be surjective. This means that at least one of the generators of $\Hom(\Omega^\Spin_3(BD_8),U(1))$ is not associated to sets of free fermions, but necessarily to some interacting spin-theory.

\subsubsection{Case 6: $G^f=\Z^f_8$}
In this case the computation of the spin selection rule is straightforward. Here we switch to the additive notation for $\Z^f_8$, so that the Hilbert space $\CH_\mathbb{I}$ now corresponds to $\CH_0$ and the $\ns$ boundary condition to $0\in \Z_8^f$. It follows that the Hilbert spaces $\CH_n$ are associated to the tori with time boundary condition equal to $\ns$ and space boundary condition equal to $\ns$ plus a twist by $n \mod 8$. In this case the element $b$ of \eqref{def:spinZ2m} that identify the  $\Spin\text{-}\Z^f_8$ structure of the correspondent mapping torus is simply $b=n\mod 8$. The spin selection rule for $\CH_{n}$ follows easily and is given in Table \ref{phases:Z8f}.
\begin{table}[h]
    \centering
    \begin{tabular}{|c||c|c|c|}
    \hline
        $n$&$B$&$\vartheta_T(n)$&$\{s\}_{\CH_n}$\\
    \hline
    \hline
        $0$&$2$&$0$&$\Z/2$\\
    \hline
        $1 \mod 2$&$8$&$\nu/2$&$\nu/16+\Z/8$\\
    \hline
        $2 \mod 4$&$4$&$0$&$\Z/4$\\
    \hline
        $4 \mod 8$&$1$&$0$&$\Z$\\
    \hline
    \end{tabular}.
    \caption{Spin selection rule for $G^f=\Z^f_8$. Here for completeness we reported also the value of the linking matrix $B=(n_g)$.}
    \label{phases:Z8f}
\end{table}

\section{Lightest operators constraints from modular consistency}
\label{sec:mod-boot}

We are now going to employ the data related to modular transformations in order to determine constraints on the lightest symmetry-preserving scalar operators for theories with the symmetry groups we discussed in Section \ref{sec:mod-transf}. 

\subsection{Modular bootstrap}
The modular bootstrap technique that we are going to employ is the so-called linear functional method. For its applicability we first need to find a set of partition functions (or combinations of them) that admit a positive expansion over some basis. Since we are interested in studying symmetry-preserving scalar operators on $\CH_\mathbb{I}$, a natural choice is given by starting to consider the sectors $Z^\Gamma$. In particular we will be interested in determining bounds for $Z^{\Gamma_0}$. As we will explain in a moment, in order to apply our bootstrap method we need to complete this set of partition functions so that it is closed under $S$ transformation. Since all the $Z^\Gamma$ admit an expansions in terms of the various $Z^{C_{G^f}}$ and that under $S$ transformation we can safely have
\begin{equation}
    Z^{C_{G^f}} \overset{S}{\longleftrightarrow} Z_{C_{G^f}}=\sum_{g\in C_{G^f}} Z_g,
    \label{S:CGf}
\end{equation}
we can use $\{Z_{C_{G^f}}\}$ to complete such set. By labelling with some index $i=0,1,\ldots,n$ the irreps $\Gamma_i$ and conjugacy classes $C_i$ so that $i=0$ correspond to the trivial ones, we define a $(2n-1)$-dimensional \emph{partition vector}
\begin{equation}
    \mathbf{Z}^T_{G^f}=(Z^{\Gamma_0},Z^{\Gamma_1},\ldots,Z^{\Gamma_n},Z_{C_1},\ldots,Z_{C_n}).
\end{equation}
The $S$ modular transformations can then be recasted into the form (restoring the dependence of the partition functions on $\tau$)
\begin{equation}
    \mathbf{Z}_{G^f}(-1/\tau,-1/\bar{\tau})- S \mathbf{Z}_{G^f}(\tau,\bar{\tau})=0,
    \label{S-transf-boot}
\end{equation}
where the matrix $S$ can be easily found from \eqref{S:CGf}.

Our choice of the partition vector $\mathbf{Z}_{G^f}$ always admits a positive expansion in terms of Virasoro characters
\begin{equation}
    \chi_0(\tau)= (1-q)\frac{q^{-\frac{c-1}{24}}}{\eta(\tau)}, \qquad  \chi_{h>0}(\tau)= \frac{q^{h-\frac{c-1}{24}}}{\eta(\tau)},
\end{equation}
where $q=e^{2\pi i \tau}$.
Indeed, recall that any TDL $\hat{g}$ commutes with the stress tensor and any $\CH_g$ can be organized into representations of the Virasoro algebra with definite conformal weights $(h,\bar{h})$. In particular, for $\CH_\mathbb{I}$ we can express
\begin{equation}
    Z(\tau,\bar{\tau})=\sum_{\Gamma,h,\bar{h}} n_{\Gamma,h,\bar{h}} \chi_h(\tau)\bar{\chi}_{\bar{h}}(\bar{\tau}), \quad n_{\Gamma,h,\bar{h}}\in \Z_+ d_\Gamma,
\end{equation}
so that for any $\Gamma$
\begin{equation}
    Z^\Gamma(\tau,\bar{\tau})=\sum_{h,\bar{h}} n_{\Gamma,h,\bar{h}} \chi_h(\tau)\bar{\chi}_{\bar{h}}(\bar{\tau}).
\end{equation}

Thus $\mathbf{Z}_{G^f}$ is a good partition vector to which applies the linear functional method. Since the $S$ modular transformation define a set of \emph{linear} equations \eqref{S-transf-boot}, by applying a linear functional $\alpha$ on them the equalities will still hold for any consistent spectrum of primaries, i.e.
\begin{equation}
    \alpha[
    \mathbf{Z}_{G^f}(-1/\tau,-1/\bar{\tau})- S \mathbf{Z}_{G^f}(\tau,\bar{\tau})]=0.
    \label{alpha-funct}
\end{equation}
By reverse, if we impose some wrong constraints that define a bad set of primaries in the spectrum, it may happen that \eqref{alpha-funct} does not hold anymore; let us be more precise. Denote as $\mathcal{Z}_j$ the entries of $\mathbf{Z}_{G^f}$ and rewrite their positive expansions in a unified notation
\begin{equation}
    \mathcal{Z}_j=\sum_{h,\bar{h}}n_{j,h,\bar{h}}\chi_h(\tau)\bar{\chi}_{\bar{h}}(\bar{\tau}).
\end{equation}
Then by applying a functional to \eqref{S-transf-boot} we find the set of $2n-1$ equations
\begin{equation}
    \alpha[\delta^j_i \chi_h(-1/\tau) \bar{\chi}_{\bar{h}}(-1/\bar{\tau})-S^j_i \chi_h(\tau) \bar{\chi}_{\bar{h}}(\bar{\tau})]\overset{!}{=}0
    \label{j-funct}
\end{equation}
that must be satisfied for any couple $(h,\bar{h})$ that is within our hypothesis on the spectrum. If instead it happens that for some $j$ we find a functional that returns non-vanishing values all of a definite sign, then we must discard such hypothesis. 

For a more in depth description of the whole process, see \cite{Collier:2016cls,Lin:2019kpn,Grigoletto:2021zyv}. Here we mention that the putative spectra that we are going to work with are constrained by the spin selection rules extracted in Section \ref{sec:mod-transf} and by changing the value of the lightest symmetry-preserving scalar operator in the untwisted Hilbert space $\CH^{(\Gamma_0)}_\mathbb{I}$. Moreover, we are going to assume that the combination of Virasoro characters $(h,\bar{h})=(0,0)$, describing the vacuum of the theory, is present only in $\CH_\mathbb{I}$.

One final technical comment is in order. Since we are going to work with a high number of variables, it is useful to reduce the dimensionality of the systems as much as possible. Indeed, this is further possible for some combinations of the symmetry group and value of the anomaly, without compromising the efficacy of the bounds found. We report in Appendix \ref{app:B} the reduced systems we considered for our cases of interest; for the general discussion see \cite{Lin:2021udi}.

\subsection{Bounds on symmetry-preserving scalar operators}
\label{sec:results}

In the following we present the numerical bounds on the upper limit of the dimension $\Delta$ of the lightest symmetry-preserving scalar operators in $\CH_\mathbb{I}$ for the range of central charges $1\le c \le 10$.
The analysis has been carried with the SDPB solver\footnote{The relevant numerical parameters that were set for the analysis are \texttt{precision=700, primalErrorThreshold=10$^{-30}$, dualErrorThreshold=10$^{-30}$, maxComplementarity=10$^{100}$, feasibleCenteringParameter=0.1, infeasibleCenteringParameter=0.3, stepLengthReduction=0.7} and $s_{\mathrm{max}}=20$.} \cite{simmonsduffin2015semidefinite,landry2019scaling}. 
In order to proceed numerically, one must truncate the functional to work with up to some finite order, e.g. upon choosing an expansion over a derivative basis, truncate it to the form
\begin{equation}
    \alpha[f]=\sum_{n+m\le \Lambda}\partial_n\bar\partial_{\bar m} f(\tau,\bar{\tau})|_{\tau=i,\bar{\tau}=-i}.
\end{equation}
In our case, due to the dimensionality of the systems we considered, we limit to $\Lambda=10$.

Since the groups of anomalies for the symmetry groups $G^f=\Z^f_2\times G$ discussed in Section \ref{sec:mod-transf} are huge, we analyze a few selected and instructive cases for each of them. We also note that up to (outer) automorphisms of $G$, our choices describe various possible anomaly combinations. This means that the bounds we found for a selected value $\nu \in \Hom(\Omega^\Spin_3(BG),U(1))$ are valid also for any $\tilde{\nu}$ such that 
\begin{equation}
    \tilde{\nu}=\hat{\phi}^*(\nu), \quad \phi \in \mathrm{Aut}(G).
\end{equation}
Moreover, the bounds for any anomaly $\nu$ are also equivalent to the bounds associated to $-\nu$. Indeed, for any theory with anomaly $\nu$ we can associate the conjugate theory with complex conjugate partition function. This will have anomaly $-\nu$, implying that the numerical bounds for $\nu$ and $-\nu$ must be the same.

In the upcoming analysis several kinks appear, suggesting the presence of particular spin-theories which saturate the bounds found. For most of these kinks the range of values of $(c,\Delta)$ that they represent imply that their description cannot be found in terms of free fermion theories (where one expects $c,\Delta\in \Z/2$). Thus, the most promising candidates of models that might describe them are given by gauged/ungauged WZW models with a dependence on the spin-structure. 
However, a first analysis shows that between the simplest models with a spin-structure dependence, e.g. of the kind $SO(N)_{2k+1}$, no particular candidate seems to be satisfying. Therefore, being these kinks beyond the critical value $\Delta_{\mathrm{mar}}\equiv 2$ and for the absence of further motivations, we do not provide their description.

\subsubsection*{Case 1: $G^f=\Z^f_2\times \Z_4$}

\begin{figure}[tb]
    \centering
    \includegraphics[scale=0.475]{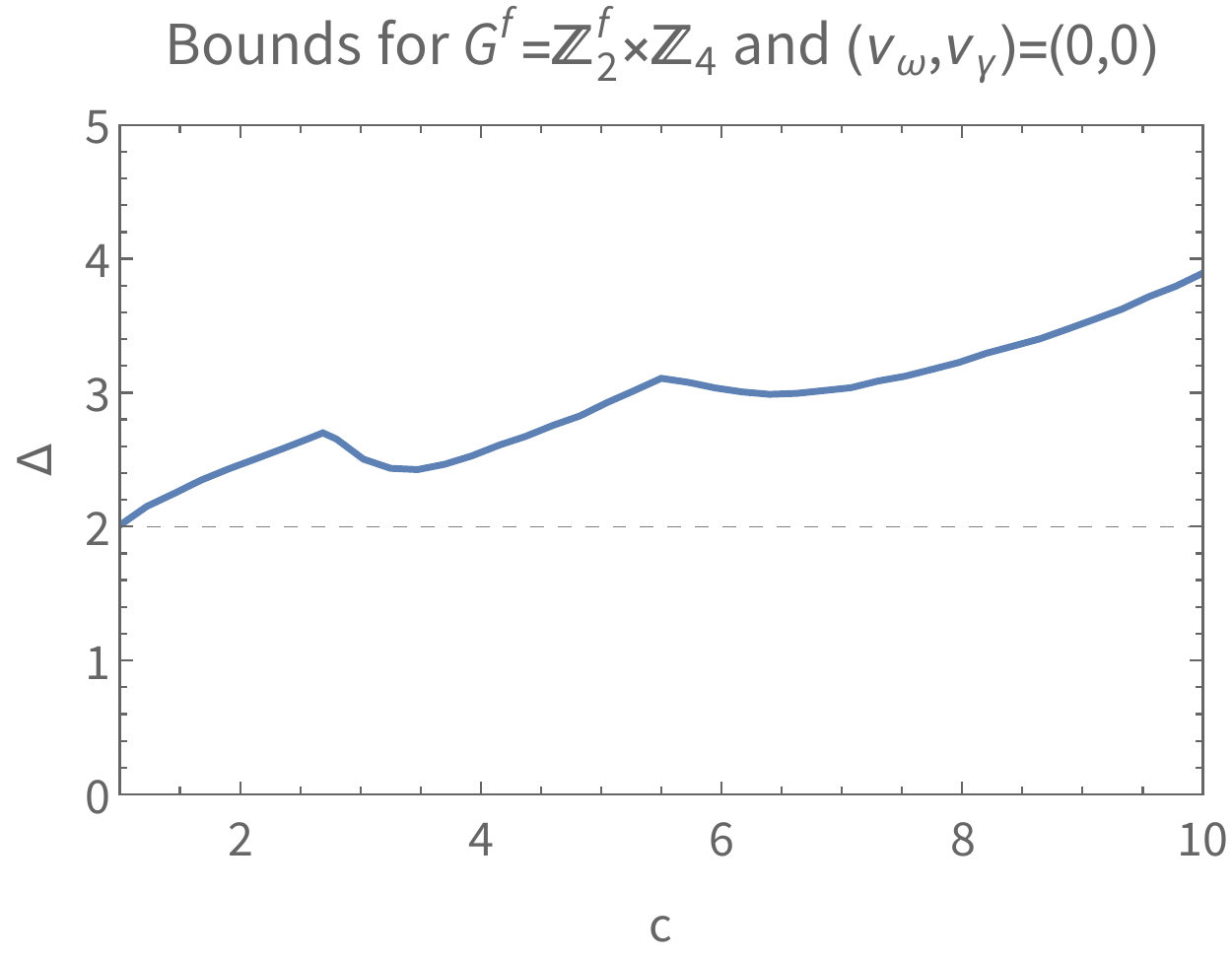}
    \qquad
    \includegraphics[scale=0.475]{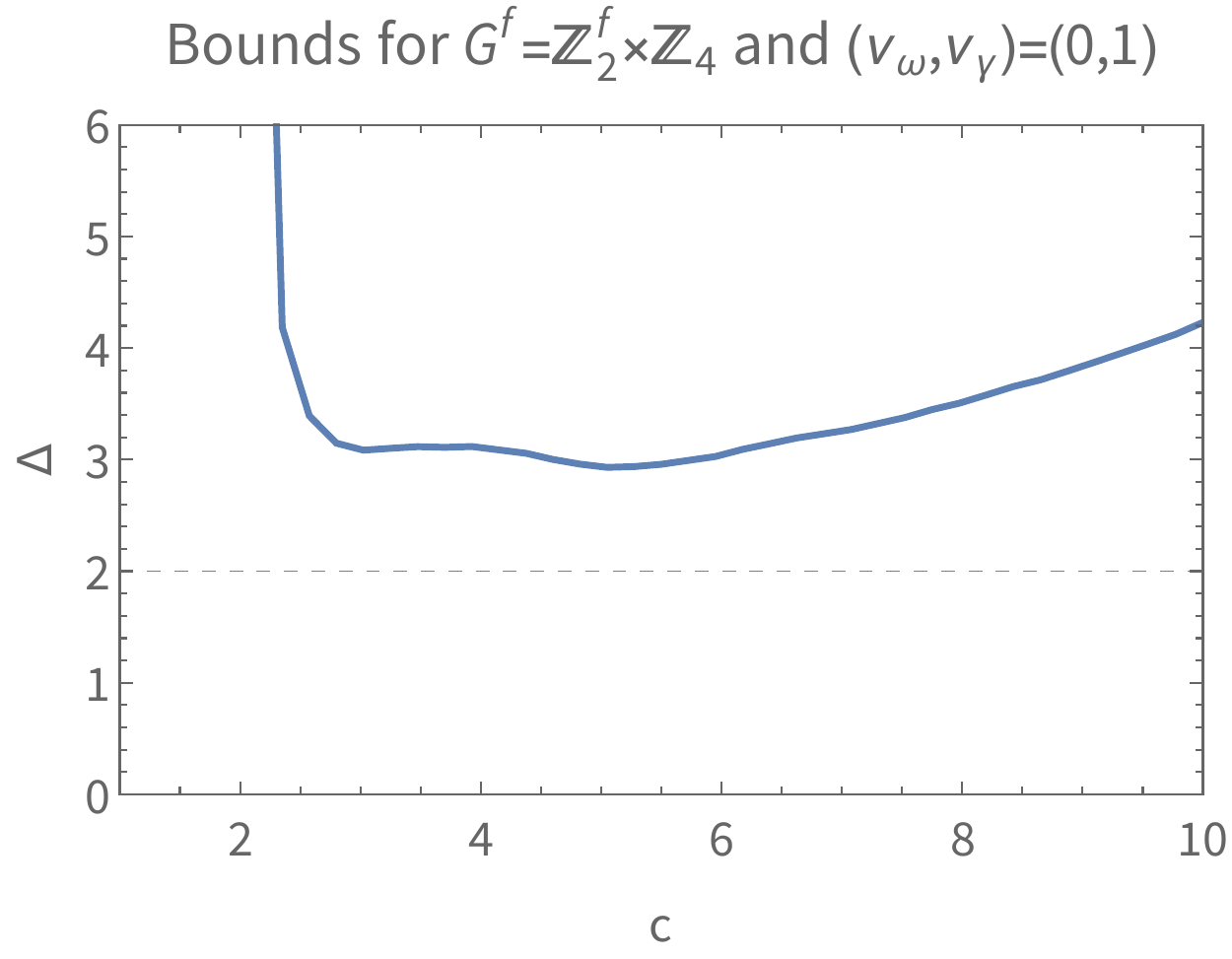}
    \caption{Upper bounds on the lightest $\Z^f_2\times\Z_4$-preserving scalar operators for anomalies $(\nu_\omega,\nu_\gamma)=(0,0)$ (left) and $(\nu_\omega,\nu_\gamma)=(0,1)$ (right).}
    \label{fig:z4v00-01}

\bigskip

    \includegraphics[scale=0.475]{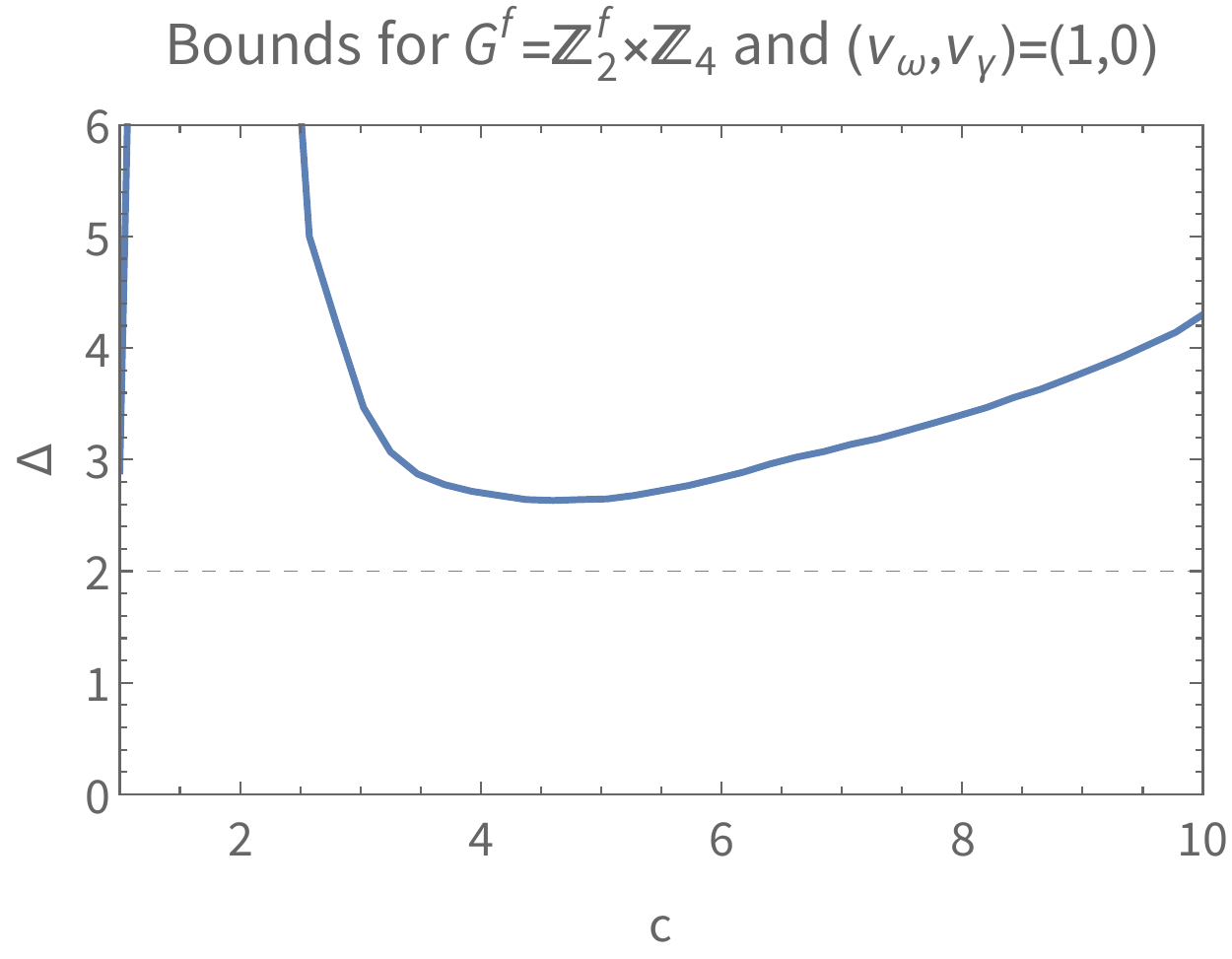}
    \qquad
    \includegraphics[scale=0.475]{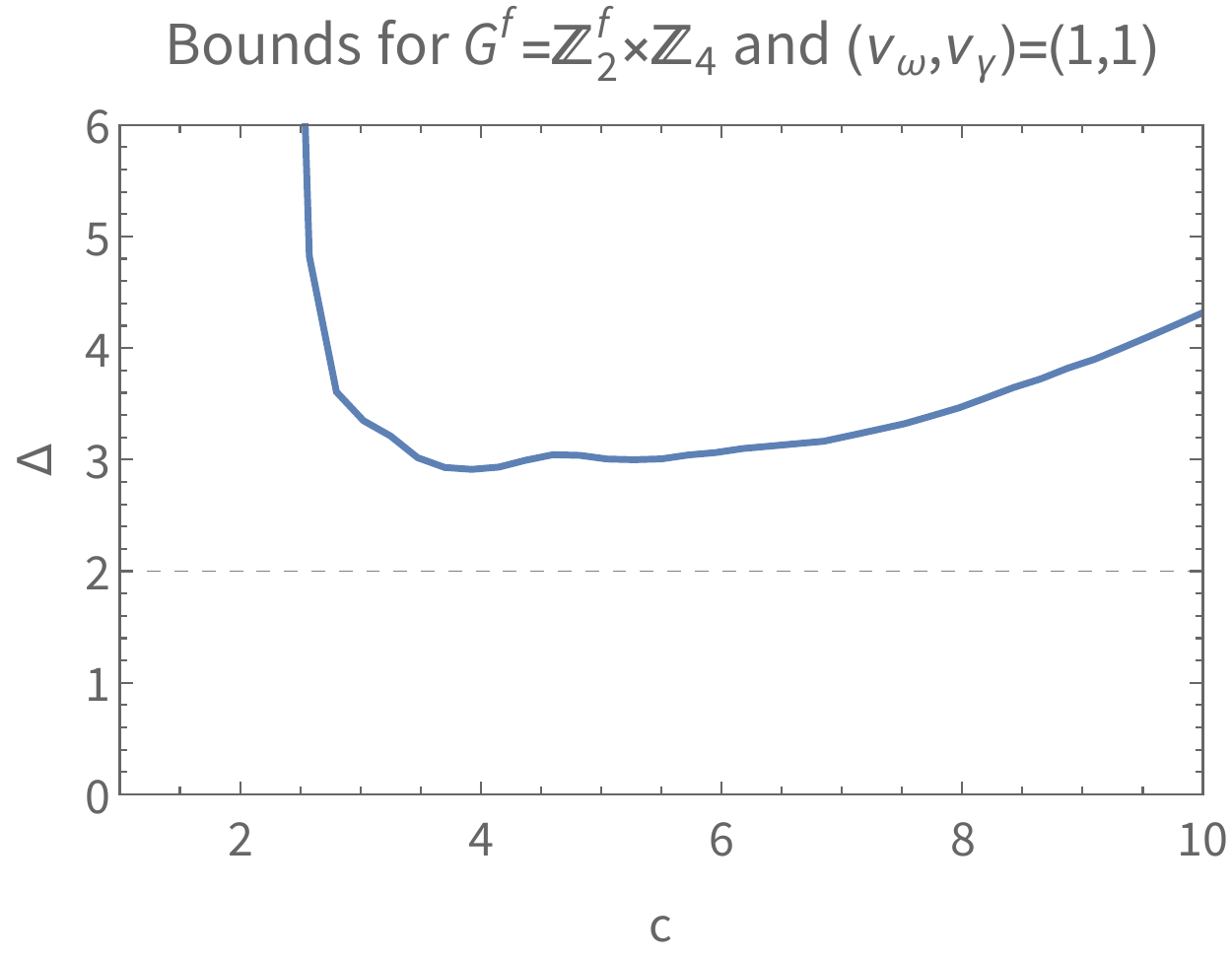}
    \caption{Upper bounds on the lightest $\Z^f_2\times\Z_4$-preserving scalar operators for anomalies $(\nu_\omega,\nu_\gamma)=(1,0)$ (left) and $(\nu_\omega,\nu_\gamma)=(1,1)$ (right).}
    \label{fig:z4v10-11}
\end{figure}

Here the values of the anomaly that were studied are
\begin{align}
    (\nu_\omega,\nu_\gamma)\in\{(0,0),\,(0,1),\,(1,0),\,(1,1)\}\subset \Z_2\times\Z_8.
\end{align}
We found smooth bounds when the anomaly is non-trivial, while for the trivial case two kinks appear around the values 
\begin{align}
    &c= 2.72 \pm 0.08 ,&& \Delta = 2.65 \pm 0.02, \\  
    &c= 5.50 \pm 0.08 ,&& \Delta = 3.10 \pm 0.02. 
\end{align}
Note that our bounds for $c>1$ are always higher that the critical value $\Delta_{\mathrm{mar}}$ for marginal operators.

\FloatBarrier
\begin{figure}
    \centering
    \includegraphics[scale=0.475]{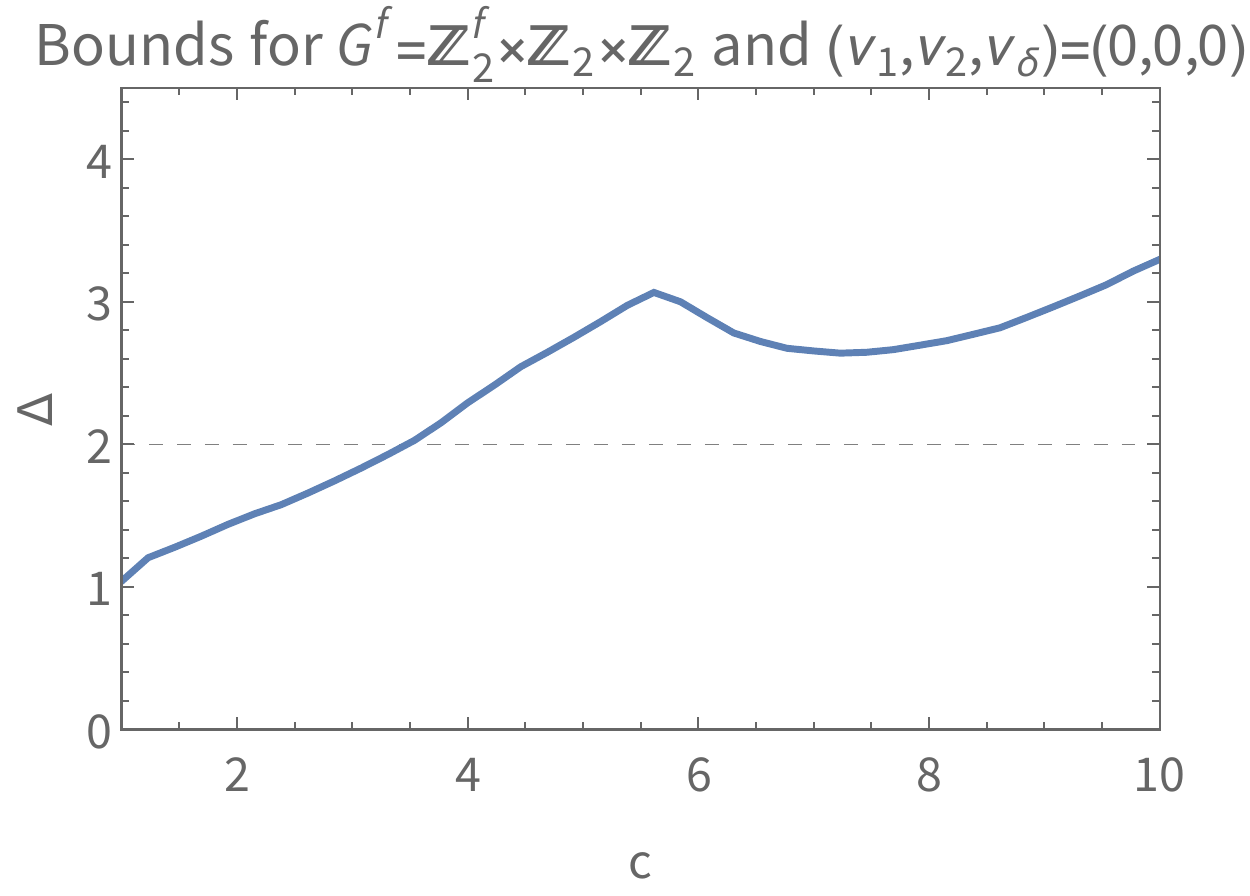}
    \qquad
    \includegraphics[scale=0.475]{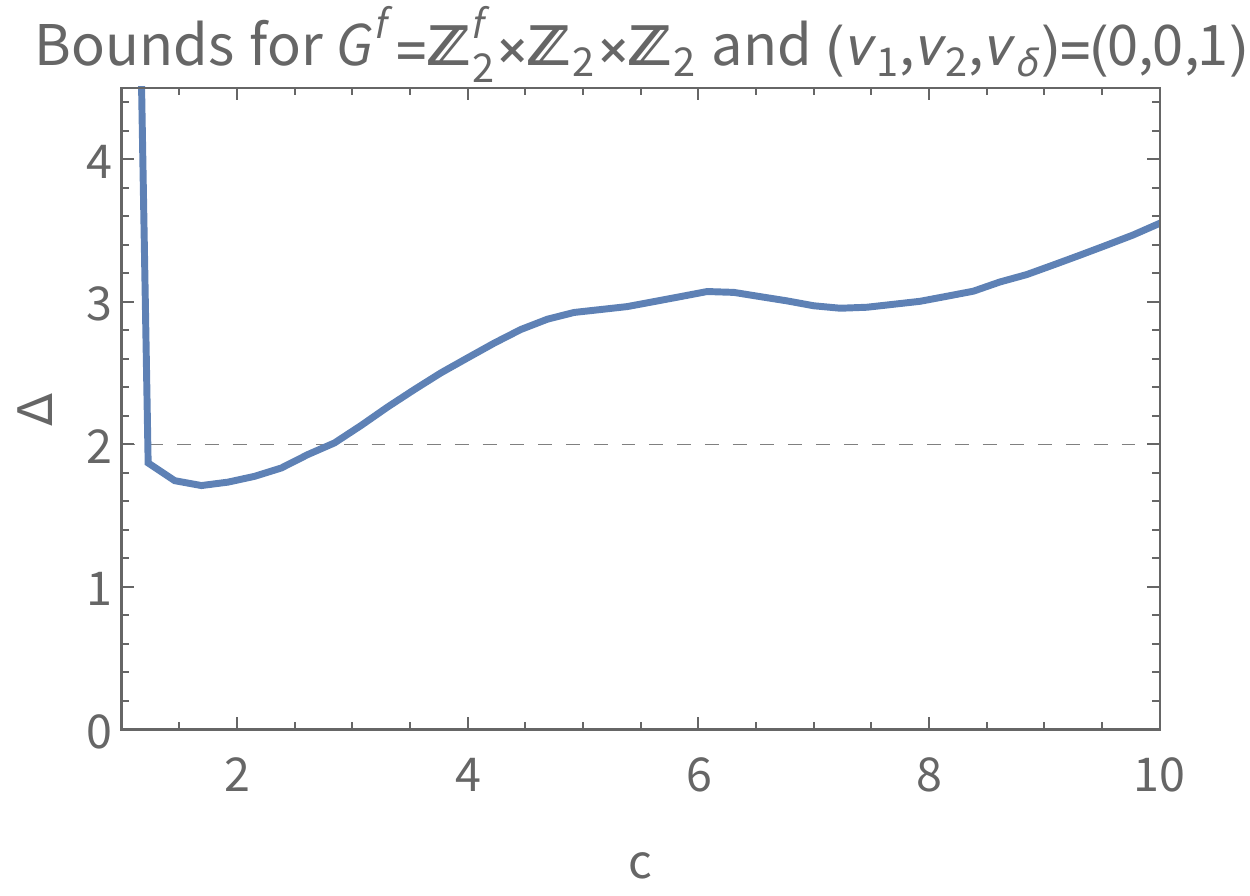}
    \caption{Upper bounds on the lightest $\Z^f_2\times\Z_2\times\Z_2$-preserving scalar operators for anomalies $(\nu_1,\nu_2,\nu_\delta)=(0,0,0)$ (left) and $(\nu_1,\nu_2,\nu_\delta)=(0,0,1)$ (right).}
    \label{fig:z2z2v000-001}

\bigskip

    \includegraphics[scale=0.475]{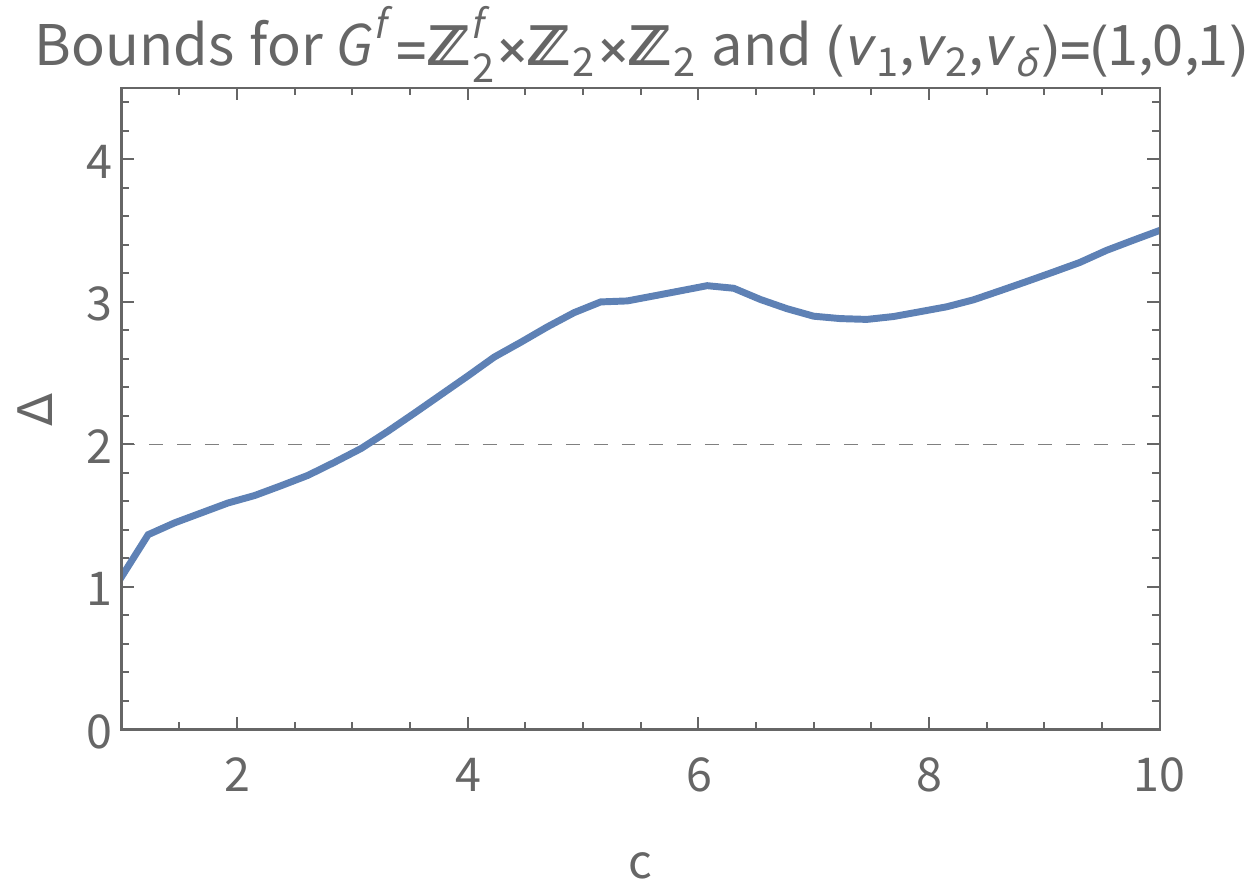}
    \qquad
    \includegraphics[scale=0.475]{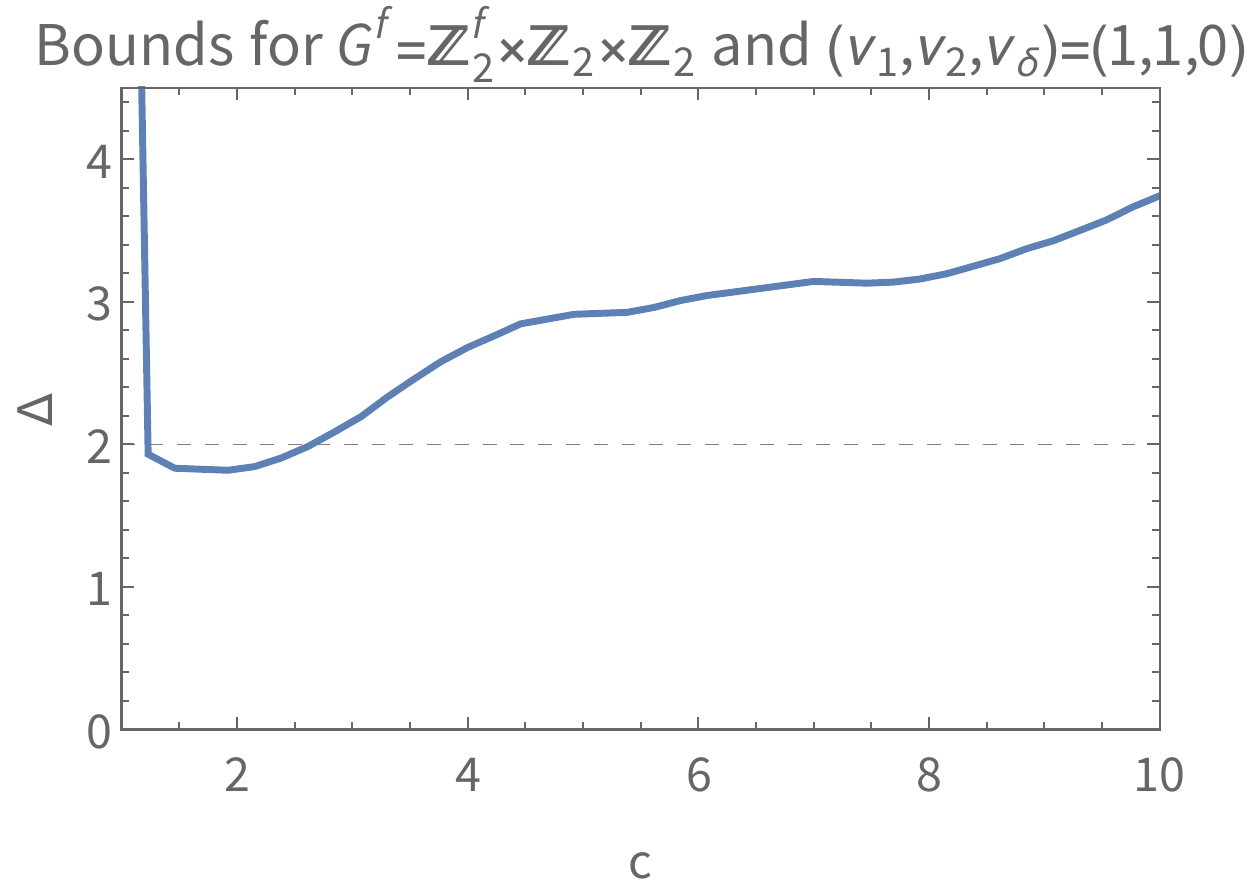}
    \caption{Upper bounds on the lightest $\Z^f_2\times\Z_2\times\Z_2$-preserving scalar operators for anomalies $(\nu_1,\nu_2,\nu_\delta)=(1,0,1)$ (left) and $(\nu_1,\nu_2,\nu_\delta)=(1,1,0)$ (right).}
    \label{fig:z2z2101-110}
\end{figure}

\subsubsection*{Case 2: $G^f=\Z^f_2\times\Z_2\times\Z_2$}

This is the only symmetry group that we analyzed and for which we have been able to find a range of central charges where it is implied the existence of relevant/marginal operators. We focused on anomalies with values 
\begin{equation}
    (\nu_1,\nu_2,\nu_\delta)\in\{(0,0,0),\, (0,0,1),\, (1,0,1),\,(1,1,0)\}\subset \Z_8\times\Z_8\times\Z_4.
\end{equation}
The ranges of central charges which admits $\Delta\le 2$ are 
\begin{align}
     &(\nu_1,\nu_2,\nu_\delta)=(0,0,0)& & 1.00 \le c \lesssim 3.55,\\
     &(\nu_1,\nu_2,\nu_\delta)=(0,0,1)& & 1.23 \lesssim c \lesssim 2.90,\\
     &(\nu_1,\nu_2,\nu_\delta)=(1,0,1)& & 1.00 \le c \lesssim 3.17,\\
     &(\nu_1,\nu_2,\nu_\delta)=(1,1,0)& & 1.23 \lesssim c \lesssim 2.68.   
\end{align}
We found also the presence of a sharp kink in the non-anomalous case at
\begin{align}
    &c= 5.61 \pm 0.08 ,&& \Delta = 3.05 \pm 0.02.
\end{align}

\subsubsection*{Case 3: $G^f=\Z^f_2 \times S_3$}

In this case the values of anomalies studied are 
\begin{equation}
    (p,q)\in\{(0,0),\, (0,1),\, (1,0),\,(1,1)\}\subset\Z_8\times\Z_3.
\end{equation}
No particular kink emerged in the analysis, but only few hills that might be worth to further analyze for the interested reader.

\begin{figure}[tb]
    \centering
    \includegraphics[scale=0.5]{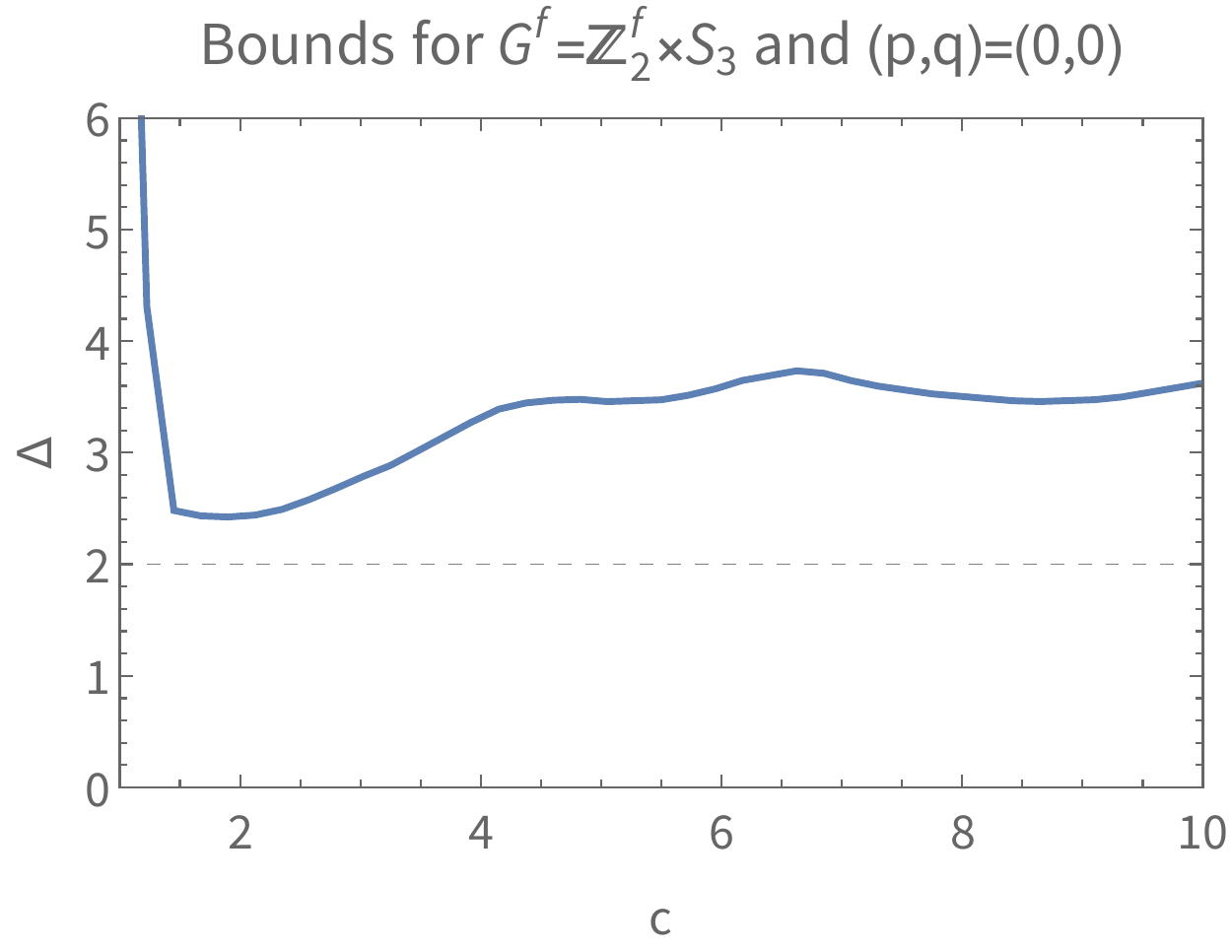}
    \qquad
    \includegraphics[scale=0.5]{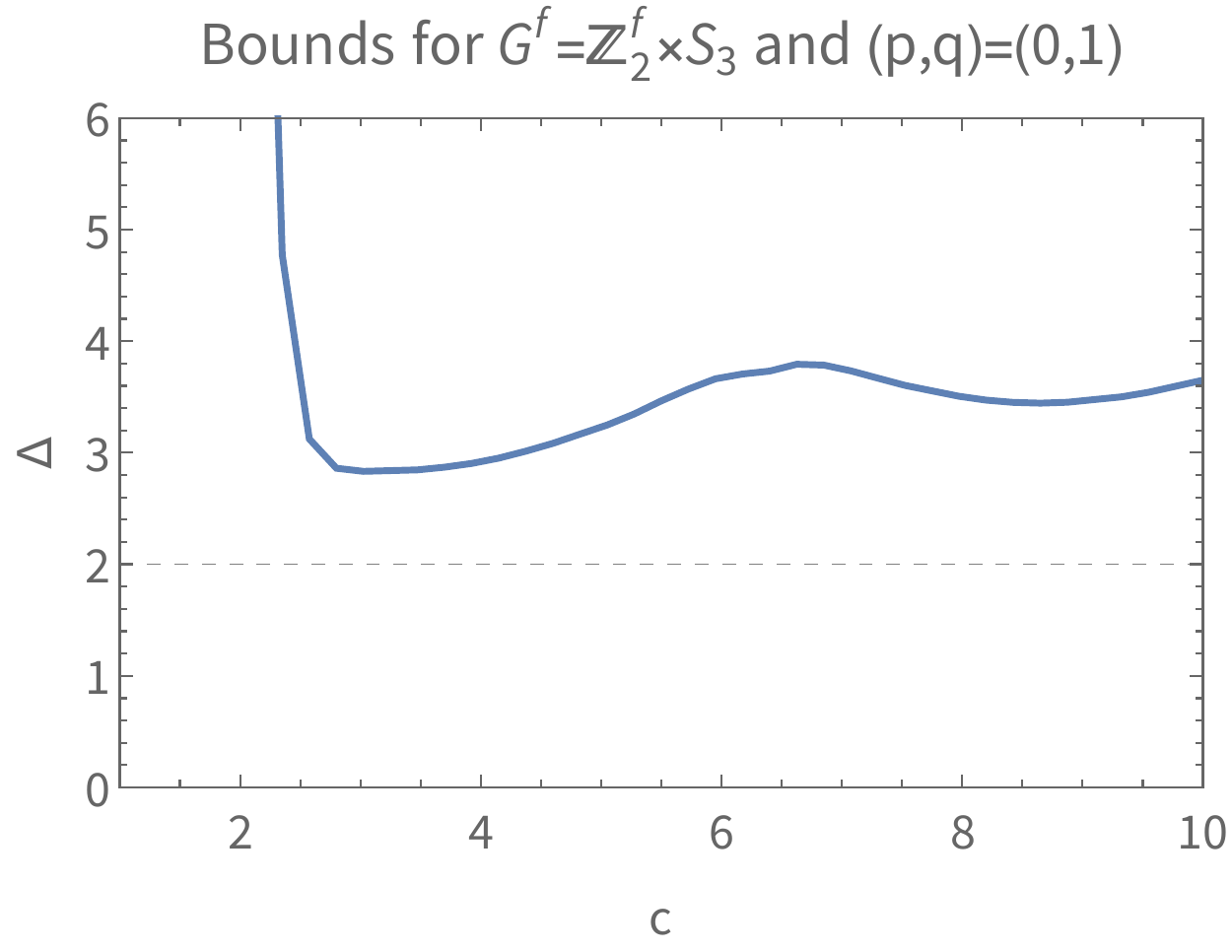}
    \caption{Upper bounds on the lightest $\Z^f_2\times S_3$-preserving scalar operators for anomalies $(p,q)=(0,0)$ (left) and $(p,q)=(0,1)$ (right).}
    \label{fig:s3v00-01}

\bigskip

    \includegraphics[scale=0.5]{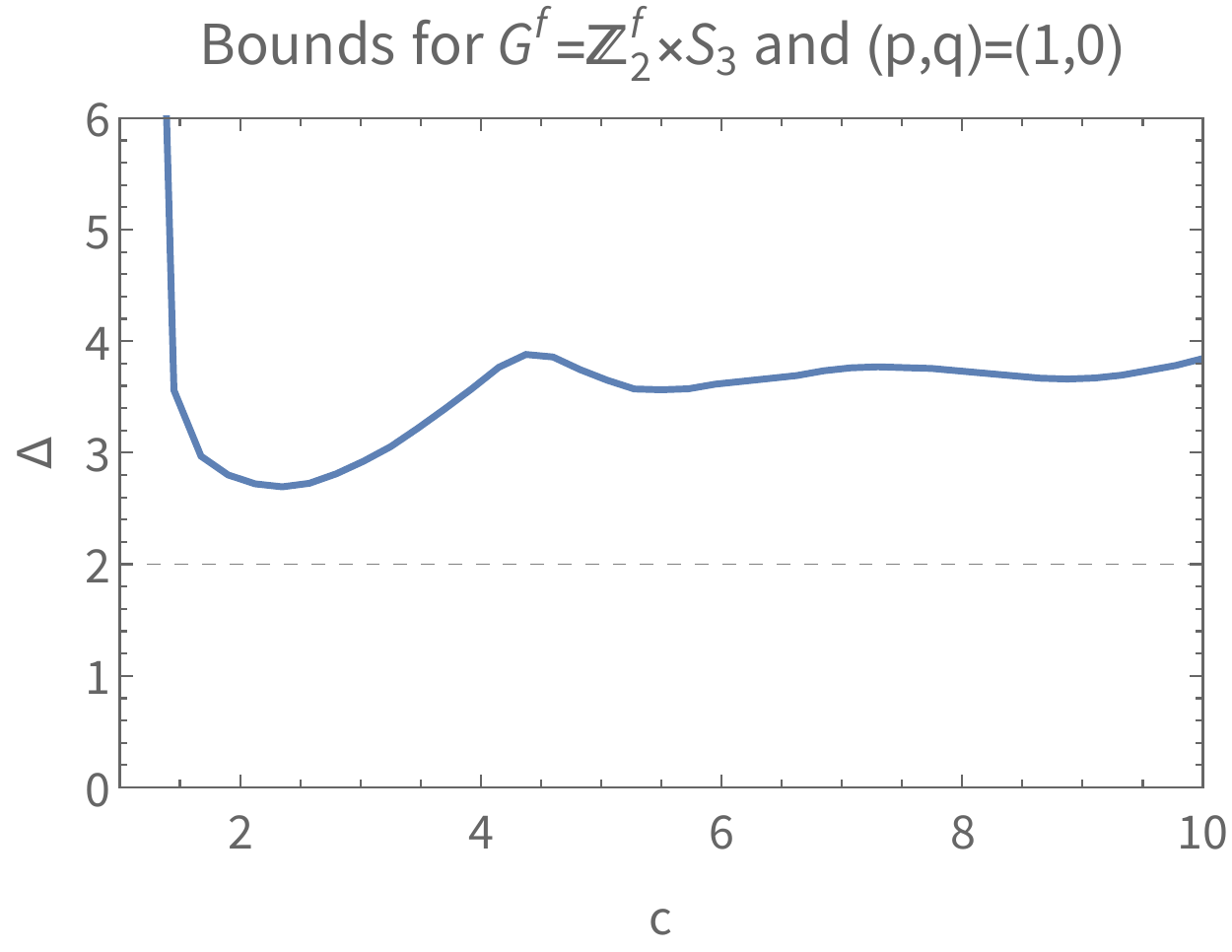}
    \qquad
    \includegraphics[scale=0.5]{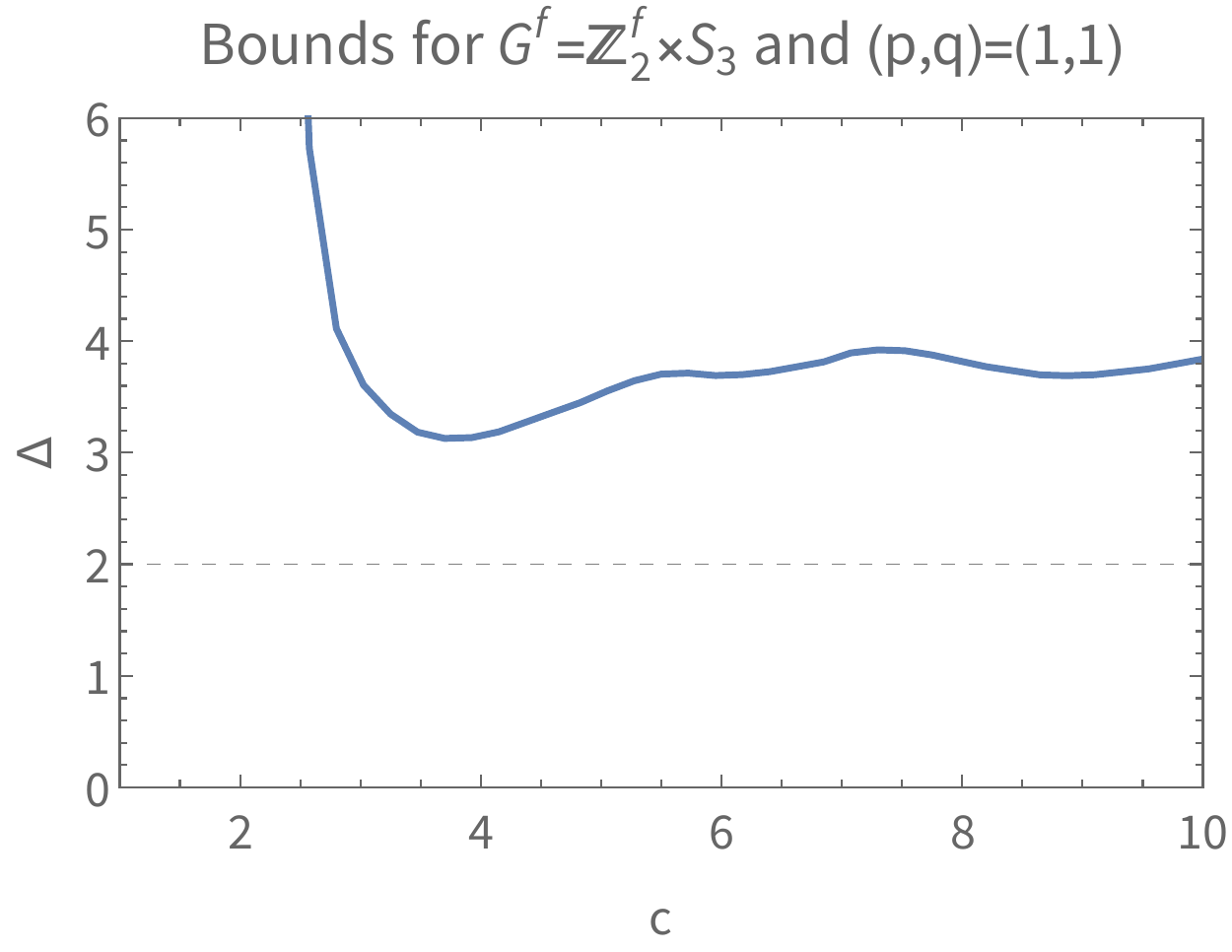}
    \caption{Upper bounds on the lightest $\Z^f_2\times S_3$-preserving scalar operators for anomalies $(p,q)=(1,0)$ (left) and $(p,q)=(1,1)$ (right).}
    \label{fig:s3v10-11}
\end{figure}

\subsubsection*{Case 4: $G^f=\Z^f_2 \times S_4$}
In this case we decided to focus on the anomalies that generate the anomaly group, namely 
\begin{equation}
    (p,q,r,t)\in\{(0,0,0,0),\, (0,0,0,1),\, (0,0,1,0),\,(0,1,0,0),\,(1,0,0,0)\}\subset \Z_8\times\Z_4\times\Z_2\times\Z_3.
    \label{anom-S4}
\end{equation}
Interestingly, we have not found any significant bound of the lightest operator for any of these cases.

\subsubsection*{Case 5: $G^f=\Z^f_2 \times D_8$}

The cases we focused on are 
\begin{equation}
    (p,q,r,t)\in\{(0,0,0,0),\, (0,0,0,1),\, (0,0,1,0),\,(1,1,0,0)\}\subset \Z_8\times\Z_8\times\Z_4\times\Z_2.
    \label{anom-D8}
\end{equation}
Here for the following values of the central charge and anomaly 
\begin{align}
    &(p,q,r,t)=(0,0,0,1), && c\approx 2.15,\\
    &(p,q,r,t)=(1,1,0,0), && c\approx 2.3,
\end{align}
the bound approaches the critical value $\Delta_{\mathrm{mar}}$. Moreover, a set of kinks appear for 
\begin{align}
    &(p,q,r,t)=(0,0,0,0)&&c= 7.00 \pm 0.08 ,&& \Delta = 4.45 \pm 0.02, \\
    &(p,q,r,t)=(1,1,0,0)&&c= 5.7 \pm 0.1 ,&& \Delta = 5.4 \pm 0.04. 
\end{align}

\begin{figure}[tb]
    \centering
    \includegraphics[scale=0.5]{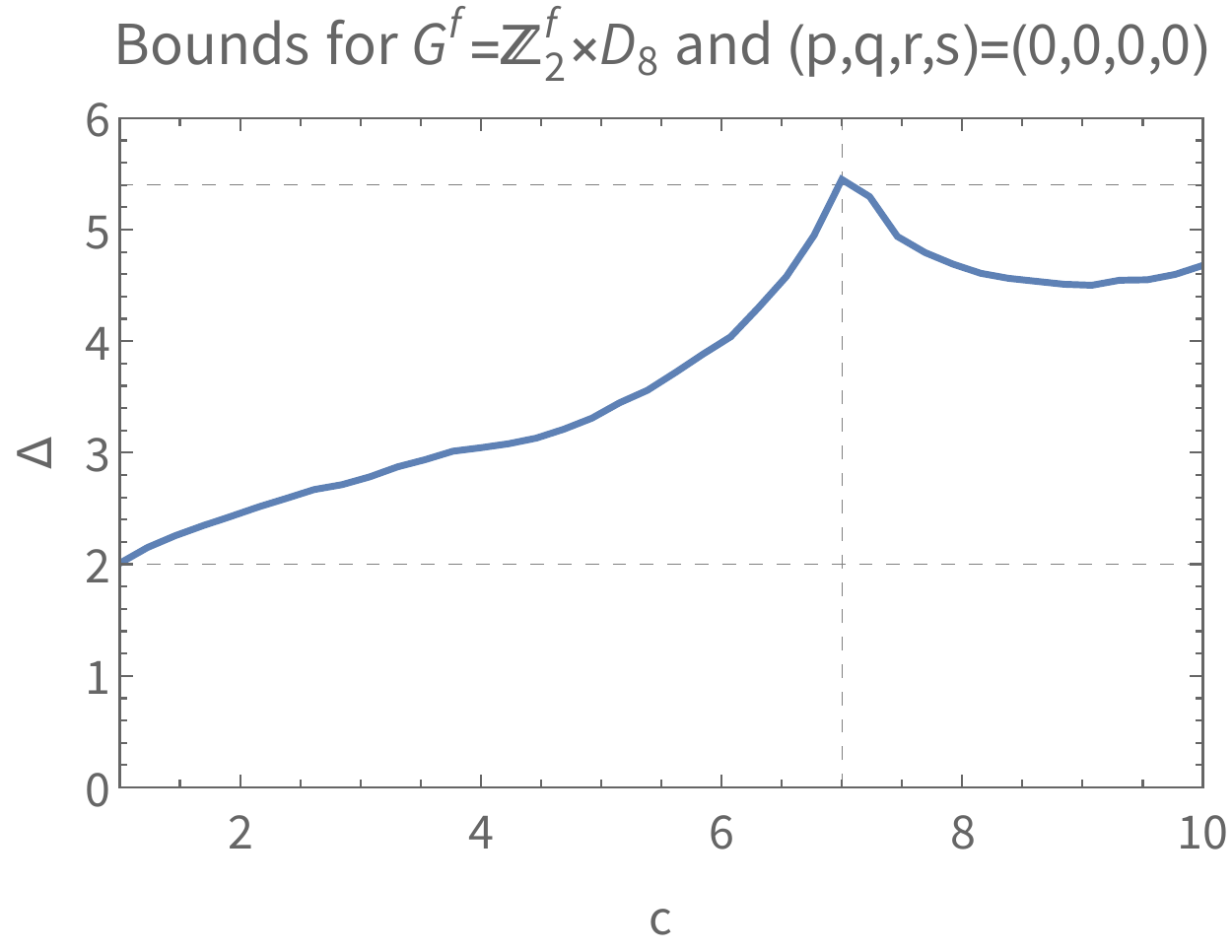}
    \qquad
    \includegraphics[scale=0.5]{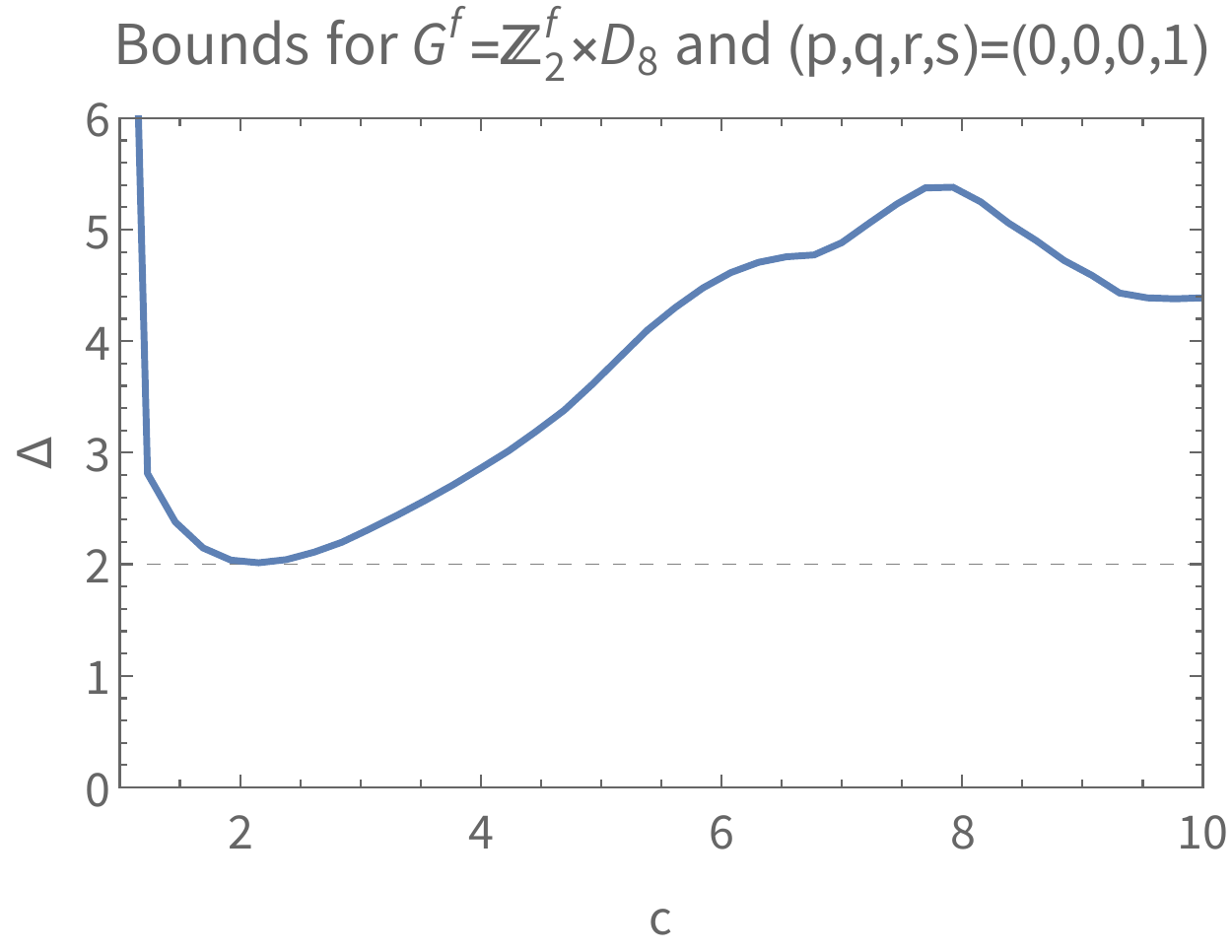}
    \caption{Upper bounds on the lightest $\Z^f_2\times D_8$-preserving scalar operators for anomalies $(p,q,r,t)=(0,0,0,0)$ (left) and $(p,q,r,t)=(0,0,0,1)$ (right).}
    \label{fig:d8v0000-0001}

\bigskip

    \includegraphics[scale=0.5]{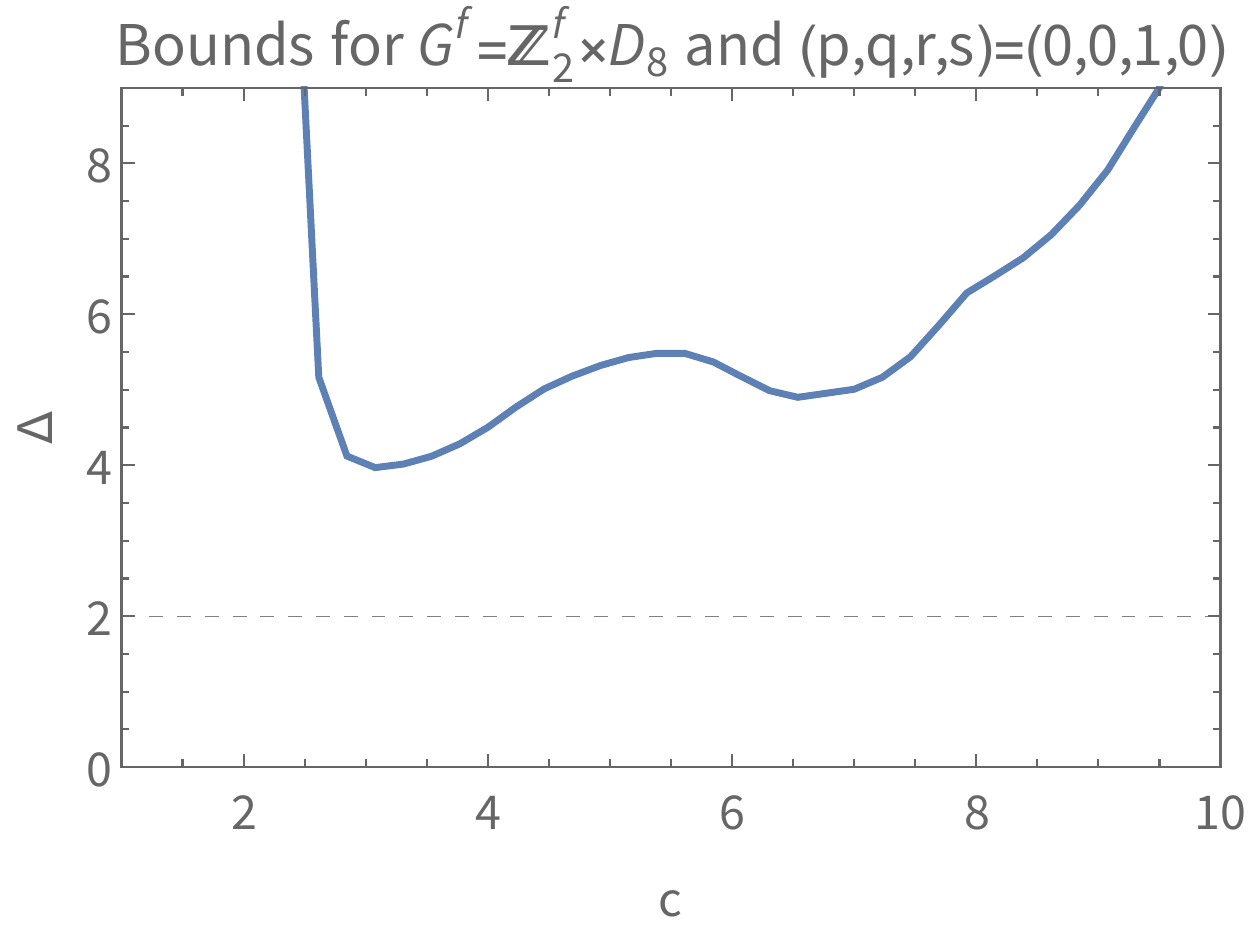}
    \qquad
    \includegraphics[scale=0.5]{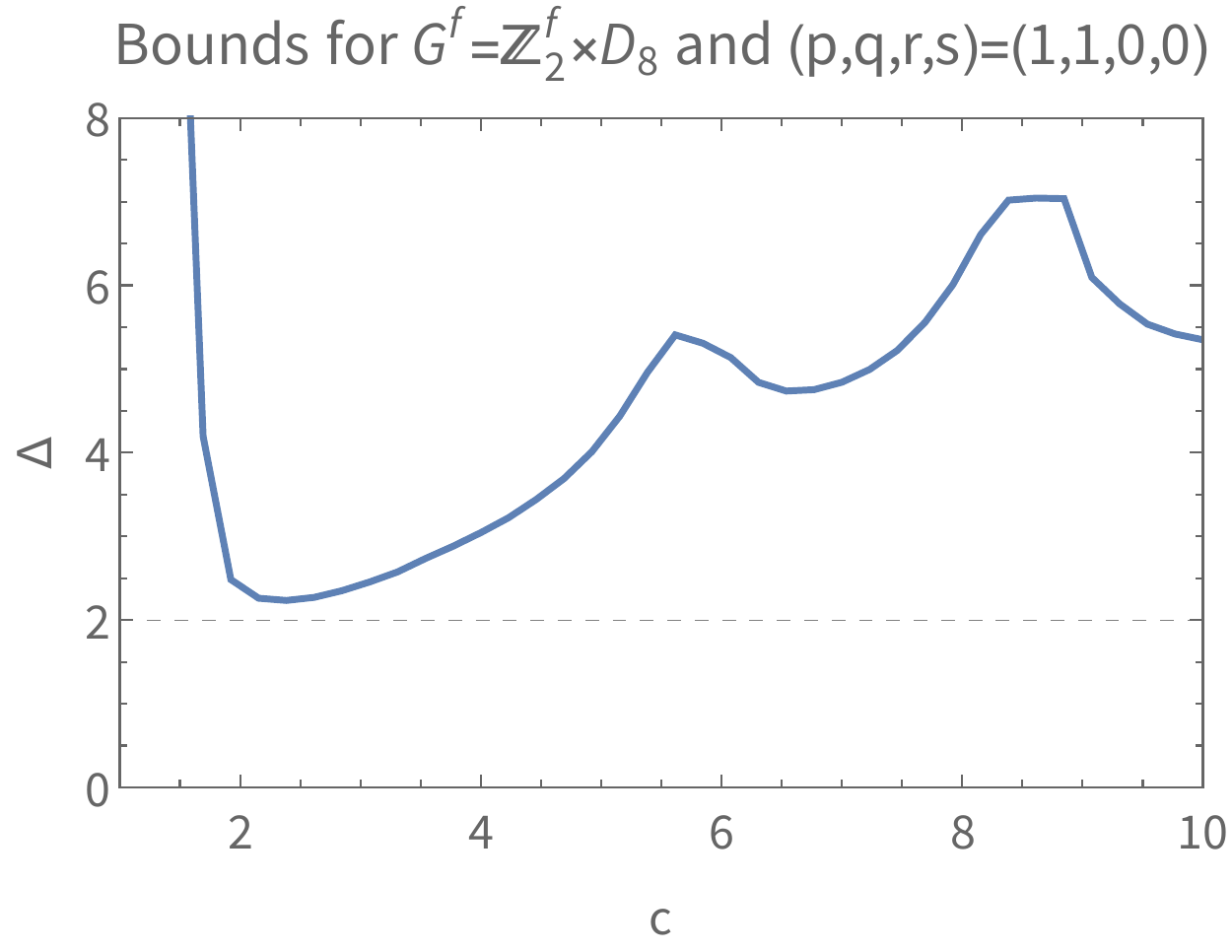}
    \caption{Upper bounds on the lightest $\Z^f_2\times D_8$-preserving scalar operators for anomalies $(p,q,r,t)=(0,0,1,0)$ (left) and $(p,q,r,t)=(1,1,0,0)$ (right).}
    \label{fig:d8v0010-1100}
\end{figure}

\FloatBarrier
\subsubsection*{Case 6: $G^f=\Z^f_8$}
No sharp kinks seem to appear for this symmetry group. Two slightly pronounced couple of valleys at 
\begin{align}
& \nu=0, && c\approx 4.0, \, 6.5,\label{rev-kink1}\\
& \nu=1, && c\approx 3.5, \, 6.0,
\label{rev-kink2}
\end{align}
are close to $\Delta_{\mathrm{mar}}$, suggesting that for functionals of higher order these points might actually describe some reversed kinks.
To this regard, a quick computation tells us that for a set of $4$ complex free fermions $\Psi_i,\,\Psi^*_i$ associated to the representation $4\crep{1}\in RO^f(\Z^f_8)$ on the left sector and $4\crep{3}$ on the right one, the lightest $\Z^f_8$-symmetry preserving scalar operators are associated to the states
\begin{equation}
    \Psi_i \Psi^*_i \overline{\Psi}_i \overline{\Psi}^*_i \left\vert 0\right\rangle, 
\end{equation}
thus describing a point $(c,\Delta)=(4,2)$ on the plot. However, no other similar descriptions have been founds for the other supposed reversed kinks \eqref{rev-kink1}-\eqref{rev-kink2}.

\begin{figure}[tb]
    \centering
    \includegraphics[scale=0.5]{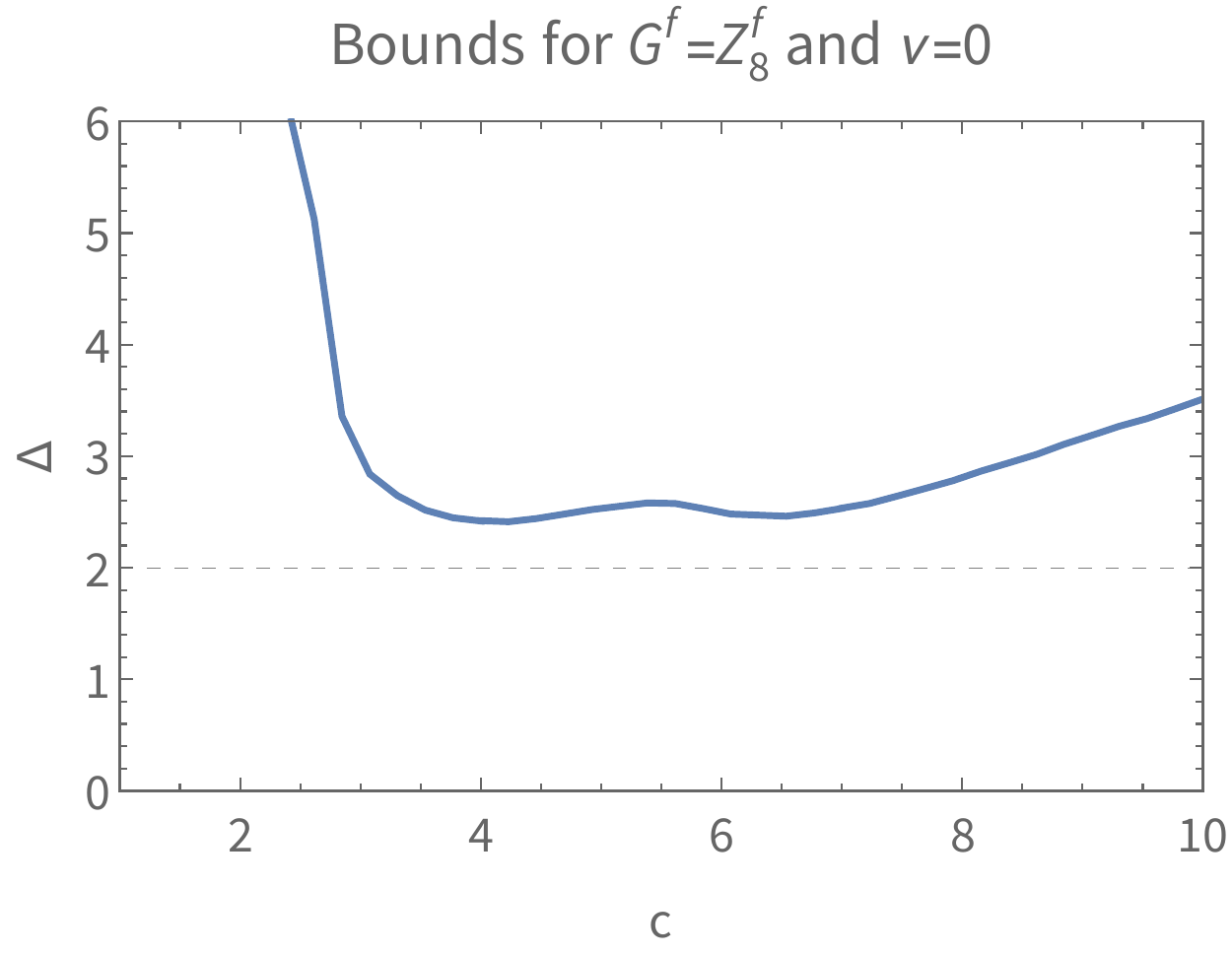}
    \qquad
    \includegraphics[scale=0.5]{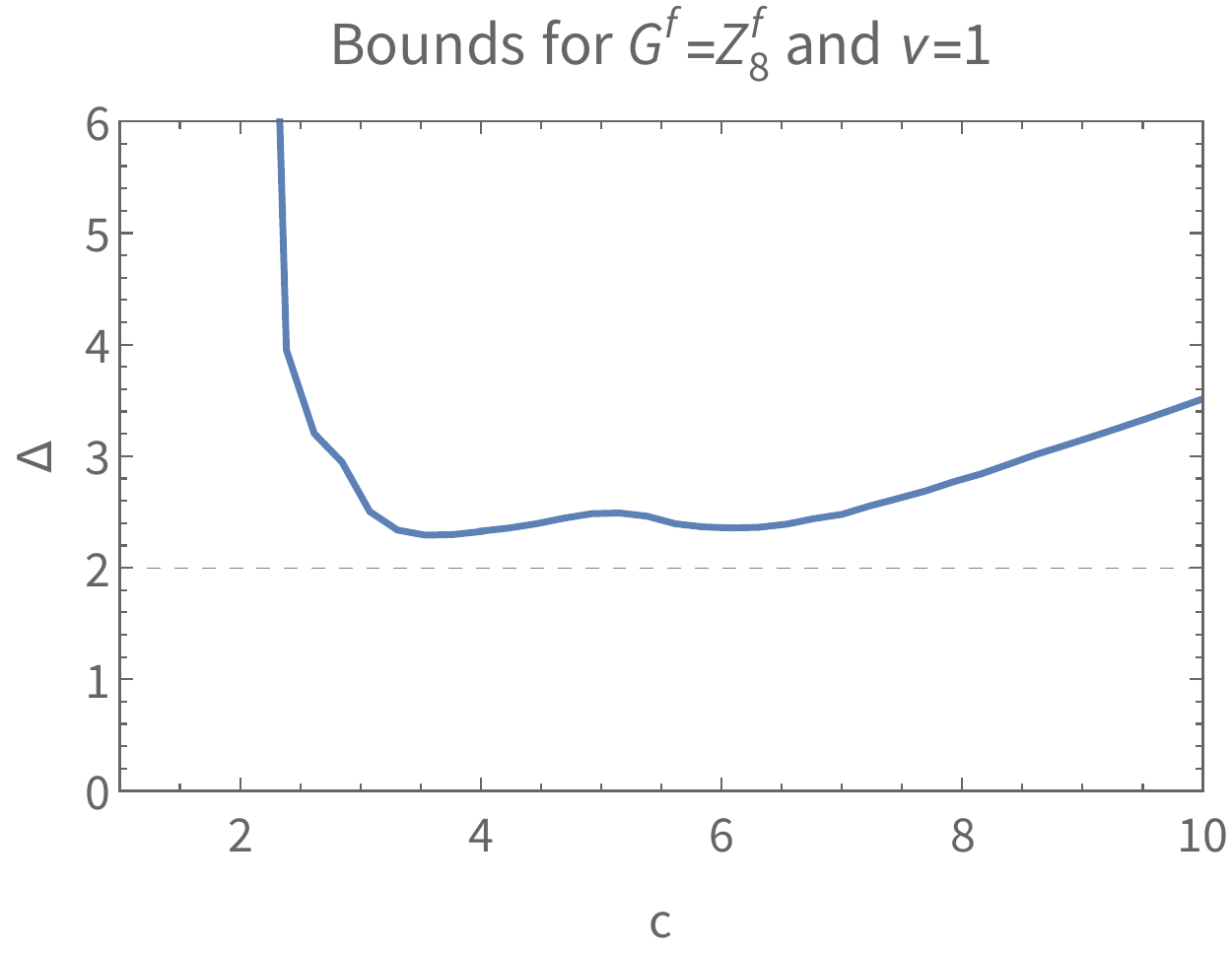}
    \caption{Upper bounds on the lightest $\Z^f_8$-preserving scalar operators for anomalies $\nu=0$ (left) and $\nu=1$ (right).}
    \label{fig:z8f}
\end{figure}

\section*{Acknowledgements}

The author would like to thank Pavel Putrov for his guidance and many useful discussions, as well as for comments on the early version of the draft. AG acknowledges support by the INFN Iniziativa Specifica ST\&FI.

\appendix
\section{Real character tables of groups}
\label{app:A}
Here we report the \emph{real}\footnote{For finite groups $G^f$ there are no indecomposable but reducible representations to worry about, since $\mathrm{char}(\mathbb{R}) \nmid |G^f|$.} character table of the groups needed for our analysis. In the non-abelian cases we focused on, we have $RO(G)\cong RU(G)$. Recall that for groups of the form $G^f=\Z^f_2\times G$ the character table of $RO(G^f)\cong RU(G^f)$ is just given by the tensor product with the character table of $G$ and $\Z^f_2$.

For $G=\Z_n$ the real irreducible representations are generally $2$-dimensional and represented by
\begin{equation}
    \crep{q}:k \longmapsto \left(\begin{array}{cc}
         \cos\left(2\pi \frac{qk}{n}\right) &-\sin\left(2\pi \frac{qk}{n}\right)  \\
         \sin \left(2\pi \frac{qk}{n}\right)& \cos\left(2\pi \frac{qk}{n}\right)
    \end{array}\right),
    \label{reps-abelian}
\end{equation}
from which one can easily read the character table. Here $q\in\Z_{[n/2]}$ and \eqref{reps-abelian} is reducible only when $n=2m$ and $q=m$, so that in that case it decomposes into $\crep{m}=2[m]$.

For $G=\Z_2\times\Z_2$, the irreps are labeled by $[n,m]=[n] \otimes [m]$, where $n,\,m\in\Z_2$ denotes the charges of the correspondent $\Z_2$ irreps. 

For the non-abelian cases, we report the character tables in Tables \ref{char:S3}, \ref{char:S4} and \ref{char:D8}.

Finally, the irreps $\Gamma=(\pm,\gamma)$ of $G^f=\Z^f_2\times G$ are given by the tensor product $[\pm]\otimes{\gamma}$ of irreps $\gamma$ of $G$ with the two irreps of $\Z^f_2$, that we will differentiate from the others by denoting them $[+]$ (the trivial representation) and $[-]$ (the sign representation).

\begin{table}[h]
\centering
\begin{tabular}{|c|c|c|c|}
        \hline    
        $S_3$&$\{()\}_1$&$\{(ab)\}_3$&$\{(abc)\}_2$\\
        \hline \hline
        $\rho_0$& 1 & 1 & 1\\
        \hline
        $\rho_a$& 1 & -1 & 1\\
        \hline
        $\rho_b$& 2 & 0 & -1\\
        \hline
    \end{tabular}
    \caption{Character table for the generators of $RO(S_3)\cong RU(S_3)$. Here the subscripts denote the size of the conjugacy classes.}    
    \label{char:S3}
    \end{table}    

    \begin{table}[h]
    \centering
    \begin{tabular}{|c|c|c|c|c|c|}
        \hline
        $S_4$&$\{()\}_1$&$\{(ab)(cd)\}_3$&$\{(abcd)\}_6$&$\{(abc)\}_8$&$\{(ab)\}_6$\\
        \hline \hline
        $\rho_0$& 1 & 1 & 1 & 1 & 1\\
        \hline
        $\rho_a$& 1 & 1 & -1 & 1 & -1\\
        \hline
        $\rho_b$& 2 & 2 & 0 & -1 & 0\\
        \hline
        $\rho_c$& 3 & -1 & -1 & 0 & 1\\
        \hline
        $\rho_d$& 3 & -1 & 1 & 0 & -1\\        
        \hline
        \end{tabular}
    \caption{Character table for the generators of $RO(S_4)\cong RU(S_4)$. Here the subscripts denote the size of the conjugacy classes.}         
    \label{char:S4}
    \end{table}    

\begin{table}[h]
    \centering
    \begin{tabular}{|c|c|c|c|c|c|}
    \hline
        $D_8$ & $\{(0,0)\}$ & $\{(2,0)\}$ & $\{(1,0),(3,0)\}$ & $\{(0,1),(2,1)\}$ & $\{(1,1),(3,1)\}$\\
        \hline \hline
        $\rho_0$& 1 & 1 & 1 & 1 & 1\\
        \hline
        $\rho_a$& 1 & 1 & -1 & 1 & -1\\
        \hline
        $\rho_b$& 1 & 1 & 1 & -1 & -1\\
        \hline
        $\rho_c$& 1 & 1 & -1 & -1 & 1\\
        \hline
        $\rho_d$& 2 & -2 & 0 & 0 & 0\\
        \hline
    \end{tabular}
    \caption{Character table for the generators of $RO(D_8)\cong RU(D_8)$.}     
    \label{char:D8}
\end{table}

\section{The reduced S matrices}
\label{app:B}

In the following we report the $S_{\mathrm{red}}$ matrix and the reduced partition vector ${\mathbf{Z}}^{\mathrm{red}}_{G^f}$ of the reduced systems used in the numerical computation. We are going to represent them in the form 
\begin{equation}
 \left(S_{\mathrm{red}}\middle\vert {\mathbf{Z}}^{\mathrm{red}}_{G^f}\right). 
\end{equation}
Here we adopt the following notation for the sets $\{Z_{C_{G^f}}\}$:
\begin{itemize}
    \item If $G^f=\Z^f_2\times G$, the conjugacy classes correspond to single elements of $G^f$. Thus we identify respectively with $Z_{+,n}$ and $Z_{-,n}$ the elements $(0,n)$ and $(1,n)$.
    \item If $G^f=\Z^f_2\times G$ with $G$ non-abelian, the conjugacy classes of $G^f$ are $\{(0,C_G)\}$ and $\{(1,C_G)\}$. Here we are going to adopt the notation $Z_{+,i}$ for the first and $Z_{-,i}$ for the latter, where $i=0,1,2,\ldots$ denote the $(i-1)$-esimal conjugacy class represented in the Tables of Appendix \ref{app:A}.
    \item For $G^f=\Z^f_8$, $Z^\crep{n}$ corresponds to the trace over $\CH_0$ with the insertion of the projector of charge $\hat{n\mod 8}$.
\end{itemize}

\subsubsection*{Case 1: $G^f=\Z^f_2\times\Z_4$}
{
\begin{equation}
\left(
\begin{array}{ccccccccccc}
 \frac{1}{8} & \frac{1}{8} & \frac{1}{8} & \frac{1}{8} & \frac{1}{8} & \frac{1}{8} & \frac{1}{8} & \frac{1}{8} & \frac{1}{8} & \frac{1}{8} & \frac{1}{8} \\
 \frac{1}{8} & \frac{1}{8} & \frac{1}{8} & \frac{1}{8} & \frac{1}{8} & \frac{1}{8} & -\frac{1}{8} & \frac{1}{8} & -\frac{1}{8} & \frac{1}{8} & -\frac{1}{8} \\
 \frac{1}{8} & \frac{1}{8} & \frac{1}{8} & \frac{1}{8} & \frac{1}{8} & \frac{1}{8} & \frac{1}{8} & \frac{1}{8} & \frac{1}{8} & -\frac{1}{8} & -\frac{1}{8} \\
 \frac{1}{8} & \frac{1}{8} & \frac{1}{8} & \frac{1}{8} & \frac{1}{8} & \frac{1}{8} & -\frac{1}{8} & \frac{1}{8} & -\frac{1}{8} & -\frac{1}{8} & \frac{1}{8} \\
 \frac{1}{4} & \frac{1}{4} & \frac{1}{4} & \frac{1}{4} & \frac{1}{4} & \frac{1}{4} & \frac{1}{4} & -\frac{1}{4} & -\frac{1}{4} & 0 & 0 \\
 \frac{1}{4} & \frac{1}{4} & \frac{1}{4} & \frac{1}{4} & \frac{1}{4} & \frac{1}{4} & -\frac{1}{4} & -\frac{1}{4} & \frac{1}{4} & 0 & 0 \\
 1 & -1 & 1 & -1 & 1 & -1 & 0 & 0 & 0 & 0 & 0 \\
 1 & 1 & 1 & 1 & -1 & -1 & 0 & 0 & 0 & 0 & 0 \\
 1 & -1 & 1 & -1 & -1 & 1 & 0 & 0 & 0 & 0 & 0 \\
 2 & 2 & -2 & -2 & 0 & 0 & 0 & 0 & 0 & 0 & 0 \\
 2 & -2 & -2 & 2 & 0 & 0 & 0 & 0 & 0 & 0 & 0 \\
\end{array}
\middle\vert
\begin{array}{c}
        Z^{(+,[0])}\\
        Z^{(-,[0])}\\
        Z^{(+,[2])}\\
        Z^{(-,[2])}\\
        Z^{(+,\crep{1})}+Z^{(+,\crep{3})}\\
        Z^{(-,\crep{1})}+Z^{(-,\crep{3})}\\
        Z_{-,0}\\
        Z_{+,2}\\
        Z_{-,2}\\
        Z_{+,1}+Z_{+,3}\\
        Z_{-,1}+Z_{-,3}\\        
    \end{array}\right)
\end{equation}
}

\subsubsection*{Case 2: $G^f=\Z^f_2\times\Z_2\times\Z_2$}
{
\begin{equation}
\left(
\begin{array}{ccccccccccccccc}
 \frac{1}{8} & \frac{1}{8} & \frac{1}{8} & \frac{1}{8} & \frac{1}{8} & \frac{1}{8} & \frac{1}{8} & \frac{1}{8} & \frac{1}{8} & \frac{1}{8} & \frac{1}{8} & \frac{1}{8} & \frac{1}{8} & \frac{1}{8} & \frac{1}{8} \\
 \frac{1}{8} & \frac{1}{8} & \frac{1}{8} & \frac{1}{8} & \frac{1}{8} & \frac{1}{8} & \frac{1}{8} & \frac{1}{8} & -\frac{1}{8} & \frac{1}{8} & -\frac{1}{8} & \frac{1}{8} & -\frac{1}{8} & \frac{1}{8} & -\frac{1}{8} \\
 \frac{1}{8} & \frac{1}{8} & \frac{1}{8} & \frac{1}{8} & \frac{1}{8} & \frac{1}{8} & \frac{1}{8} & \frac{1}{8} & \frac{1}{8} & -\frac{1}{8} & -\frac{1}{8} & \frac{1}{8} & \frac{1}{8} & -\frac{1}{8} & -\frac{1}{8} \\
 \frac{1}{8} & \frac{1}{8} & \frac{1}{8} & \frac{1}{8} & \frac{1}{8} & \frac{1}{8} & \frac{1}{8} & \frac{1}{8} & -\frac{1}{8} & -\frac{1}{8} & \frac{1}{8} & \frac{1}{8} & -\frac{1}{8} & -\frac{1}{8} & \frac{1}{8} \\
 \frac{1}{8} & \frac{1}{8} & \frac{1}{8} & \frac{1}{8} & \frac{1}{8} & \frac{1}{8} & \frac{1}{8} & \frac{1}{8} & \frac{1}{8} & \frac{1}{8} & \frac{1}{8} & -\frac{1}{8} & -\frac{1}{8} & -\frac{1}{8} & -\frac{1}{8} \\
 \frac{1}{8} & \frac{1}{8} & \frac{1}{8} & \frac{1}{8} & \frac{1}{8} & \frac{1}{8} & \frac{1}{8} & \frac{1}{8} & -\frac{1}{8} & \frac{1}{8} & -\frac{1}{8} & -\frac{1}{8} & \frac{1}{8} & -\frac{1}{8} & \frac{1}{8} \\
 \frac{1}{8} & \frac{1}{8} & \frac{1}{8} & \frac{1}{8} & \frac{1}{8} & \frac{1}{8} & \frac{1}{8} & \frac{1}{8} & \frac{1}{8} & -\frac{1}{8} & -\frac{1}{8} & -\frac{1}{8} & -\frac{1}{8} & \frac{1}{8} & \frac{1}{8} \\
 \frac{1}{8} & \frac{1}{8} & \frac{1}{8} & \frac{1}{8} & \frac{1}{8} & \frac{1}{8} & \frac{1}{8} & \frac{1}{8} & -\frac{1}{8} & -\frac{1}{8} & \frac{1}{8} & -\frac{1}{8} & \frac{1}{8} & \frac{1}{8} & -\frac{1}{8} \\
 1 & -1 & 1 & -1 & 1 & -1 & 1 & -1 & 0 & 0 & 0 & 0 & 0 & 0 & 0 \\
 1 & 1 & -1 & -1 & 1 & 1 & -1 & -1 & 0 & 0 & 0 & 0 & 0 & 0 & 0 \\
 1 & -1 & -1 & 1 & 1 & -1 & -1 & 1 & 0 & 0 & 0 & 0 & 0 & 0 & 0 \\
 1 & 1 & 1 & 1 & -1 & -1 & -1 & -1 & 0 & 0 & 0 & 0 & 0 & 0 & 0 \\
 1 & -1 & 1 & -1 & -1 & 1 & -1 & 1 & 0 & 0 & 0 & 0 & 0 & 0 & 0 \\
 1 & 1 & -1 & -1 & -1 & -1 & 1 & 1 & 0 & 0 & 0 & 0 & 0 & 0 & 0 \\
 1 & -1 & -1 & 1 & -1 & 1 & 1 & -1 & 0 & 0 & 0 & 0 & 0 & 0 & 0 \\
\end{array}
\middle\vert
\begin{array}{c}
        Z^{(+,[0,0])}\\
        Z^{(-,[0,0])}\\
        Z^{(+,[0,1])}\\
        Z^{(-,[0,1])}\\
        Z^{(+,[1,0])}\\
        Z^{(-,[1,0])}\\
        Z^{(+,[1,1])}\\
        Z^{(-,[1,1])}\\ 
        Z_{-,0,0}\\
        Z_{+,0,1}\\
        Z_{-,0,1}\\
        Z_{+,1,0}\\
        Z_{-,1,0}\\
        Z_{+,{1,1}}\\
        Z_{-,{1,1}}\\        
\end{array}\right)
\end{equation}
}

\subsubsection*{Case 3: $G^f=\Z^f_2\times S_3$}
{
\begin{equation}
\left(
\begin{array}{ccccccccccc}
 \frac{1}{12} & \frac{1}{12} & \frac{1}{12} & \frac{1}{12} & \frac{1}{12} & \frac{1}{12} & \frac{1}{12} & \frac{1}{12} & \frac{1}{12} & \frac{1}{12} & \frac{1}{12} \\
 \frac{1}{12} & \frac{1}{12} & \frac{1}{12} & \frac{1}{12} & \frac{1}{12} & \frac{1}{12} & -\frac{1}{12} & \frac{1}{12} & -\frac{1}{12} & \frac{1}{12} & -\frac{1}{12} \\
 \frac{1}{12} & \frac{1}{12} & \frac{1}{12} & \frac{1}{12} & \frac{1}{12} & \frac{1}{12} & \frac{1}{12} & -\frac{1}{12} & -\frac{1}{12} & \frac{1}{12} & \frac{1}{12} \\
 \frac{1}{12} & \frac{1}{12} & \frac{1}{12} & \frac{1}{12} & \frac{1}{12} & \frac{1}{12} & -\frac{1}{12} & -\frac{1}{12} & \frac{1}{12} & \frac{1}{12} & -\frac{1}{12} \\
 \frac{1}{3} & \frac{1}{3} & \frac{1}{3} & \frac{1}{3} & \frac{1}{3} & \frac{1}{3} & \frac{1}{3} & 0 & 0 & -\frac{1}{6} & -\frac{1}{6} \\
 \frac{1}{3} & \frac{1}{3} & \frac{1}{3} & \frac{1}{3} & \frac{1}{3} & \frac{1}{3} & -\frac{1}{3} & 0 & 0 & -\frac{1}{6} & \frac{1}{6} \\
 1 & -1 & 1 & -1 & 1 & -1 & 0 & 0 & 0 & 0 & 0 \\
 3 & 3 & -3 & -3 & 0 & 0 & 0 & 0 & 0 & 0 & 0 \\
 3 & -3 & -3 & 3 & 0 & 0 & 0 & 0 & 0 & 0 & 0 \\
 2 & 2 & 2 & 2 & -1 & -1 & 0 & 0 & 0 & 0 & 0 \\
 2 & -2 & 2 & -2 & -1 & 1 & 0 & 0 & 0 & 0 & 0 \\
\end{array}
\middle\vert
\begin{array}{c}
        Z^{(+,\rho_0)}\\
        Z^{(-,\rho_0)}\\
        Z^{(+,\rho_a)}\\
        Z^{(-,\rho_a)}\\
        Z^{(+,\rho_b)}\\
        Z^{(-,\rho_b)}\\
        Z_{-,0}\\
        Z_{+,1}\\
        Z_{-,1}\\
        Z_{+,2}\\
        Z_{-,2}\\        
\end{array}\right)
\end{equation}
}

\subsubsection*{Case 4: $G^f=\Z^f_2\times S_4$}
{
\begin{equation}
\left(
\begin{array}{ccccccccccccccccccc}
 \frac{1}{48} & \frac{1}{48} & \frac{1}{48} & \frac{1}{48} & \frac{1}{48} & \frac{1}{48} & \frac{1}{48} & \frac{1}{48} & \frac{1}{48} & \frac{1}{48} & \frac{1}{48} & \frac{1}{48} & \frac{1}{48} & \frac{1}{48} & \frac{1}{48} & \frac{1}{48} & \frac{1}{48} & \frac{1}{48} & \frac{1}{48} \\
 \frac{1}{48} & \frac{1}{48} & \frac{1}{48} & \frac{1}{48} & \frac{1}{48} & \frac{1}{48} & \frac{1}{48} & \frac{1}{48} & \frac{1}{48} & \frac{1}{48} & -\frac{1}{48} & \frac{1}{48} & -\frac{1}{48} & \frac{1}{48} & -\frac{1}{48} & \frac{1}{48} & -\frac{1}{48} & \frac{1}{48} & -\frac{1}{48} \\
 \frac{1}{48} & \frac{1}{48} & \frac{1}{48} & \frac{1}{48} & \frac{1}{48} & \frac{1}{48} & \frac{1}{48} & \frac{1}{48} & \frac{1}{48} & \frac{1}{48} & \frac{1}{48} & \frac{1}{48} & \frac{1}{48} & -\frac{1}{48} & -\frac{1}{48} & \frac{1}{48} & \frac{1}{48} & -\frac{1}{48} & -\frac{1}{48} \\
 \frac{1}{48} & \frac{1}{48} & \frac{1}{48} & \frac{1}{48} & \frac{1}{48} & \frac{1}{48} & \frac{1}{48} & \frac{1}{48} & \frac{1}{48} & \frac{1}{48} & -\frac{1}{48} & \frac{1}{48} & -\frac{1}{48} & -\frac{1}{48} & \frac{1}{48} & \frac{1}{48} & -\frac{1}{48} & -\frac{1}{48} & \frac{1}{48} \\
 \frac{1}{12} & \frac{1}{12} & \frac{1}{12} & \frac{1}{12} & \frac{1}{12} & \frac{1}{12} & \frac{1}{12} & \frac{1}{12} & \frac{1}{12} & \frac{1}{12} & \frac{1}{12} & \frac{1}{12} & \frac{1}{12} & 0 & 0 & -\frac{1}{24} & -\frac{1}{24} & 0 & 0 \\
 \frac{1}{12} & \frac{1}{12} & \frac{1}{12} & \frac{1}{12} & \frac{1}{12} & \frac{1}{12} & \frac{1}{12} & \frac{1}{12} & \frac{1}{12} & \frac{1}{12} & -\frac{1}{12} & \frac{1}{12} & -\frac{1}{12} & 0 & 0 & -\frac{1}{24} & \frac{1}{24} & 0 & 0 \\
 \frac{3}{16} & \frac{3}{16} & \frac{3}{16} & \frac{3}{16} & \frac{3}{16} & \frac{3}{16} & \frac{3}{16} & \frac{3}{16} & \frac{3}{16} & \frac{3}{16} & \frac{3}{16} & -\frac{1}{16} & -\frac{1}{16} & -\frac{1}{16} & -\frac{1}{16} & 0 & 0 & \frac{1}{16} & \frac{1}{16} \\
 \frac{3}{16} & \frac{3}{16} & \frac{3}{16} & \frac{3}{16} & \frac{3}{16} & \frac{3}{16} & \frac{3}{16} & \frac{3}{16} & \frac{3}{16} & \frac{3}{16} & -\frac{3}{16} & -\frac{1}{16} & \frac{1}{16} & -\frac{1}{16} & \frac{1}{16} & 0 & 0 & \frac{1}{16} & -\frac{1}{16} \\
 \frac{3}{16} & \frac{3}{16} & \frac{3}{16} & \frac{3}{16} & \frac{3}{16} & \frac{3}{16} & \frac{3}{16} & \frac{3}{16} & \frac{3}{16} & \frac{3}{16} & \frac{3}{16} & -\frac{1}{16} & -\frac{1}{16} & \frac{1}{16} & \frac{1}{16} & 0 & 0 & -\frac{1}{16} & -\frac{1}{16} \\
 \frac{3}{16} & \frac{3}{16} & \frac{3}{16} & \frac{3}{16} & \frac{3}{16} & \frac{3}{16} & \frac{3}{16} & \frac{3}{16} & \frac{3}{16} & \frac{3}{16} & -\frac{3}{16} & -\frac{1}{16} & \frac{1}{16} & \frac{1}{16} & -\frac{1}{16} & 0 & 0 & -\frac{1}{16} & \frac{1}{16} \\
 1 & -1 & 1 & -1 & 1 & -1 & 1 & -1 & 1 & -1 & 0 & 0 & 0 & 0 & 0 & 0 & 0 & 0 & 0 \\
 3 & 3 & 3 & 3 & 3 & 3 & -1 & -1 & -1 & -1 & 0 & 0 & 0 & 0 & 0 & 0 & 0 & 0 & 0 \\
 3 & -3 & 3 & -3 & 3 & -3 & -1 & 1 & -1 & 1 & 0 & 0 & 0 & 0 & 0 & 0 & 0 & 0 & 0 \\
 6 & 6 & -6 & -6 & 0 & 0 & -2 & -2 & 2 & 2 & 0 & 0 & 0 & 0 & 0 & 0 & 0 & 0 & 0 \\
 6 & -6 & -6 & 6 & 0 & 0 & -2 & 2 & 2 & -2 & 0 & 0 & 0 & 0 & 0 & 0 & 0 & 0 & 0 \\
 8 & 8 & 8 & 8 & -4 & -4 & 0 & 0 & 0 & 0 & 0 & 0 & 0 & 0 & 0 & 0 & 0 & 0 & 0 \\
 8 & -8 & 8 & -8 & -4 & 4 & 0 & 0 & 0 & 0 & 0 & 0 & 0 & 0 & 0 & 0 & 0 & 0 & 0 \\
 6 & 6 & -6 & -6 & 0 & 0 & 2 & 2 & -2 & -2 & 0 & 0 & 0 & 0 & 0 & 0 & 0 & 0 & 0 \\
 6 & -6 & -6 & 6 & 0 & 0 & 2 & -2 & -2 & 2 & 0 & 0 & 0 & 0 & 0 & 0 & 0 & 0 & 0 \\
\end{array}
\middle\vert
\begin{array}{c}
        Z^{(+,\rho_0)}\\
        Z^{(-,\rho_0)}\\
        Z^{(+,\rho_a)}\\
        Z^{(-,\rho_a)}\\
        Z^{(+,\rho_b)}\\
        Z^{(-,\rho_b)}\\
        Z^{(+,\rho_c)}\\
        Z^{(-,\rho_c)}\\
        Z^{(+,\rho_d)}\\
        Z^{(-,\rho_d)}\\        
        Z_{-,0}\\
        Z_{+,1}\\
        Z_{-,1}\\
        Z_{+,2}\\
        Z_{-,2}\\
        Z_{+,3}\\
        Z_{-,3}\\
        Z_{+,4}\\
        Z_{-,4}\\
\end{array}\right)
\end{equation}
}

\subsubsection*{Case 5: $G^f=\Z^f_2\times D_8$}
{
\begin{equation}
\left(
\begin{array}{ccccccccccccccccccc}
 \frac{1}{16} & \frac{1}{16} & \frac{1}{16} & \frac{1}{16} & \frac{1}{16} & \frac{1}{16} & \frac{1}{16} & \frac{1}{16} & \frac{1}{16} & \frac{1}{16} & \frac{1}{16} & \frac{1}{16} & \frac{1}{16} & \frac{1}{16} & \frac{1}{16} & \frac{1}{16} & \frac{1}{16} & \frac{1}{16} & \frac{1}{16} \\
 \frac{1}{16} & \frac{1}{16} & \frac{1}{16} & \frac{1}{16} & \frac{1}{16} & \frac{1}{16} & \frac{1}{16} & \frac{1}{16} & \frac{1}{16} & \frac{1}{16} & -\frac{1}{16} & \frac{1}{16} & -\frac{1}{16} & \frac{1}{16} & -\frac{1}{16} & \frac{1}{16} & -\frac{1}{16} & \frac{1}{16} & -\frac{1}{16} \\
 \frac{1}{16} & \frac{1}{16} & \frac{1}{16} & \frac{1}{16} & \frac{1}{16} & \frac{1}{16} & \frac{1}{16} & \frac{1}{16} & \frac{1}{16} & \frac{1}{16} & \frac{1}{16} & \frac{1}{16} & \frac{1}{16} & -\frac{1}{16} & -\frac{1}{16} & \frac{1}{16} & \frac{1}{16} & -\frac{1}{16} & -\frac{1}{16} \\
 \frac{1}{16} & \frac{1}{16} & \frac{1}{16} & \frac{1}{16} & \frac{1}{16} & \frac{1}{16} & \frac{1}{16} & \frac{1}{16} & \frac{1}{16} & \frac{1}{16} & -\frac{1}{16} & \frac{1}{16} & -\frac{1}{16} & -\frac{1}{16} & \frac{1}{16} & \frac{1}{16} & -\frac{1}{16} & -\frac{1}{16} & \frac{1}{16} \\
 \frac{1}{16} & \frac{1}{16} & \frac{1}{16} & \frac{1}{16} & \frac{1}{16} & \frac{1}{16} & \frac{1}{16} & \frac{1}{16} & \frac{1}{16} & \frac{1}{16} & \frac{1}{16} & \frac{1}{16} & \frac{1}{16} & \frac{1}{16} & \frac{1}{16} & -\frac{1}{16} & -\frac{1}{16} & -\frac{1}{16} & -\frac{1}{16} \\
 \frac{1}{16} & \frac{1}{16} & \frac{1}{16} & \frac{1}{16} & \frac{1}{16} & \frac{1}{16} & \frac{1}{16} & \frac{1}{16} & \frac{1}{16} & \frac{1}{16} & -\frac{1}{16} & \frac{1}{16} & -\frac{1}{16} & \frac{1}{16} & -\frac{1}{16} & -\frac{1}{16} & \frac{1}{16} & -\frac{1}{16} & \frac{1}{16} \\
 \frac{1}{16} & \frac{1}{16} & \frac{1}{16} & \frac{1}{16} & \frac{1}{16} & \frac{1}{16} & \frac{1}{16} & \frac{1}{16} & \frac{1}{16} & \frac{1}{16} & \frac{1}{16} & \frac{1}{16} & \frac{1}{16} & -\frac{1}{16} & -\frac{1}{16} & -\frac{1}{16} & -\frac{1}{16} & \frac{1}{16} & \frac{1}{16} \\
 \frac{1}{16} & \frac{1}{16} & \frac{1}{16} & \frac{1}{16} & \frac{1}{16} & \frac{1}{16} & \frac{1}{16} & \frac{1}{16} & \frac{1}{16} & \frac{1}{16} & -\frac{1}{16} & \frac{1}{16} & -\frac{1}{16} & -\frac{1}{16} & \frac{1}{16} & -\frac{1}{16} & \frac{1}{16} & \frac{1}{16} & -\frac{1}{16} \\
 \frac{1}{4} & \frac{1}{4} & \frac{1}{4} & \frac{1}{4} & \frac{1}{4} & \frac{1}{4} & \frac{1}{4} & \frac{1}{4} & \frac{1}{4} & \frac{1}{4} & \frac{1}{4} & -\frac{1}{4} & -\frac{1}{4} & 0 & 0 & 0 & 0 & 0 & 0 \\
 \frac{1}{4} & \frac{1}{4} & \frac{1}{4} & \frac{1}{4} & \frac{1}{4} & \frac{1}{4} & \frac{1}{4} & \frac{1}{4} & \frac{1}{4} & \frac{1}{4} & -\frac{1}{4} & -\frac{1}{4} & \frac{1}{4} & 0 & 0 & 0 & 0 & 0 & 0 \\
 1 & -1 & 1 & -1 & 1 & -1 & 1 & -1 & 1 & -1 & 0 & 0 & 0 & 0 & 0 & 0 & 0 & 0 & 0 \\
 1 & 1 & 1 & 1 & 1 & 1 & 1 & 1 & -1 & -1 & 0 & 0 & 0 & 0 & 0 & 0 & 0 & 0 & 0 \\
 1 & -1 & 1 & -1 & 1 & -1 & 1 & -1 & -1 & 1 & 0 & 0 & 0 & 0 & 0 & 0 & 0 & 0 & 0 \\
 2 & 2 & -2 & -2 & 2 & 2 & -2 & -2 & 0 & 0 & 0 & 0 & 0 & 0 & 0 & 0 & 0 & 0 & 0 \\
 2 & -2 & -2 & 2 & 2 & -2 & -2 & 2 & 0 & 0 & 0 & 0 & 0 & 0 & 0 & 0 & 0 & 0 & 0 \\
 2 & 2 & 2 & 2 & -2 & -2 & -2 & -2 & 0 & 0 & 0 & 0 & 0 & 0 & 0 & 0 & 0 & 0 & 0 \\
 2 & -2 & 2 & -2 & -2 & 2 & -2 & 2 & 0 & 0 & 0 & 0 & 0 & 0 & 0 & 0 & 0 & 0 & 0 \\
 2 & 2 & -2 & -2 & -2 & -2 & 2 & 2 & 0 & 0 & 0 & 0 & 0 & 0 & 0 & 0 & 0 & 0 & 0 \\
 2 & -2 & -2 & 2 & -2 & 2 & 2 & -2 & 0 & 0 & 0 & 0 & 0 & 0 & 0 & 0 & 0 & 0 & 0 \\
\end{array}
\middle\vert
\begin{array}{c}
        Z^{(+,[0])}\\
        Z^{(-,[0])}\\
        Z^{(+,\rho_a)}\\
        Z^{(-,\rho_a)}\\
        Z^{(+,\rho_b)}\\
        Z^{(-,\rho_b)}\\
        Z^{(+,\rho_c)}\\
        Z^{(-,\rho_c)}\\
        Z^{(+,\rho_d)}\\
        Z^{(-,\rho_d)}\\        
        Z_{-,0}\\
        Z_{+,1}\\
        Z_{-,1}\\
        Z_{+,2}\\
        Z_{-,2}\\
        Z_{+,3}\\
        Z_{-,3}\\
        Z_{+,4}\\
        Z_{-,4}\\
\end{array}\right)
\end{equation}
}

\subsubsection*{Case 6: $G^f=\Z^f_8$}
\begin{equation}
\left(
\begin{array}{ccccccc}
 \frac{1}{8} & \frac{1}{8} & \frac{1}{8} & \frac{1}{8} & \frac{1}{8} & \frac{1}{8} & \frac{1}{8} \\
 \frac{1}{2} & \frac{1}{2} & \frac{1}{2} & \frac{1}{2} & 0 & 0 & -\frac{1}{2} \\
 \frac{1}{4} & \frac{1}{4} & \frac{1}{4} & \frac{1}{4} & 0 & -\frac{1}{4} & \frac{1}{4} \\
 \frac{1}{8} & \frac{1}{8} & \frac{1}{8} & \frac{1}{8} & -\frac{1}{8} & \frac{1}{8} & \frac{1}{8} \\
 4 & 0 & 0 & -4 & 0 & 0 & 0 \\
 2 & 0 & -2 & 2 & 0 & 0 & 0 \\
 1 & -1 & 1 & 1 & 0 & 0 & 0 \\
\end{array}
\middle\vert
\begin{array}{c}
        Z^{[0]}\\
        Z^\crep{1}+Z^\crep{3}+Z^\crep{5}+Z^\crep{7}\\
        Z^{\crep{2}}+Z^\crep{6}\\
        Z^\crep{4}\\
        Z_1+Z_3+Z_5+Z_7\\
        Z_2+Z_6\\
        Z_4
\end{array}\right)
\end{equation}

\clearpage

\bibliography{refr}
\bibliographystyle{JHEP}

\end{document}